\documentclass[iop,apj,twocolappendix,numberedappendix,appendixfloats]{emulateapj}
\usepackage{color}
\usepackage[export]{adjustbox}
\usepackage{scrextend}
\usepackage[colorlinks,citecolor=blue]{hyperref}

\newcommand{\ha}{\rm\,H\alpha}
\newcommand{\fhawin}{\rm\,f_{H\alpha,WIN}(x,y)}
\newcommand{\fhawincor}{\rm\,f_{H\alpha,WINcor}(x,y)}

\newcommand{\Ks}{\rm\,Ks}
\newcommand{\ktd}{\rm\,KMOS^{3D}}
\newcommand{\zrange}{\rm\,0.7 \lesssim z \lesssim 2.7}
\newcommand{\Msol}{\rm\,M_{\odot}} 
\newcommand{\Mstellar}{M_{*}}

\newcommand{\Mpc}{\rm\,Mpc} 
\newcommand{\kpc}{\rm\,kpc}

\newcommand{\kmsMpc}{\rm\,km\,s^{-1}\,Mpc^{-1}}
\newcommand{\kms}{\rm\,km\,s^{-1}}

\newcommand{\blue}{\textcolor{blue}}

\newcommand{\imfit}{{\sc imfit}}
\newcommand{\galfit}{{\sc galfit}}
\newcommand{\reff}{\rm\,r_{e}}
\newcommand{\reffcirc}{\rm\,r_{e,circ}}

\newcommand{\reffH}{\rm\,r_{e(F160W)}}

\newcommand{\reffcorr}{\rm\,r_{e({\rm r}6500)}}

\newcommand{\sigmastatreff}{\rm\,\sigma_{stat} (r_{e})}
\newcommand{\sigmareff}{\rm\,\sigma (r_{e})}
\newcommand{\sigmastatreffcirc}{\rm\,\sigma_{stat} (r_{e,circ})}
\newcommand{\sigmareffcirc}{\rm\,\sigma (r_{e,circ})}

\newcommand{\reffha}{\rm\,r_{e(H\alpha)}}
\newcommand{\reffsf}{\rm\,r_{e(SF)}}
\newcommand{\reffmstar}{\rm\,r_{e(M_*)}}
\newcommand{\nsers}{\rm\,n_{Sersic}}
\newcommand{\fwhmpsf}{\rm\,FWHM_{PSF,major}}
\newcommand{\fwhmpsfcirc}{\rm\,FWHM_{PSF,circ}}
\newcommand{\deltapsfcirc}{\rm\,\Delta FWHM_{PSF,circ}}
\newcommand{\kubeviz}{{\sc kubeviz}}
\newcommand{\spark}{{\sc SPARK}}
\newcommand{\esorex}{{\sc esorex}}

\newcommand{\sextractor}{{\sc SExtractor}}
\newcommand{\fgas}{M_{\rm{molgas}}/M_*}
\newcommand{\deltaMS}{\rm\,\delta(MS)}
\newcommand{\drdM}{\rm\,\frac{dlog(\reff)}{dlog(\Mstellar)}}

\shorttitle{$\ha$ Sizes and Regulation of $\ktd$ galaxies}
\shortauthors{Wilman et al.}

\begin{document}

\title{The Regulation of Galaxy Growth along the Size-Mass
  Relation by star-formation, as traced by H$\alpha$ in $\rm KMOS^{3D}$ 
  galaxies at $\zrange$ \footnotemark[$\dagger$]} \footnotetext[$\dagger$]{Based
  on observations obtained at the Very Large Telescope (VLT) of the
  European Southern Observatory (ESO), Paranal, Chile (ESO program IDs
  092.A-0091, 093.A-0079, 094.A-0217, 095.A-0047, 096.A-0025,
  097.A-0028, 098.A-0045, 099.A-0013}

\author{David~J.~Wilman\altaffilmark{1,2}, Matteo
  Fossati\altaffilmark{3,2,1,\footnotemark[$\ddagger$]\footnotetext[$\ddagger$]{email: matteo.fossati@durham.ac.uk}}, J. Trevor Mendel\altaffilmark{4,5,2},
  Roberto Saglia\altaffilmark{2,1},
  Emily Wisnioski\altaffilmark{4,5,2}, Stijn Wuyts\altaffilmark{6,2}, 
  Natascha F{\"o}rster Schreiber\altaffilmark{2}, Alessandra Beifiori\altaffilmark{1,2},
  Ralf Bender\altaffilmark{1,2},
  Sirio Belli\altaffilmark{2}, Hannah {\"U}bler\altaffilmark{2}, Philipp Lang\altaffilmark{7,2}, Jeffrey C.C. Chan\altaffilmark{8,2}, Rebecca L.
  Davies\altaffilmark{2}, Erica J.
  Nelson\altaffilmark{9,2}, Reinhard
  Genzel\altaffilmark{2}, Linda J. Tacconi\altaffilmark{2}, Audrey
  Galametz\altaffilmark{10,2}, Richard I. Davies\altaffilmark{2}, Dieter
  Lutz\altaffilmark{2}, Sedona Price\altaffilmark{2}, Andreas
  Burkert\altaffilmark{1,2}, Ken-ichi Tadaki\altaffilmark{11,2}, Rodrigo
  Herrera-Camus\altaffilmark{12,2}, Gabriel
  Brammer\altaffilmark{13}, Ivelina Momcheva\altaffilmark{14}, Pieter
  van Dokkum\altaffilmark{15}}

\altaffiltext{1}{Universit\"{a}ts-Sternwarte M\"{u}nchen,
Scheinerstrasse 1, 81679 M\"{u}nchen, Germany}
\altaffiltext{2}{Max-Planck-Insitut f\"{u}r extraterrestrische Physik,
Giessenbachstrasse, 85748 Garching, Germany}
\altaffiltext{3}{Centre for Extragalactic Astronomy and Institute for Computational Cosmology, Durham University, South Road, Durham, DH1 3LE, UK}
\altaffiltext{4}{Research School of Astronomy and Astrophysics, Australian National University, Canberra, ACT 2611, Australia}
\altaffiltext{5}{3ARC Centre of Excellence for All Sky Astrophysics in 3 Dimensions (ASTRO 3D)}
\altaffiltext{6}{Department of Physics, University of Bath, Claverton Down, Bath, BA2 7AY, UK}
\altaffiltext{7}{Max Planck Institute for Astronomy (MPIA), K\"{o}nigstuhl 17, D-69117, Heidelberg, Germany}
\altaffiltext{8}{Department of Physics and Astronomy, University of California Riverside, 900 University Avenue, Riverside, CA 92521, USA}
\altaffiltext{9}{Harvard-Smithsonian Center for Astrophysics, 60 Garden St, Cambridge, MA 02138}
\altaffiltext{10}{Observatoire de Genève, Université de Genève, 51 Ch. des Maillettes, CH-1290 Versoix, Switzerland}
\altaffiltext{11}{National Astronomical Observatory of Japan, 2-21-1 Osawa, Mitaka, Tokyo 181-8588, Japan 0000-0001-9728-8909}
\altaffiltext{12}{Astronomy Department, Universidad de Concepci\'on, Barrio Universitario, Concepci\'on, Chile}
\altaffiltext{13}{Cosmic Dawn Center, Niels Bohr Institute, University of Copenhagen, Juliane Maries Vej 30, DK-2100 Copenhagen, Denmark}
\altaffiltext{14}{Space Telescope Science Institute, Baltimore, MD, United States}
\altaffiltext{15}{Astronomy Department, Yale University, 52 Hillhouse Avenue, New Haven, CT 06511, USA}

\begin{abstract}

We present half-light sizes measured from $\ha$ emission
  tracing star-formation in 281
  star-forming galaxies from the $\ktd$ survey at
  $\zrange$.
  Sizes are derived by fitting 2D exponential disk models, 
  with bootstrap errors averaging 20\%. $\ha$ sizes are a median (mean) of $1.19$
  ($1.26$) times larger than those of the stellar continuum 
  -- which due to radial dust gradients places an upper limit on the growth in stellar size via star formation -- with just $\sim 43\%$ intrinsic scatter. 
  At fixed continuum size the $\ha$ size shows no residual trend with stellar mass, star
  formation rate, redshift or morphology. The only significant residual trend is with the excess obscuration of $\ha$ by dust, at fixed continuum obscuration. 
  The scatter in continuum size at fixed stellar mass is likely driven by the scatter in halo spin parameters.
  The stability of the ratio of $\ha$ size to continuum size demonstrates a high degree of stability in halo spin and in the transfer of angular momentum to the disk 
  over a wide range of physical conditions and cosmic time. This may require local regulation by feedback processes. 
  The implication of our results, as we demonstrate using a toy
  model, is that our upper limit on star-formation driven growth is sufficient only to
  evolve star-forming galaxies approximately {\it along} the observed size-mass
  relation, consistent with the size growth of galaxies at constant cumulative co-moving number density. 
  To explain the observed evolution of the size-mass relation of star-forming disk galaxies
  other processes, such as 
  the preferential quenching of compact galaxies or galaxy mergers, may be required.
\end{abstract}


\section{Introduction} \label{sec:intro}

Most star-forming galaxies in the Universe above stellar masses of $\Mstellar \sim 10^{9}\Msol$ have most of their stars in disks
\citep[e.g.][]{van-der-Wel14a,Wuyts11}.  These are stable,
rotationally-supported structures which, in the absence of dramatic
events such as major mergers, survive at
least for the Hubble time with disk galaxies still dominant in the
local star-forming population. Galaxy disks typically have radial surface brightness
profiles well described by a declining exponential function (although
an improved description of many local stellar disks is a broken
exponential law, becoming either steeper or more shallow beyond a
break radius, \citealt{Erwin08}). Disks exist not only in the stellar
component but also in the gas which feeds them and, although the
thickness of stellar disks and turbulence in gas disks can vary with
time, basic disk structures with dominant rotational support exist to
high redshifts \citep[e.g.][]{Genzel06,Forster-Schreiber06,Forster-Schreiber09,Kassin12,Livermore15}, up to at least $z \sim 3$ \citep{Turner17}, and are dominant among the
high mass population by $z\sim 2.2$ \citep{Wisnioski15, Stott16},
with evidence that they are common even in the most compact
\citep{Wisnioski18} and passively evolving old galaxies at that redshift
\citep[e.g.][]{McGrath08, van-der-Wel11, Chang13, Newman15, Toft17, Hill19}. 

The structure of massive star-forming galaxies is made not only of rotating disks but also by central dispersion dominated bulges \citep{Lang14}. These could be the result of violent star formation from low angular momentum cold gas in the center of galaxies, although they can also form during merger events.  Submillimeter
observations reveal very high rates of highly obscured star-formation
at the centre of massive galaxies at high redshift \citep[][]{Tadaki17}. Such events appear to co-exist with more extended and less obscured star formation \citep[e.g.][]{Chen17}, such that star-forming disks tend to retain an exponential profile, even in the presence of a bulge or bar.

Observations in the local
Universe indicate that stars form predominantly from the dense and cool molecular gas component, with star-formation surface density well correlated to the molecular gas surface density \citep{Bigiel08}, with a
slope close to unity in the disk regime, implying a constant timescale for the depletion of molecular gas by star-formation. Interestingly, the star-formation also appears to track the existing stars, to first order. For example, there exists a relation between the local density of star-formation and that of stars \citep{Gonzalez-Delgado16}.
This reflects on the spatial extent of these components, such that
in terms of half-light sizes, the size of the star-forming disks are found to be extremely similar to that of the stellar disk in the local Universe \citep{Fossati13}. 

As gas accretes onto a galaxy it still carries much of the angular
momentum from the cosmic filaments which feed the galaxy and its halo
\citep{Fall18}.  Smooth accretion of gas with a consistent axis of angular momentum leads to the formation of gas disks. 
While the mean specific angular momentum of disks is similar to that of their halo \citep{Burkert16}, the distribution within any single galaxy of angular momentum from
newly accreted halo gas is expected to extend to both lower and
higher values than found in typical galaxy disks
\citep[e.g.][]{Dalcanton97, van-den-Bosch01, Dutton09}. The high angular momentum material can be transported to large radii where it will exist in a
diffuse atomic or ionized component unable to form new stars, while the low angular momentum material can be removed in energetic supernovae-driven winds. Such winds are particularly effective at removing material from low mass and compact galaxies \citep{Dutton09} but can be delay the evolution of higher mass galaxies via high redshift ejection and re-incorporation  \citep[e.g.][]{Hirschmann13}.

The existence at $0<z \lesssim 3$ of a Main Sequence
(MS) of star-formation, relating the galaxy star-formation rate (SFR)
to the stellar mass with a small scatter \citep[$\sim 0.3$ dex, e.g.][]{Noeske07,Whitaker14,Schreiber15,Gavazzi15a} implies that the combined processes of gas accretion and
star-formation must be smooth and stable over the relatively short
timescales to which we are sensitive with typical star-formation
indicators. Moreover these small variations in star-formation rate at
fixed stellar mass seem to have no measurable dependence on the galaxy size, but are rather driven by the molecular content of galaxies or its depletion rate \citep{Saintonge11,Tacconi18}. The mass of the cold gas reservoir is also the main driver of the cosmic evolution of the star formation activity, with gas rich galaxies at $z\sim1-2$, forming stars much more rapidly than in the local Universe \citep{Madau14, Whitaker14}.

The relationship between the local density of star
formation and of stellar mass found in the local Universe appears to extend to at 
least $z\sim 1$ \citep{Wuyts13}.  Half-light sizes in the $\ha$ emission line tracing
unobscured star-formation are similar to or slightly larger than the
size in continuum light in both individual highly star-forming galaxies
\citep{Nelson12} and in the stacked averages for normally star-forming galaxies \citep{Nelson16a}.  Using 3D-HST slitless spectroscopic
data \citep{van-dokkum11, Brammer12, Skelton14, Momcheva16}, \citet{Nelson16a} show
that the stacked average $\ha$ profiles of star-forming galaxies with
higher or lower than normal SFR for their stellar mass are
self-similar, changing only in normalization and not half-light size.
The sizes of molecular gas disks themselves are not easily measured at high redshift in normally star-forming galaxies.  Where measured, they appear similar in extent to the stellar or star-forming disks
\citep{Tacconi13, Bolatto15} while in the highly 
star-forming, high mass population the situation 
is more complex: highly compact dust emission can 
co-exist with more extended emission from tracers of molecular gas 
such as CO \citep{Calistro-Rivera18}.

$\ktd$ is a unique 75 night guaranteed time program with the ESO Very Large Telescope (VLT) with the second generation
instrument KMOS (K-band Multi-Object Spectrograph,
\blue{Sharples12,14}) targeting the $\ha+[NII]$ emission line complex in $\sim740$ galaxies selected to have a magnitude $\Ks < 23$ and in the range $\zrange$ (\citealt[][-- hereafter W19 --,]{Wisnioski19} \citealt{Wisnioski15}).
The multiplexing capabilities of KMOS allow us to target more galaxies and with deeper observations than was possible with single object IFUs such as SINFONI \citep[e.g. the SINS survey][]{Forster-Schreiber09}, and compliments contemporary work on smaller numbers of objects featuring the high spatial resolution available with adaptive optics \citep[e.g.][]{Forster-Schreiber18}. 

In this paper we use $\ktd$ data
to map $\ha$ and measure $\ha$ disk sizes in individual star-forming galaxies across a wide range in redshift and SFR. 
We examine whether the
stacked results of \citet{Nelson16a} apply for individual galaxies, 
whether size growth via star-formation is correlated with the stellar mass and star-formation rate or is driven by other fundamental parameters. across a wide baseline in redshift including the peak of the cosmic star formation activity. 
$\ktd$ offers several advantages compared to 3D-HST for a study of this nature: it is significantly deeper, its spectral resolution allows us to resolve the $\ha+[NII]$ emission line complex, and observations in the YJ to $\Ks$ band allow us to trace $\ha$ emission over a larger redshift range.  Being a seeing-limited ground-based survey, this goes at the expense of spatial resolution.

Following our brief introduction to the $\ktd$ survey in
Section~\ref{sec:survey}, Sections~\ref{sec:datared} to \ref{sec:sample} give a detailed account of how we go from raw KMOS data to accurate size measurements of $\ktd$ galaxies with well calibrated errors. Readers primarily interested in our results on -- and interpretation of -- the size growth of star-forming galaxies may skip to Section~\ref{sec:results}. For readers interested in the technical steps, we describe the basic data reduction in
Section~\ref{sec:datared} and the generation of $\ha$ maps and profiles in Section~\ref{sec:maps}. 
In Section~\ref{sec:sample} we describe the
flagging procedures used to verify our sample and show that it is not
biased with respect to normally star-forming MS galaxies.  We also release to the community the size measurements derived in this work.
We then examine which parameters control the $\ha$
size of $\ktd$ galaxies in Section~\ref{sec:results} and in Section~\ref{sec:interp} discuss what
this means for our understanding of how galaxies grow in size through
star-formation.  Our key conclusions are presented in
Section~\ref{sec:conclusions}. Throughout this paper we assume a flat 
$\Lambda$CDM cosmology with $H_0 = 70 \kmsMpc$,
$\Omega_m=0.3$ and $\Omega_\Lambda = 0.7$.

\section{The $\ktd$ Survey}\label{sec:survey}

$\ktd$ takes advantage of the unique multiplexing and
spatially-resolved near infrared (NIR) spectroscopic capabilities of
KMOS as well as the large collecting area of the 8.2~m VLT mirror,
targeting the $\ha$+[NII] emission line complex in galaxies at
$\zrange$. This provides simultaneous flux and kinematic maps of the
ionized gas for up to 24 galaxies in one exposure by deploying 24
configurable arms in the $\rm 7.2\arcmin$ field of view, each hosting a
$\rm 2.8\times 2.8 \arcsec$ integral field unit (IFU).

The first year of the $\ktd$ survey was described by
\citep{Wisnioski15}. $\ktd$ targets galaxies selected from the 3D-HST
grism \citep{Brammer12, Skelton14, Momcheva16} and CANDELS imaging
\citep{Grogin11,Koekemoer11} surveys with the Hubble Space Telescope
(HST) in the COSMOS, GOODS-South and UDS deep fields accessible from
Paranal. Targets
are selected to have a magnitude $\Ks < 23$, and a known spectroscopic
or grism redshift
(grism redshifts from 3D-HST have an accuracy of $\rm \approx
1000\kms$ \citealt{Momcheva16, Fossati17}) for which the spectrum around
$\ha$ should be relatively free of atmospheric OH lines, and visible in
the KMOS YJ, H or K-bands. We apply no prior selection on star
formation rate or $\ha$ flux in order to avoid selection bias and
sample the full range of galaxies down to our detection limits. Due to
the unique multiplexing capabilities of KMOS, we are able to
observe galaxies from $\sim 3-30$ hours by re-targeting objects with
weak detections to improve the signal to noise ratio.

Observations for $\ktd$ were carried out from October 2013 until April
2018 following an object-sky-object (OSO) observation pattern such that 
each object exposure is adjacent to a sky exposure in the same IFU. Three 
IFUs were placed on stars to trace the variable spatial Point Spread 
Function (PSF) and throughput from exposure to exposure, leading to a 
simultaneous observations of up to 21 galaxies per exposure.

In this work we consider all data taken up until April 2017, comprising 645 galaxies targeted for observations of $\ha$ and [NII].  Data were taken in a range of observing conditions, with PSF minor axis FWHM ranging between $\rm 0.3\arcsec$ and $\rm 0.92\arcsec$ and a
median of $\rm 0.456\arcsec$.  The final KMOS3D data set is fully described in \blue{(W19)}.
Star-formation rates (SFR) used in this paper are computed using the data and method described by \citet{Wuyts11}, based on infrared, UV
and optical observations and thus independent of our $\ha$
measurements.

\section{Data Reduction}\label{sec:datared}

Our reductions in this paper are intermediate between the early data reduction described by \citet{Wisnioski15} and that described in the data release paper \blue{(W19)}. We refer to \blue{W19} for much of the reduction procedure noting where implementation of specific steps differ. In particular we describe in detail the steps which optimise the background subtraction and astrometry, and which were tailored to allow a robust extraction of $\ha$ profiles and sizes of galaxies. 

\subsection{Basic reduction}\label{sec:framered}
All our basic calibration steps, with the exception of the sky and background subtraction, are identical to those described by \blue{W19}. We make use of the Software Package for Astronomical Reductions with KMOS (\spark) code which works within the ESO pipeline execution tool (\esorex), supplemented with some custom tools written in the {\sc idl} and {\sc python} languages. This includes masking of bad pixels and flattening at the detector level; reconstruction of data cubes including a refined wavelength calibration using sky lines and a heliocentric correction; followed by a correction for the spatial illumination uniformity, and flux calibration using standard star observations. During this last step we used the flux from stars observed in the same setup as the galaxies to correct for frame to frame variations in the throughput. Bad frames are inspected and removed. 
Skyline subtraction was applied using the standard method in SPARK which subtracts an adjacent sky frame with skylines scaled to optimally match the science observation \citep{Davies07}.

Once individual frames are generated. it is essential to subtract a residual background level per frame: not doing so results in a factor of three reduction in continuum signal to noise in the final co-adds, primarily due to the significant variations in instrumental, sky (e.g. twilight and moon illumination) and thermal (especially in K-band) background between object and adjacent sky frames. Instrumental variations include a readout channel dependent effect, which can vary frame to frame. To account for this effect, we derive and subtract a background value for each of the readout channels of the detectors.

\subsection{Astrometric Registration, Improved Background Subtraction and Generation of Combined Cubes}\label{sec:fitimage}

After the reconstruction of individual frames, {\it Partial} combined cubes, defined to be the co-add of the data taken for a given galaxy within a given observing setup (commonly one per observing run), are generated assuming astrometric shifts between frames equal to the average of the measured shifts for the three stars included in the same setup (this accounts for the telescope dithering and the gradual drift of the KMOS arm positions). 

We also generate 100 bootstrap cubes obtained by randomly resampling the input frames for each partial combine. We use these cubes for the propagation of uncertainties.
In this work, we make use of a modified combined noise cube with the aim to obtain a robust estimate of the spectral uncertainty close to the edges of the cubes where few exposures are available (given that \spark\ estimates variance from the distribution of values in each exposure). 
We derive a single variance spectrum per cube using the \spark\ variance estimate in spaxels to which at least 75\% of the total number of exposures have contributed. We then scale this spectrum by the exposure time of each pixel in the cube.

To achieve the best signal to noise ratio and image quality for our data in the final cubes we 
further process the individual frames to obtain a flat background and an accurate registration of the astrometry between frames observed in different runs. 
To do this we generate images of each galaxy by collapsing the KMOS partial combine datacubes along the
wavelength axis. The galaxy continuum is well detected for most sources brighter than our
$\Ks=23$ magnitude limit, with an increasing fraction of non-detections in continuum close to this limit ($\sim 11\%$ in the range $\Ks=22.5-23$). 

At this stage, the partial combine cubes retain a residual, negative background caused by 
the overestimation of background levels in individual frames due to the contribution of the source.
Its magnitude in bright sources is $\lesssim 10\%$ of the variation in background level between exposures and decreases in fainter sources, but it is systematic and limits the depth of our final mosaics.
Thus, we derive an additive correction to the background as described below.

We first convolve the HST image, selected in the nearest available
band (WCF3 F125W for KMOS YJ, WFC3 F160W for KMOS H and K) with
a multi Gaussian kernel to optimally convolve the
HST PSF to that of our KMOS PSF image for that galaxy. Each model
solution is defined by an astrometric offset, a normalising flux scale factor and an additive
background correction per readout channel which contributes in the partially combined 
datacube. Each model image is generated by projecting the convolved HST image onto the 
KMOS pixel grid and cropping to the KMOS field of view, and then adding the
background correction image. We use the {\sc mpfit} non-linear least
squares fitting algorithm to find the minimum chi-squared solution
\citep{Markwardt09}\footnote{https://www.physics.wisc.edu/$\sim$craigm/idl/fitting.html}.
To ensure we do not get stuck in a local minimum, we iterate over the initial guess for 
the astrometric centroid on a grid of 1 pixel
resolution, allowing centroids within $\pm 30\%$ of the FOV from the image centre, selecting
the solution which gives us the global minimum chi-squared.  The full procedure is repeated for each bootstrap cube (except the initial guess for the astrometric centroid is now fixed) to help evaluate errors and degeneracy in the astrometric registration.

Of 166 objects with multiple setups, the median residual shift is $\sim 1.33$ KMOS pixels ($\sim 0.27$\arcsec) with $\sim27\%$ of shifts above 2 KMOS pixels ($0.4$\arcsec), ranging as high as $4.35$ pixels ($0.87$\arcsec).\footnote{W19 reports a lower fraction of large shifts because in that paper we divided the list of exposures into smaller units with smaller shifts for the final data release.} Not accounting for such shifts artificially blurs the galaxy by an average of $\sim$2\kpc\ and up to $\sim$7\kpc. These shifts are caused by the variations in the calibration parameters of individual KMOS arms, which are periodically tweaked by the observatory to ensure that the arm positioning remains within specifications. Fits to the individual partial combines are visually inspected and a new list of astrometric shifts is derived by combining the frame to frame shifts measured using PSF stars and the setup to setup shifts from the fits of partial combines. We also subtract the best-fit background image from each individual exposure contributing to a given partial combine cube.

With the updated list of astrometric shifts, we combine all frames contributing to a single object and we re-generate the bootstrap cubes. This produces our final {\it total} combined datacubes. At this stage we also derive the instrumental resolution for each cube and its associated PSF image. For a detailed description of these procedures we refer the reader to \blue{W19}.   
No correction is applied during the fit of partial combines to the absolute astrometry. This
is done by fitting the total combines in order to have the deepest KMOS images register onto the 
HST astrometry. This last fit does not include a background level as the background 
has already been flattened during the previous step. 

\begin{figure}
  \centering
  \includegraphics[width=0.95\columnwidth]{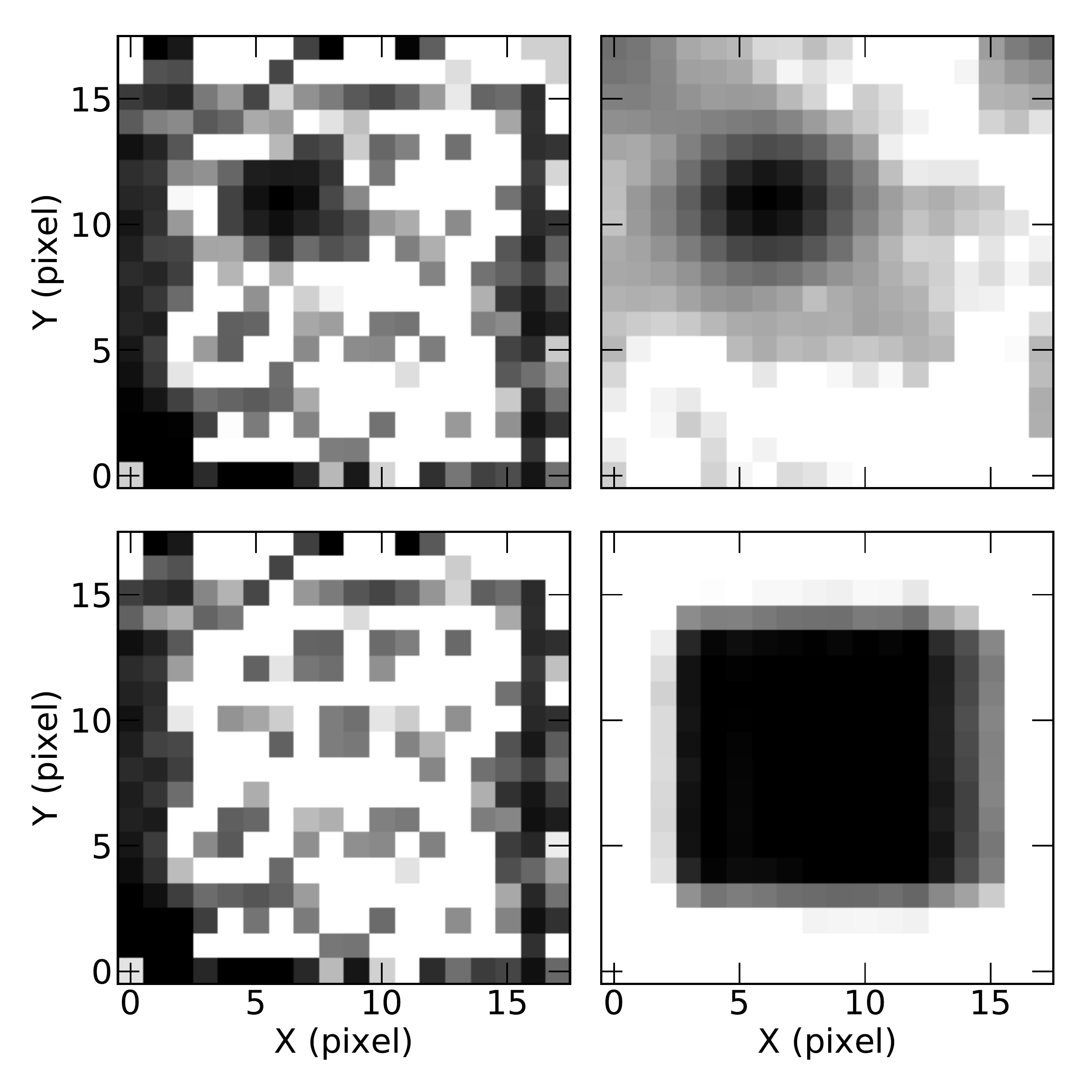}
  \caption{Example of astrometric registration fitting procedure. The
    collapsed KMOS continuum image (top-left) is fit to the
    PSF-convolved and resampled CANDELS image in the nearest band,
    including a shift (top-right), leaving the residual image
    (bottom-left). Fitting weights are applied (bottom-right) which
    down-weight the outer part of the image (subject to cosmetic
    effects and lower total exposure). For this example case
    (COS4\_12148) a large shift of 2.5 pixels each of X and Y is
    required.}
    \label{fig:fitimage}
\end{figure}

Each astrometric solution is now visibly inspected by
looking at the object centroid in the collapsed KMOS image, and in the model HST image. In 615 of 645 cases the automated solution is good, as in the example case shown in Figure~\ref{fig:fitimage}. These imply a median shift of $\sim 1$ KMOS pixel ($0.2\arcsec$) with a tail extending to $>5$ pixels ($>1\arcsec$) and a median bootstrap error of $\sim 0.1$pixel, with errors up to $\sim 1$ pixel. This is consistent with the shifts computed for the same object observed in multiple setups and with the expected positioning accuracy of the KMOS arms.  In 26 cases we apply a manual shift, of which for 15 it was necessary to inspect also the $\ha$ image. We used this image to confirm the low signal to noise
continuum centre finding consistency with the $\ha$ image in all cases.
Only four targets are not visible in continuum or $\ha$, for these no
astrometric correction is applied. Manual astrometric solutions are usually accurate within a pixel, with the exception of some low S/N continuum or $\ha$-based centroids which can be less accurate (up to $\sim 2$ pixels).

The astrometric correction derived above is applied, registering datacubes and bootstrap cubes to the HST astrometry. These astrometrically and background corrected cubes are considered our final datacube products. 

\section{Generation of maps and profiles}\label{sec:maps}

\subsection{Emission Line Measurement}

To fit the $\ha+$[NII] emission line complex we utilize our versatile {\sc idl}-based emission-line fitting software \kubeviz\footnote{ \kubeviz\ is made publicly available at \url{https://github.com/matteofox/kubeviz} and \url{http://www.mpe.mpg.de/\~dwilman/kubeviz} and is easily adaptable to new instruments. }. 
\citep[see e.g.][]{Fumagalli14, Fossati16}. \kubeviz\ can be operated in interactive or batch modes, and provides the user with full access to the options provided by {\sc mpfit} which fits the continuum and emission lines. Gaussian fits to emission lines automatically account for the (known) spectral resolution of the instrument as a function of wavelength.

We derived accurate maps of emission line flux down to low surface brightness levels, as well as velocity maps and masks of good kinematic fits as described below.

\subsubsection{Kinematic Fits}

{Our first fit is mostly aimed at generating velocity and dispersion maps. 
To improve the S/N per pixel, the flux, noise and bootstrap cubes are median smoothed along spatial axes with a top-hat smoothing kernel of $3\times3$ spaxels. We fit the spectral continuum underlying the $\ha + [NII]$ emission line complex, assuming a constant value, independently computed for each spaxel, $\rm C_{x,y}$. This is the inverse-variance weighted average value within spectral continuum windows defined to either side of the $\ha$ line, corresponding to between $\rm 2000\kms$ and $5000\kms$ in rest-frame velocity offset (thus excluding [NII] and [SII] lines). The spectral region from 12680-12710\AA\ containing the strongest part of the atmospheric O2 feature, and regions within 10\AA\ of either end of the spectrum are excluded. 
We generate continuum-subtracted cubes $\rm CS_{x,y,\lambda}$ by subtracting the continuum for all spaxels from the flux cube, $\rm F_{x,y,\lambda}$, in symbols:
\begin{equation}
\rm CS_{x,y,\lambda} = F_{x,y,\lambda} - C_{x,y}.
\end{equation}
}

{We simultaneously derive kinematic and flux information for the $\ha$, $[NII]\lambda6583$ and $[NII]\lambda6548$ emission lines by fitting the inverse-variance weighted continuum-subtracted spectra for each spaxel.
 We fit a single Gaussian line profile for each emission line which accounts for the redshifting and instrumental line broadening of the specific KMOS observation, returning the rest-frame velocity and intrinsic dispersion of the ionized gas. 
Since, for this step, we are interested in robustly detected emission lines, these fits are constrained such that lines have a minimum of zero flux}. Multiple lines all share a single velocity and dispersion, and the ratio of flux in the two [NII] lines is fixed to the value from atomic physics \citep[3.071,][]{Storey00}. As a result, output 2-D maps are generated for each fit parameter (line flux, velocity, dispersion).

{The full continuum and emission line fitting process is repeated for each bootstrap cube, generating 100 bootstrap realisations of the fitting parameters. These are used to generate images where each spaxel represents the probability of non-zero $\ha$ line flux $\rm P_{f_{\ha}>0}$ and positive non-zero dispersion $\rm P_{\sigma>0}$. $\rm P_{f_{\ha}>0}$ represents a detection significance of the flux per spaxel, and $\rm P_{\sigma>0}$ that the emission line is significantly resolved. High values $\rm P_{\sigma>0}\gtrsim0.9$ provide a good indication that {\sc mpfit} has picked up a real feature in the spectrum rather than a noise or skyline residual spike, and correlates well with regions where the velocity map is relatively smooth. Intrinsic velocity dispersions of $\rm \sigma \gtrsim 25\kms$ are usually well resolved.} 

\subsubsection{Masking}

{We then generate an automated spaxel mask to identify spaxels with trustworthy kinematic fits. Within this mask, a good (unmasked) spaxel must meet the following conditions:
\begin{equation}\label{equ:mask}
\begin{array}{lllll}
  & \rm (f_{\ha}>0.) & \rm and & \rm (0 < \sigma \leq \sigma_{max}) & \rm and \\   
  & \rm (P_{f_{\ha}>0} \geq 0.95) &  \rm and & \rm (P_{\sigma>0} \geq 0.9)\\
\end{array} 
\end{equation}
where we set $\rm \sigma_{max} = 250\kms$ to exclude broad line features, given that our primary goal at this stage is to define the velocity map. This mask is applied to the velocity map of example galaxy U4\_25642 in the left-hand panel of Figure~\ref{fig:velocity}.

\begin{figure}
  \centering
  \includegraphics[scale=0.27]{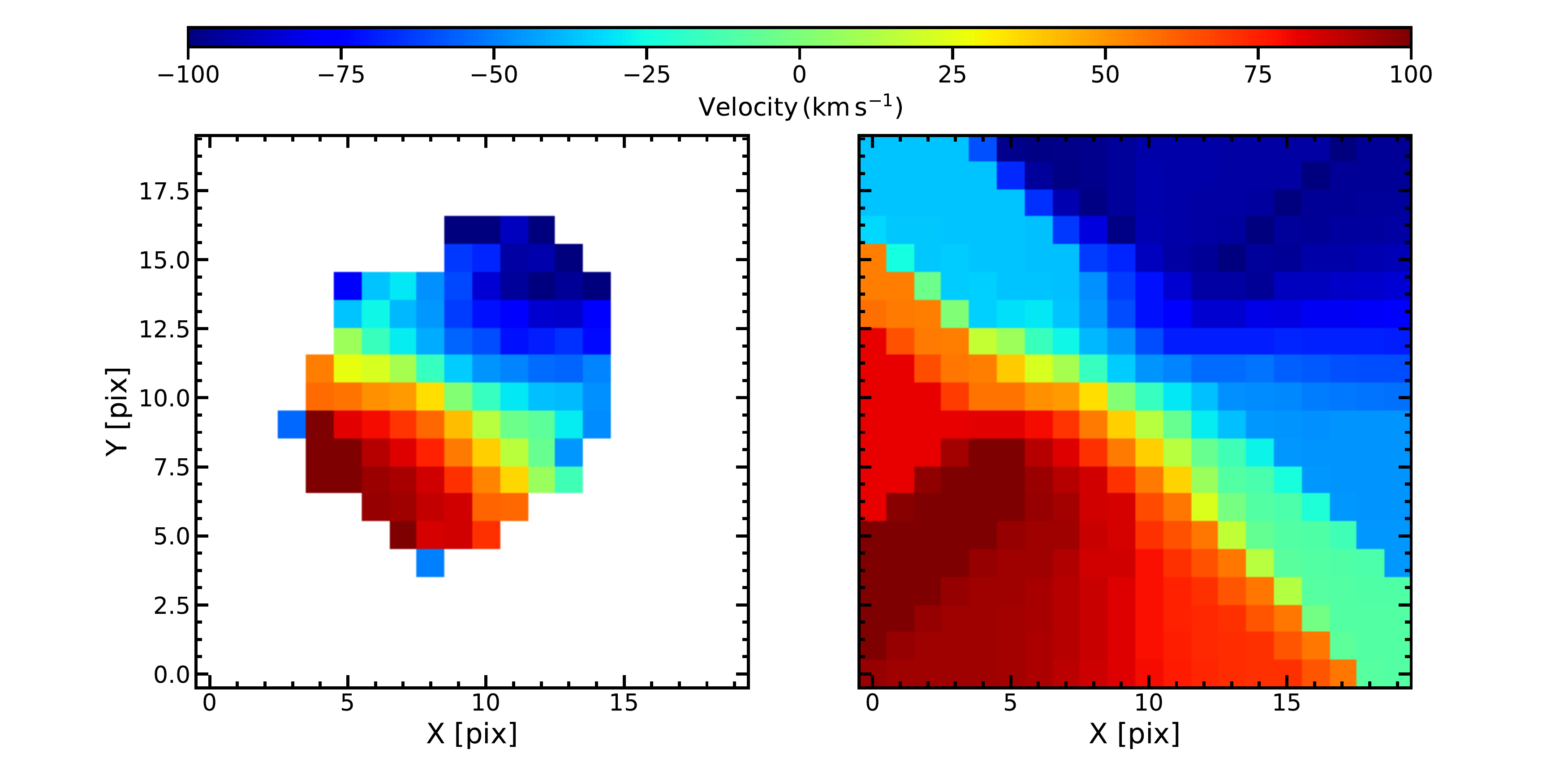}
            \caption{Example masked velocity field (left) and grown velocity map (right) for galaxy U4\_25642. Blank pixels in the left hand image are masked, using the pre-sigma clipping mask discussed in the text. Additional automated sigma-clipping and (in some cases) additional manual editing of the mask are then applied, followed by smoothing and growth by extrapolation of the masked velocity map to generate the grown velocity map as described in the text.}
    \label{fig:velocity}
\end{figure}

We then apply some further steps designed to throw out potential outlier spaxels in the velocity map. The sigma-clipped mean and rms velocities are computed, and any spaxel with velocity outside the range $\rm mean \pm 3 \times rms$ is thrown out. Isolated unmasked spaxels are removed and the remaining are smoothed with a 
 $3 \times 3$ top-hat filter.  This leaves us with a conservative mask and a smooth velocity map. While conservative, our maps are consistent with those presented in other $\ktd$ papers focused on galaxy kinematics \citep[see e.g.][]{Wisnioski15}}.

{Galaxies with $\rm <3$ valid spaxels are dropped, as are poor fits established by inspecting objects with $\rm <10$ valid spaxels. This relatively small minimum number of spaxels is sufficient to establish a zero-point and lack of chaotic variation in the velocity maps of compact galaxies which is sufficient for our purpose.
This leaves 462 galaxies. All velocity maps are visually inspected, resulting in the removal of a further seven cases (leaving 455 galaxies) and the manual correction of 117 masks. These fix cases in which the rotation curve gets truncated by the sigma-clipping procedure, or there are systematic fits to sky features which are usually spatially offset from the galaxy, and are not sigma-clipped.}

{We then extrapolate the rest-frame velocity map to the edges of the KMOS field of view to generate a {\it grown velocity map}, $dv_{rest}(x,y)$. This involves setting the velocity of masked spaxels to the average value of their neighbours, starting with those with unmasked neighbours and iterating up until the point that the whole area is filled. While this extrapolation cannot pick up changes in the rotation curve in the outer, low surface brightness parts of the galaxy, it involves minimal assumptions. This procedure is repeated for each bootstrap iteration independently. The right-hand panel of Figure~\ref{fig:velocity} shows the resultant grown velocity map for U4\_25642. }

\subsubsection{Deep emission line flux maps}

{To derive our final $\ha$ flux maps we reapply continuum fitting and subtraction, this time to the {\it unsmoothed} cubes. Then we integrate the flux within each spaxel of the continuum-subtracted cube, centred at the wavelength specified for $\ha$, assuming the ``grown'' velocity map. In the observed-frame this is $\lambda_{cen}(x,y) =
    \lambda_{\ha}.(1+dv_{rest}(x,y)/c)\times(1+z)$ with a window width of $\pm
    200\times (1+z) \kms$. This window is sufficiently wide to encompass offsets from
    the true velocity caused by the extrapolation process into the outer disk \citep[see e.g.][]{Lang17}, without losing too much flux in the wings of all but the broadest emission lines. A broader window would lead to reduced signal-to-noise. This narrow-band extraction generates a map of $\ha$ flux down to
    regions of low signal-to-noise where parametric fitting
    fails or becomes unreliable. Our
    window-integrated $\ha$ map is computed as:
\begin{equation}
\rm \fhawin = \Delta\lambda \cdot \Sigma_{\lambda_{upper}(x,y)}^{\lambda_{lower}(x,y)} CS_{x,y,\lambda}   
\end{equation}
with the bounds in the sum given by $\lambda_{cen}(x,y) \pm (200\kms/c)\times
(1+z)$, and $\rm \Delta \lambda$ being the spectral step (in \AA) of the datacubes.  This procedure is again repeated for each of the 100 bootstrap realizations.
}

{Finally we correct these flux maps $\fhawin$ for flux lost
    outside the $\rm \pm 200\kms$ window from the tails of broad lines.
    An alternate mask is used to define regions for which the velocity
    and dispersion from Gaussian fitting is usable for this purpose: this is equivalent to
    equation~\ref{equ:mask} but with a relaxed upper limit on
    dispersion $\sigma$: in this case $\rm \sigma_{max} = 1000\kms$ for
    $\rm S/N_{\ha} \geq 4$ and $\rm \sigma_{max} = 400\kms$ for lower
    $\rm S/N_{\ha} < 4$ (thresholds selected from visual inspection of
    fits to broad lines). Isolated, unmasked spaxels are thrown out,
    and manual edits made to our earlier masks are reapplied to this
    mask. The dispersion measured by the fit is used to compute
    the fraction of emission flux which falls outside our $\rm
    \pm200\kms$ window, and this correction is applied to the flux map:
\begin{equation} \label{eq:hafluxmap}
  \rm \fhawincor = \fhawin / c_{\sigma_{200}}
\end{equation}
with $\rm c_{\sigma_{200}}$ the two-tailed cumulative distribution
function for a Gaussian, evaluated at $\rm \sigma_{200} =
\sigma/200\kms$. The correction is usually small except in regions
containing very broad lines ($c_{\sigma_{200}} =
0.955$, $0.576$, $0.197$ for velocity dispersions  
$\sigma = 100\kms$, $250\kms$, $1000\kms$ respectively). 
Outside the mask we do not have reliable data to make a correction. However, these regions usually correspond to the outer and low surface brightness parts of galaxies where the dispersions are
typically low $\lesssim 100\kms$ \citet{Wisnioski15} and so any
correction would be very small.

This correction is also applied to the bootstrap realizations, and we
consider $\fhawincor$ our best estimate of the flux map for the $\ha$
emission line, from now on simply known as the $\ha$ flux map or
image.}

\subsection{Image Fitting in 2D}

We now model the radial surface brightness profiles by fitting the
images of our galaxies in 2D using the Levenburg-Marquardt solver from the image fitting code 
\imfit\footnote{http://www.mpe.mpg.de/~erwin/code/imfit/} \citep{Erwin15}.  
Our goal is to quantify the
distribution of continuum flux (in particular the F160W band tracing
older stars), and $\ha$ flux (tracing ongoing star-formation) obtained with the spectral windows method defined above. We fit
continuum profiles with a Sersic profile \citep{Sersic68} and $\ha$
profiles with a pure exponential, motivated by the results of \citet{Nelson16a}. 
Despite radial and / or azimuthal variations from
a pure exponential profile, the best-fit solution will account for the
mean surface brightness at fixed radius, and as such
provides a measure of the average star-formation at that radius, smoothing out
temporal fluctuations due to bursty star-formation on local scales and
the short-lived nature of the $\ha$ emission.
This is demonstrated by the accuracy of the exponential fit viewed as 1D
azimuthally-averaged radial profiles.

We fit Sersic models convolved to the HST PSF, as derived by the 3D-HST team \citep{Skelton14},
to postage stamps
of each galaxy in F160W and F125W bands 
Fitting F160W data,
the centroid, ellipticity and position angle are left free in addition
to the effective ($=$ half-light) semi-major axis radius $\reff$, Sersic index
$n_{\rm Sersic}$ and normalizing surface brightness.
Initial guesses
for fit parameters are taken from the fits of \citet[][, hereafter \blue{vdW14}]{van-der-Wel14} using
the \galfit-software \citep{Peng10a}\footnote{We intialize the
  centroids to the best fit ones from \galfit, not \sextractor\ as
  published, private communication with A. Van der Wel.}. To estimate
initial parameters for galaxies which were not fit by \galfit\ (flags
of $\geq2$ in the \blue{vdW14} catalog), we use the \sextractor\
parameters from \citet{Skelton14} and empirical relations between
\sextractor\ and \galfit\ parameters for size, axis ratio ($q = 1 -
\epsilon$ where $\epsilon$ is the ellipticity) and position angle. This
includes empirical fits for size and axis ratio to those objects which
were fit well by \galfit, using the \sextractor\ major and minor axis
size parameters {\sc A\_IMAGE} and {\sc B\_IMAGE}:

\begin{equation} 
\begin{array}{l}
\reff = 1.25\times\sqrt{A\_IMAGE^2 - 0.15^2}\\
q = {\sqrt{B\_IMAGE^2 - 0.15^2}}/{\sqrt{A\_IMAGE^2 - 0.15^2}}\\
\end{array}
\label{equation:sextractortogalfit}
\end{equation}

\begin{figure*}
  \centering
  \includegraphics[width=1.00\textwidth]{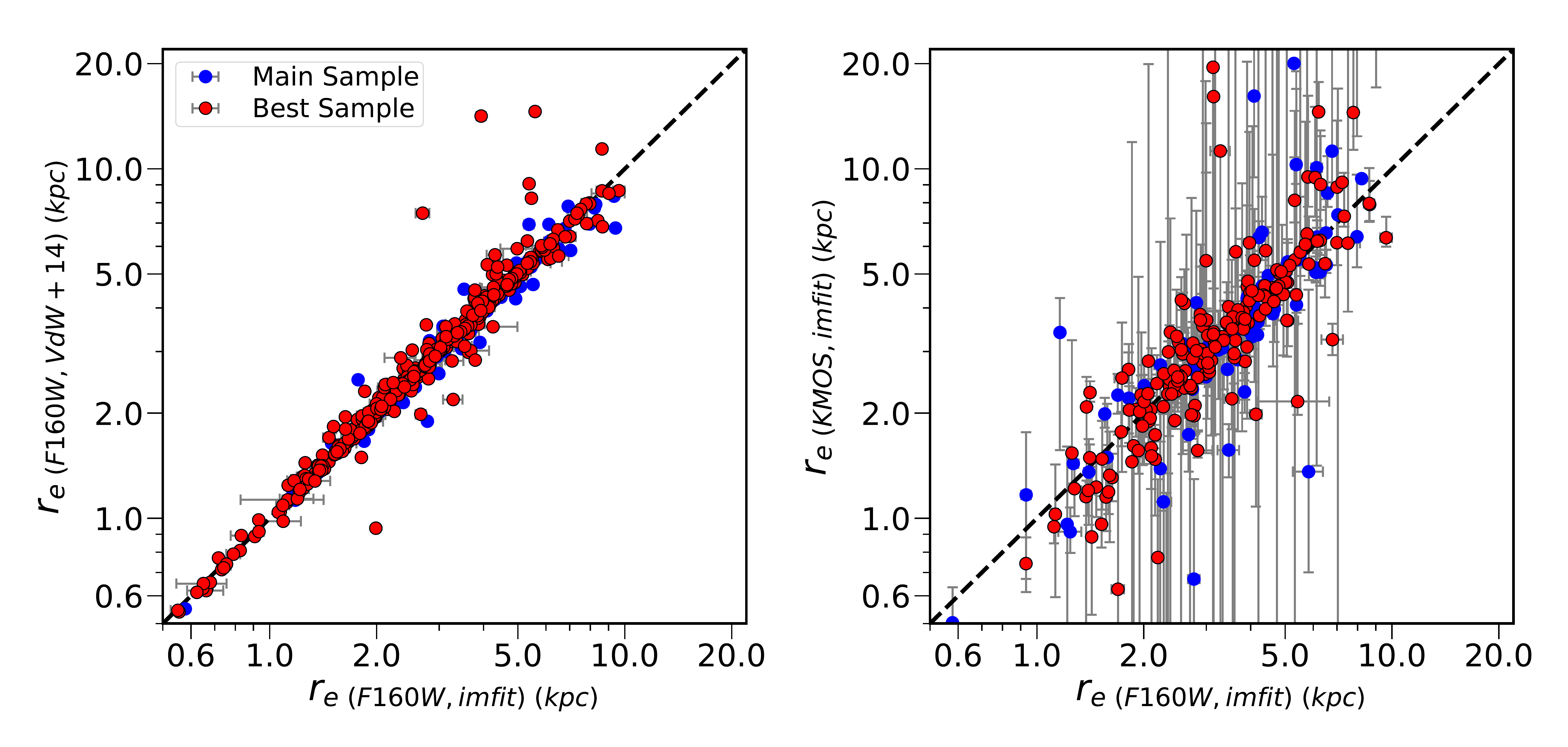}
            \caption{Measurement of intrinsic half-light size for valid
              fits (see text). Left: \imfit-based fits to F160W-band
              CANDELS data at native HST resolution (this paper)
              compared to \galfit\ fits from \blue{vdW14}. 
              Medians are within $0.2\%$ for main and best
              samples, and $68$, $95\%$ of galaxies within
              $^{+3}_{-5}\%$, $\pm21\%$ respectively.  Right: \imfit\
              fits at full HST resolution vs those from the collapsed
              KMOS continuum (with bootstrap errors, see
              Section~\ref{sec:sizeconsistency}.  Fits are remarkably
              consistent: Medians are within $1\%$, $68\%$ of galaxies
              are within $\pm23\%$, and outlying points tend to have
              large errorbars.}
    \label{fig:fitsf160w}
\end{figure*}

To avoid biased fits due to neighbouring galaxies, we simultaneously
fit all neighbouring galaxies within 5\arcsec\ and less than 3.5
magnitudes fainter than the primary source in both bands (F160W and
F125W), or within 2\arcsec\ and less than 5 magnitudes fainter. The
left-hand panel of Figure~\ref{fig:fitsf160w} demonstrates the good
agreement of effective radii fit using \imfit\ and \galfit. The few
outliers mainly move along the degeneracy between $\nsers$ and $\reff$
and tend to include multiple simultaneously fit objects of similar
magnitude.

We also fit the KMOS continuum image and bootstrap realizations (a
resistant weighted mean along the wavelength axis as described in
Section~\ref{sec:fitimage}) with the same set of constraints, resulting
in similar fits, though the lower signal to noise leads to a larger
scatter about the 1:1 relation (right-hand panel of
Figure~\ref{fig:fitsf160w}). 

Finally we fit the $\ha$ flux image (precisely, $\fhawincor$) and
bootstrap realizations. The $\ha$ disk is modelled with a simple
exponential profile convolved with the KMOS PSF and with the KMOS pixel
size and field of view. We only fit data from spaxels
with at least 20$\%$ of the nominal number of exposures for each object.\footnote{For four objects in our sample, this threshold would result in spaxels with less than 5 individual exposures, in this case we used the latter value to define the spaxels to be fit.}. Unlike continuum fits, we only fit the primary galaxy (as
in all but a few cases the redshift of any photometric neighbours puts
any emission line outside our windows for $\ha$ or continuum). The
centroid, ellipticity and position angle are fixed to those measured at
HST native resolution in F160W band to ensure that we are geometrically
tracing the same disk in star-formation as we see in stars. The half-light 
radius starting guess is fixed to that of the stars: we confirm that this 
has no influence on our results by re-fitting each source (excluding the 
bootstraps) with the slower differential evolution solver method which 
does not require initial estimates for the parameters. For all 
galaxies in our analysis (see Section~\ref{sec:flags}) the resultant 
sizes are identical.
A minority of galaxies do not host such star-forming disks but do still host
$\ha$ emission tracing some other component (e.g. outflows) which is
not only physically disassociated to the disk , but does not share its inclination and geometry.  
Cases which are not well modelled by the exponential disk model are flagged as described in Section~\ref{sec:flags}.
We further investigate the impact of our assumption of an
exponential profile for the $\ha$ emission by fitting Sersic models to these images. 
We find that, for most of the sources in our analysis, the best fit 
Sersic index is close to unity and the effective radii are consistent with those 
from the exponential fits. Quantitatively we find consistency between $\reff$ from 
Sersic and exponential fits within 1, 2, and 3$\sigma$ for 65\%, 75\%, and 86\% of the 
galaxies in our sample, respectively. We also find that the trends presented in 
Section \ref{sec:results} are unchanged within the uncertainties. However, 
a substantial number of $\ha$ images whose Sersic fits hit the
fitting limits for the Sersic index, leading to more discrepant values 
for $\reff$ when compared to the exponential fits. For these reasons, 
in this work, we use the results of the exponential fits to the $\ha$ images.

Appendix \ref{sec:sizeconsistency} describes how we test and derive final errors on size measurements, concluding that not only are the continuum sizes consistent with those from HST as shown in Figure~\ref{fig:fitsf160w} but that deviations are, statistically, very consistent with our  errors, and that the accuracy of our PSF model is not the dominant source of error. 

\subsection{Major Axis Radial Profiles}

For our current purposes we are not
interested in azimuthal variations. Therefore to establish that the
radial flux profiles are indeed well characterized by the model and
fit, we also extract one-dimensional radial surface brightness profiles
in elliptical annuli, aligned to the galaxy's best fit ellipticity and
position angle. In practice these are computed by embedding the images
in a larger grid, weighting the individual image pixels by their
effective contribution per radial bin, and then integrating the flux
within differential elliptical bins corresponding to a given
major-axis-equivalent radius. Equivalent profiles generated from each
bootstrap image provide estimates of the error on the profile. We also
generate profiles for the best \imfit\ fit model profiles in the
same way -- the projection of the model image onto elliptical radial
apertures, and of the PSF (placed at the centroid of the galaxy).  This
enforces the same pixelization as exists in the data itself for direct
comparison. In Appendix \ref{sec:profilesgallery} we show a gallery of $\ha$ 
profiles spanning a range of size, redshift and observed surface brightness, 
to highlight the quality of the data and of the fitting procedure.

\begin{figure}
\centerline{\includegraphics[width=0.5\textwidth]{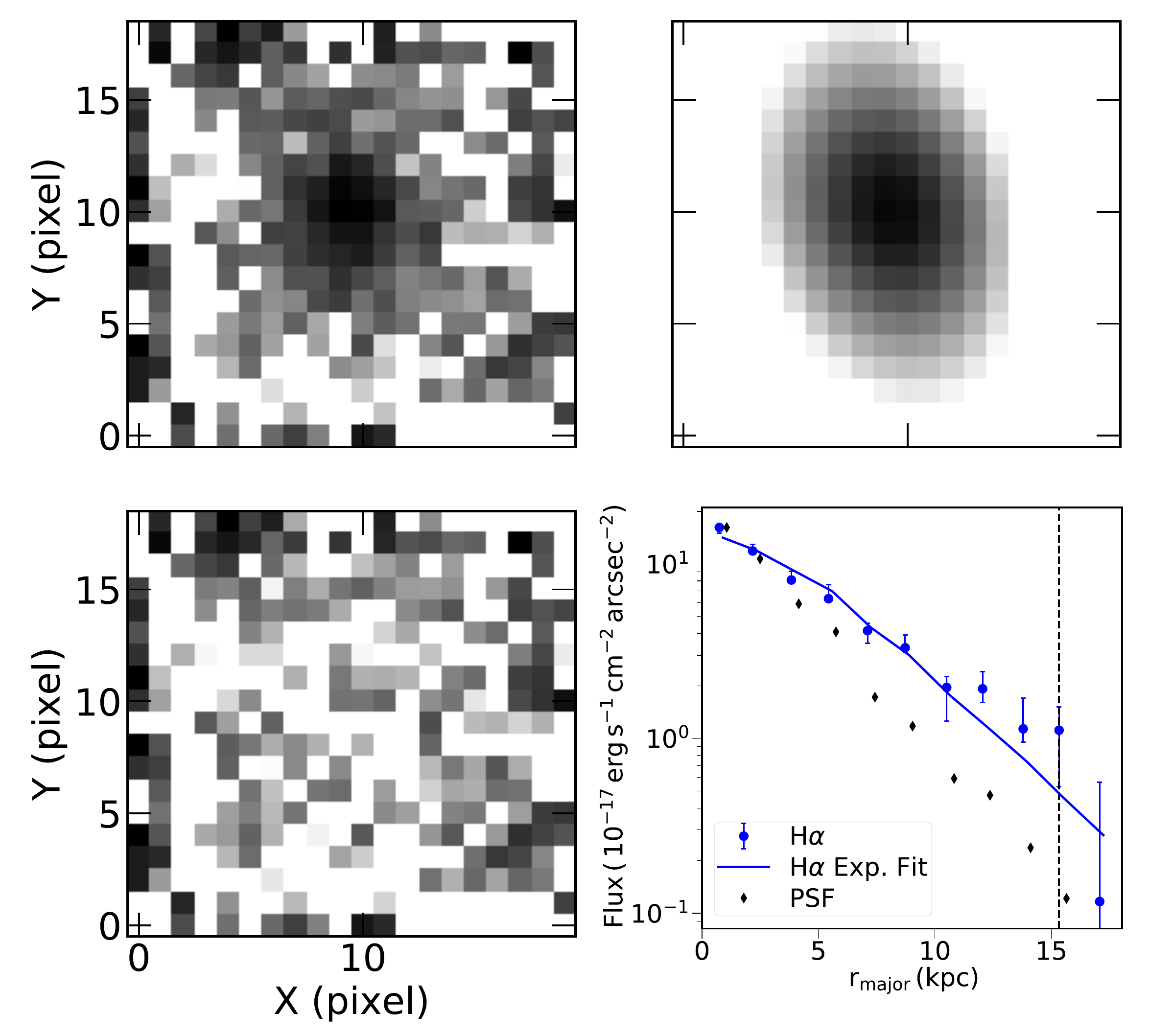}}
            \caption{Example \imfit\ fit to the data for galaxy GS3\_11606. Top-left: log-scaled $\ha$ image; top-right: Best fit model exponential galaxy convolved with PSF, from \imfit\ (log-scaled); Bottom-left: Residual image (linear scaling); Bottom-right: Log-scaled radial profile, extracted using elliptical apertures from both data (blue points with $1-\sigma$ bootstrap errors) and best fit model (blue solid line) images. For comparison, the PSF image is also extracted in the same apertures (black diamonds). The residual image and 1D profile demonstrate a good fit for this galaxy, with a best fit size of $\reff=3.74^{+0.6}_{-0.45}~\kpc$. The vertical dashed line indicates the radius where the major axis first crosses the edge of the KMOS field of view.}
    \label{fig:exampleimfit}
\end{figure}

\section{Sample}\label{sec:sample}

\subsection{Flagging}\label{sec:flags}

Our analysis requires size estimates which accurately reflect the true
profiles of continuum and $\ha$. Starting with our astrometrically
calibrated, $\ha$-detected and masked sample with valid velocity grown
maps (457 galaxies), we need to weed out objects with strong skyline
contamination or poorly fit continuum / $\ha$ profiles. We do this via
a series of steps during which two authors (MF and DJW) independently visually inspected the data, removing objects not satisfying a series of requirements. 

This results into a sample including all the objects for the analysis of the $\ha$ and F160W sizes (the $\ha$ sample hereafter), which we further split into a MAIN and a BEST sample, where the BEST sample includes only
the best \imfit\ fits and sky subtraction (for $\ha$) while the MAIN samples include all reasonable fits. 

In detail, we first we inspected the data for atmospheric skyline residual contamination which is evaluated by simultaneously
inspecting the inverse-variance weighted summed spectrum within the
mask of good Gaussian fits, the variance spectrum, and the
inverse-variance weighted summed spectrum from outside the mask. The comparison of spectra inside and outside the mask serves to establish which spectral features are associated to the galaxy spectrum and which are spurious features commonly associated to high variance residuals from sky-subtraction.
83 of 457 galaxies have strong
skyline contamination and will be excluded from further analysis while
101 have a weaker contamination (skyline residuals are sub-dominant compared to the underlying $\ha$ emission, or lie just outside the $\ha$ wavelength range). The latter are included in the
analysis as part of the $\ha$ MAIN sample, but we examine their influence on our results by excluding them from the $\ha$ BEST sample. 

We also exclude 22 of the remaining galaxies due to close pairs for
which the paired galaxy also appears in the $\ha$ image, two galaxies
with PSF axis ratios $>1.5$ and six galaxies for which the apparent, measured $\ha$ flux is incoherently spread across the FOV. 

In general, our fits are deemed to be good if: a) the magnitude of
fractional residuals and $\chi^2$ in the image plane is visually defined to be
small; b) the 1-D extracted profiles
from data and best fit are visually in close agreement (and, 
generally within the
errorbars in radial bins)\footnote{Note: not all radial bins have to be
  consistent with the fit: our flag represents an {\it average} and is
  meant to indicate if the size measurement is likely to be biased.}; c) \imfit\ converged on a
best-fit which did not hit the parameters limits. For Sersic fits to
the continuum, the parameters are limited to the range:
$\rm 0.2<n_{\rm Sersic}<8.0$, surface brightness $\rm I(\reff) > 0$, $\rm 0.01\kpc
<\reff<99\kpc$ and for exponential fits to $\ha$, central surface
brightness $\rm 0<I_0<10.I_{peak}$ where $I_{peak}$ is the maximum surface
brightness per pixel, and $\rm 0.01\kpc <\reff<99\kpc$. 
399 of 457 galaxies meet these criteria, of which 207 fits are excellent. 

At native CANDELS resolution in F160W, 559 of 645 galaxies are well fit with a single Sersic profile. 
We also require best fit CANDELS F160W ellipticities
$\rm \epsilon<0.7$ (which corresponds to an inclination of $i\gtrsim72.5$
in an infinitely thin disk) because for the edge-on cases the intrinsic disk
thickness cannot be ignored.

Merging all these criteria, our final $\ha$ MAIN sample
contains 281 galaxies, with 89 in the $\ha$ BEST sample. For the
remainder of the paper we will discuss the results derived for the
$\ha$ MAIN sample only, having tested that there are no qualitative
changes if we restrict to the $\ha$ BEST sample.

Of the final $\ha$ MAIN (BEST) galaxy sample, 42 (11) have clear
broad lines, and 38 (13) have known AGN. We do not remove these
from the samples as these galaxies have been screened for a good fit
exponential profile, suggesting that the outflow / AGN does not
dominate the profile. Results are examined with and
without these objects, with no notable difference to our conclusions. 

The samples for comparison of CANDELS
F160W and KMOS continuum sizes are slightly different, requiring good (excellent) fits in
both continuum bands and ellipticities $\rm \epsilon<0.7$. 
This results in sample sizes of 288
and 193 for the continuum MAIN and continuum BEST samples respectively.
Galaxy sizes and errors for the $\ha$ MAIN sample are provided in Appendix~\ref{sec:publicdata} for the on-line version of the article.

\subsection{Sample Bias}\label{sec:selectionbias}

\begin{figure}
  \centerline{\includegraphics[width=0.5\textwidth]{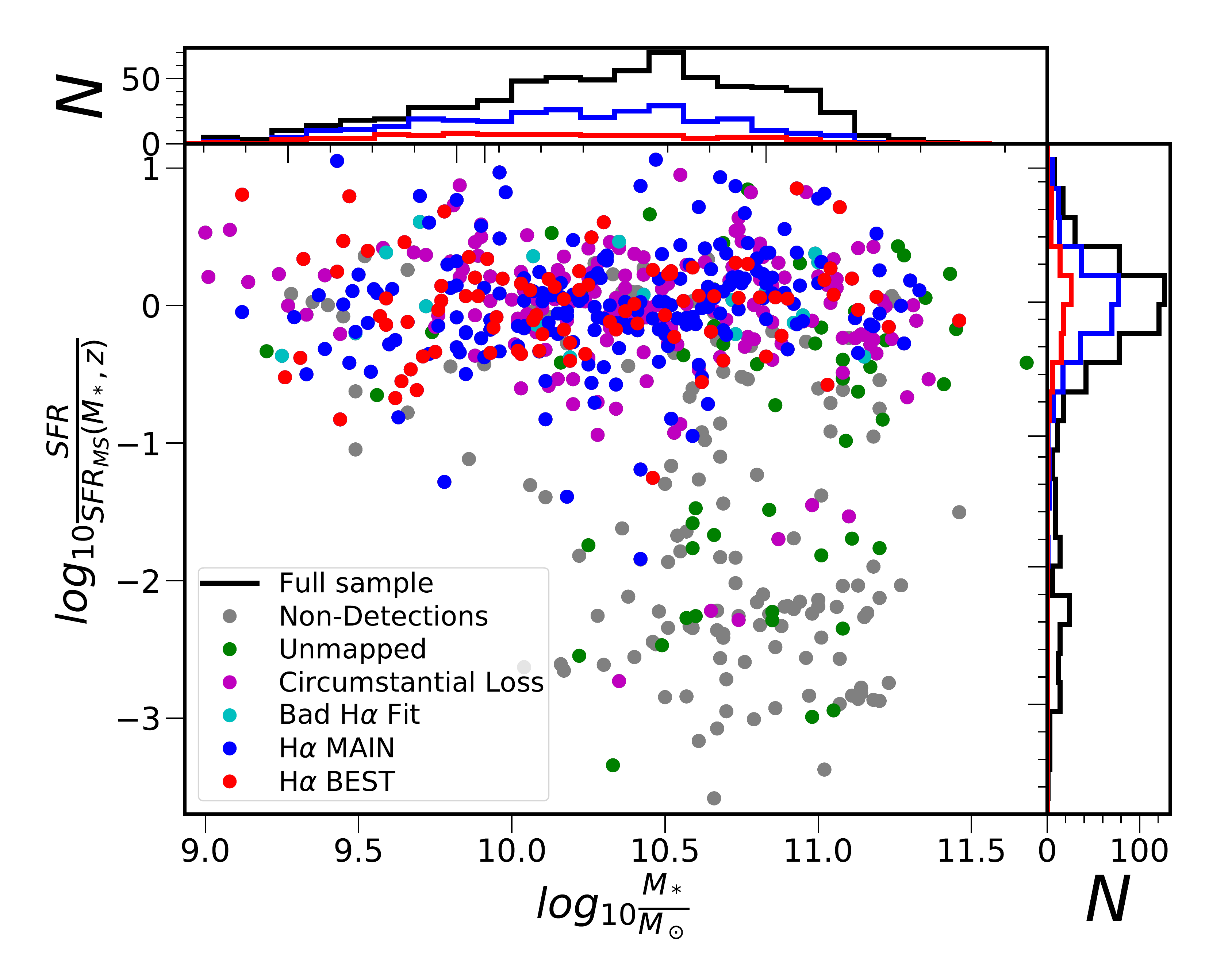}}
  \centerline{\includegraphics[width=0.5\textwidth]{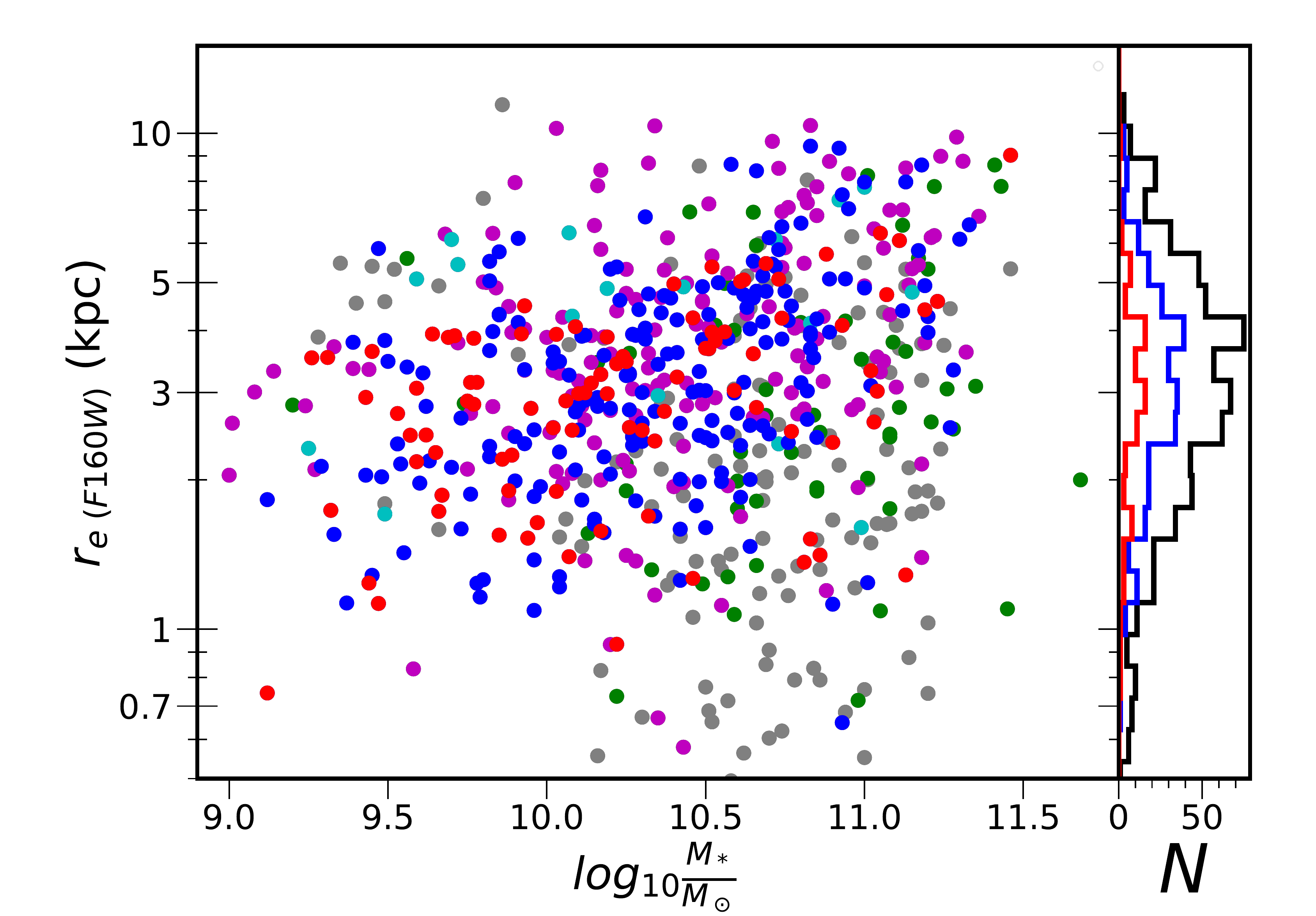}}
  \centerline{\includegraphics[width=0.5\textwidth]{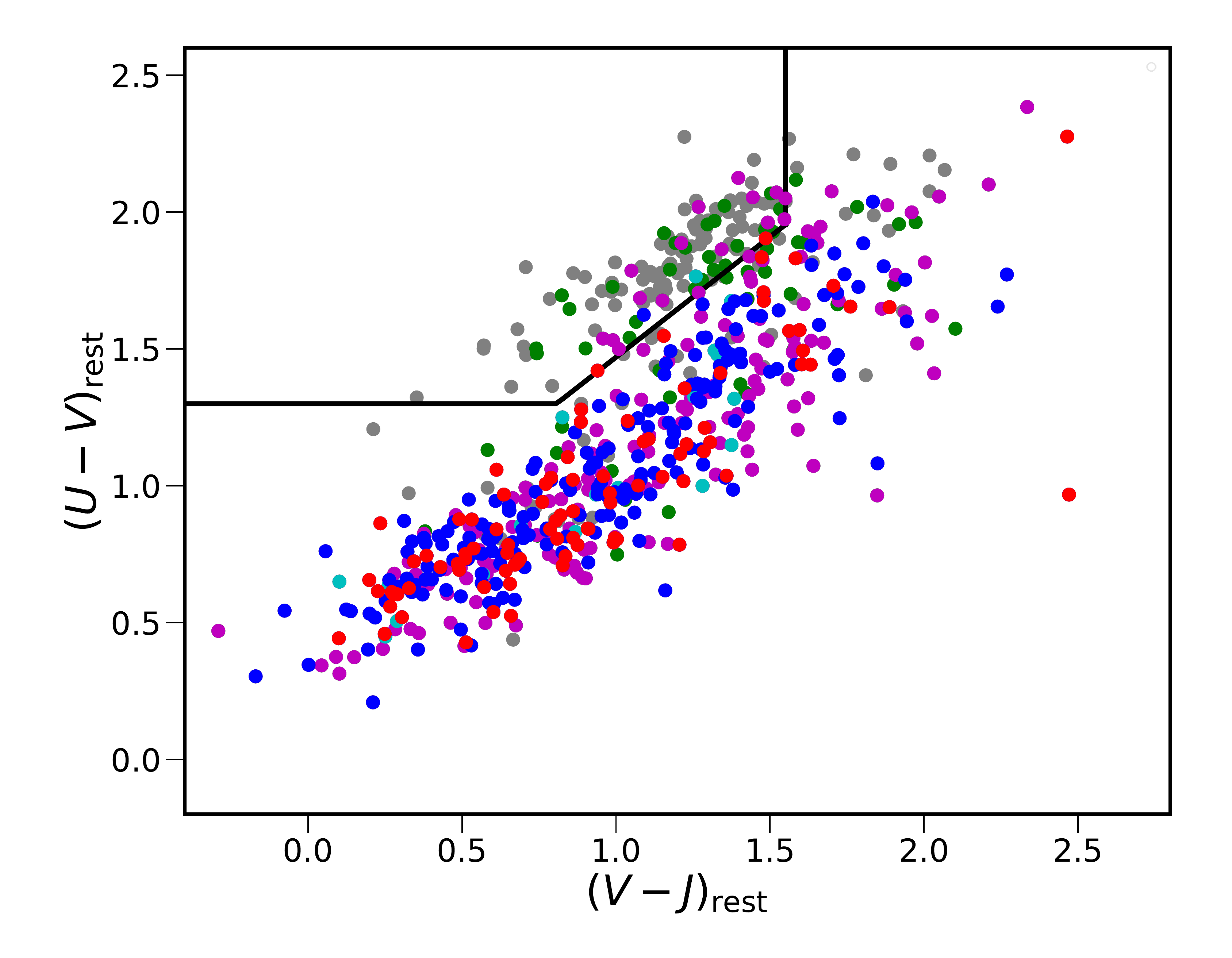}}
  \caption{Upper panel: Offset in SFR of each $\ktd$ galaxy from the
    main sequence at its stellar mass and redshift, as defined by
    \citet{Whitaker14}; middle panel: continuum (F160W)
    size versus stellar mass; and lower panel: $\rm U-V$ vs $\rm V-J$
    colour-colour space (passive galaxies are to the top-right of the black demarcation line).  The histograms show the marginalized distributions for the full, $\ha$ BEST and $\ha$
    MAIN samples in stellar mass, SFR, and galaxy size. Galaxies in the $\ha$ BEST and $\ha$
    MAIN samples occupy the whole main sequence, falling off to low SFR
    and into the passive region of the colour-colour space. 
    These samples are therefore representative of normal main sequence galaxies, with no significant
    detection bias.}
  \label{fig:sfrmassdetection}
\end{figure}

Figure~\ref{fig:sfrmassdetection} examines whether any bias might
be introduced through galaxies remaining undetected or unmapped in
$\ha$ or otherwise not included in the $\ha$ MAIN or BEST samples.  The
upper panel shows the difference between a galaxy SFR and
the main-sequence SFR of a galaxy at the same mass and redshift using the
prescription derived by \citet{Whitaker14}. We show only galaxies
observed by $\ktd$: see \citet{Wisnioski15}, \citet{Wuyts16}, and \blue{W19} for the impact of our $\Ks$-band selection in this plane.
All 645 target $\ktd$ galaxies are shown in Figure~\ref{fig:sfrmassdetection}.
The middle panel shows the
location of the same galaxies in the F160W half-light size - stellar
mass plane. 
The lower panel shows the galaxies in the ``UVJ'' ($\rm U-V$ vs $\rm V-J$)
color-color plane \citep{Williams09}, which separates passive galaxies from star-forming galaxies (at  the black demarcation line). Using two colors it is possible to disentangle passive galaxies (top left) from optically red, dusty star-forming galaxies (top right).

Figure~\ref{fig:sfrmassdetection} shows that the fraction of galaxies with detected $\ha$ is high for main sequence galaxies, then drops rapidly to low SFR and redder colors. 
Galaxies designated ``unmapped" in Figure~\ref{fig:sfrmassdetection} can have either low signal to noise ratio or chaotic kinematics / dominant
broad line components, and it is impossible to trace a dominant star-forming disk-like component. 
Combining non-detections and unmapped galaxies accounts for 83\% of UVJ passive galaxies and 98\% of galaxies more than 1.0 dex below the main sequence, but only 14\% of UVJ star-forming galaxies and 11\% of galaxies less than 0.3 dex below the main sequence. 
Of this, the unmapped population contributes an increasing fraction at high mass, reaching $\sim 20\%$ for $\log_{10}({\Mstellar}/{\Msol})>10.9$ galaxies within 1 dex of the main sequence. {These very massive objects are more difficult to map in $\ha$ due to their lower specific star formation rates and higher dust extinctions, which in turn make their redshift determinations more uncertain. However, the number of these objects is relatively low and we found no significant difference in the distributions of size and offset from the MS for the $\ha$ MAIN sample if we split above and below $\log_{10}({\Mstellar}/{\Msol})=10.5$. The $\ha$ MAIN sample probes well into the dusty 
star-forming region of the UVJ diagram, which is a key feature of the $\ktd$ survey design \citep[see][]{Wisnioski15,Wisnioski19} that helps in reducing selection biases for our sample.}
Along the main sequence, most galaxies which are not in the $\ha$
MAIN sample are dropped for circumstantial reasons which does not
introduce any selection bias (magenta points), i.e. mostly due to
significant sky line residuals or high ellipticity. The cyan points denote the few galaxies
with clean, mappable emission line signal for which the exponential fit
to $\ha$ emission was flagged as bad. There are 18 such objects of
which the majority display $\ha$ emission offset from the continuum and
some are clear mergers. These
galaxies tend to have large continuum sizes for their stellar mass
(lower panel), but given their small number and, for some of them,
uncertain continuum sizes, we consider there to be no notable bias
against normal disk-like extended star-forming disks in our sample.
Finally, we note that there is no notable difference in the properties
of $\ha$ MAIN and $\ha$ BEST galaxies supporting our decision to focus on the $\ha$ MAIN sample.

\section{Results}\label{sec:results}

\subsection{$\ha$ size correlations with continuum size and stellar mass}\label{sec:sizesizemass}

\begin{figure}
\centerline{
  \includegraphics[width=0.5\textwidth]{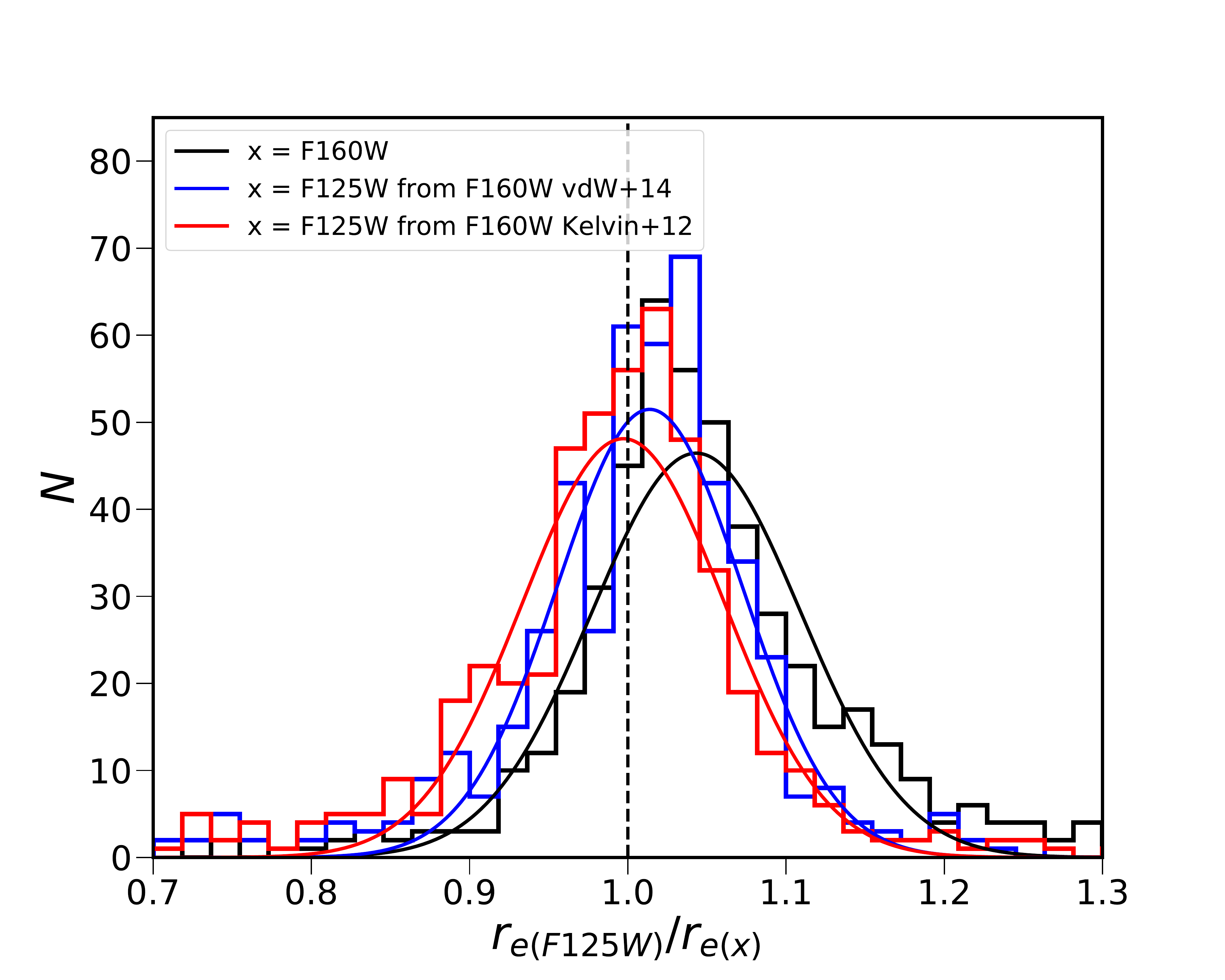}}
\caption{Histograms of the ratio of the best fit half-light size in the
  F125W band to that in the F160W band for galaxies in the {\sc
    Continuum MAIN} sample, and a Gaussian best fit (black lines). The blue and red lines show the ratio with observed F160W sizes corrected to F125W with the fitting functions of \blue{vdW14}, and \citet{Kelvin12}, respectively. The black dashed line marks where the size ratio is unity.}
  \label{fig:reratioJH}
\end{figure}

Armed with accurate galaxy half-light (size) measurements in continuum
(tracing stars) and in $\ha$ (tracing star-formation), we now examine how
the extent of star-forming gas relates to other known galaxy properties.
In particular, we are interested in how closely the distribution of new
stars, as traced by $\ha$, follows the distribution of old stars. 
Before answering this question we turn our attention to the best tracer 
for the size of old stars. \citet{Kelvin12}, and \blue{vdW14} showed that
the star-forming galaxies have negative color gradients, implying that 
their size is smaller at longer rest-frame wavelengths. This is shown in the black histogram of Figure \ref{fig:reratioJH},
where we plot the ratio of sizes in F125W to F160W. The center of the Gaussian fit assumes a value 1.042 for the size ratio. \citet{Kelvin12}, and \blue{vdW14} provided fitting 
functions for the wavelength dependence of the observed size, which we
applied to our observed F160W sizes to correct them to F125W. We note that 
the \citet{Kelvin12} correction is only a function of observed wavelength, 
while the \blue{vdW14} correction depends also on the galaxy stellar mass. The 
blue and red histograms and Gaussian fits show that these corrected sizes 
match the observed F125W sizes much better than the uncorrected data with 
average ratios of 1.013 and 0.997 for \blue{vdW14} and \citet{Kelvin12}
respectively. Due to the simpler nature of the correction proposed by
\citet{Kelvin12}, and its excellent accuracy in correcting the sizes in our
sample, we use this fitting function to correct the observed F160W sizes to
rest-frame 6500\AA. This rest-frame wavelength has multiple advantages:
first and foremost it is close to the rest-frame wavelength probed by F160W
in the center of our redshift range, and therefore a roughly equal number
of galaxies are corrected to a shorter and longer wavelength. Moreover, 
it is close to the rest-frame wavelength of the $\ha$ emission, mimicking
the observing strategy of narrow band surveys in the local Universe (where the
continuum size is evaluated from a filter close in wavelength to the narrow
band filter used for the $\ha$ observations, as done by e.g. \citealt{Fossati13, Boselli15}).
We add scatter to the correction applied to individual
galaxies by randomly sampling a Gaussian function with $\sigma$ equal to the
standard deviation of the Gaussian fit shown in Figure \ref{fig:reratioJH}.
Hereafter we will use this corrected continuum size as a tracer for the size
of the old stars and we label it $r_{e(\rm{r}6500)}$, unless otherwise noted.

\begin{figure*}
\centerline{
  \includegraphics[width=0.5\textwidth]{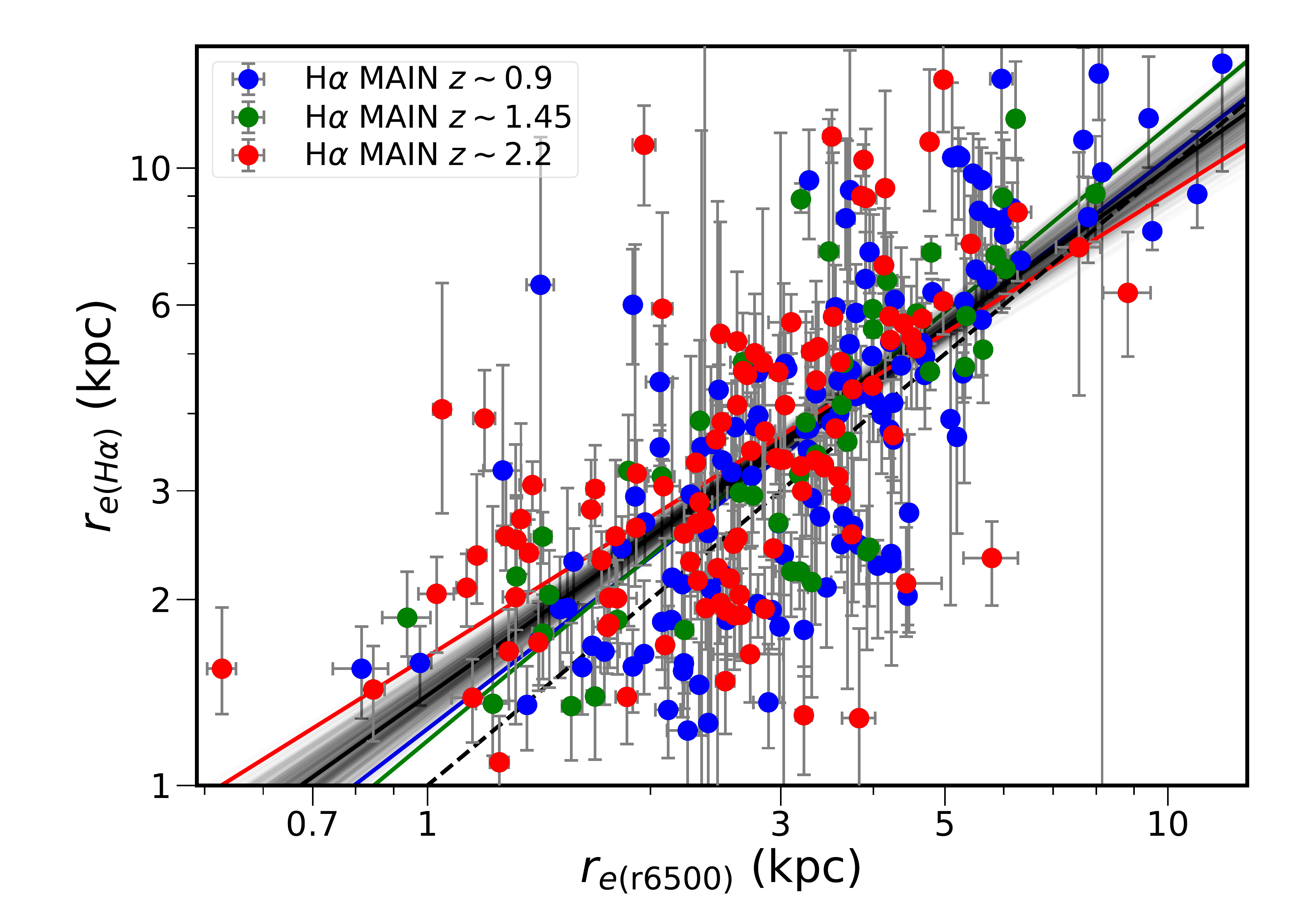}
  \includegraphics[width=0.5\textwidth]{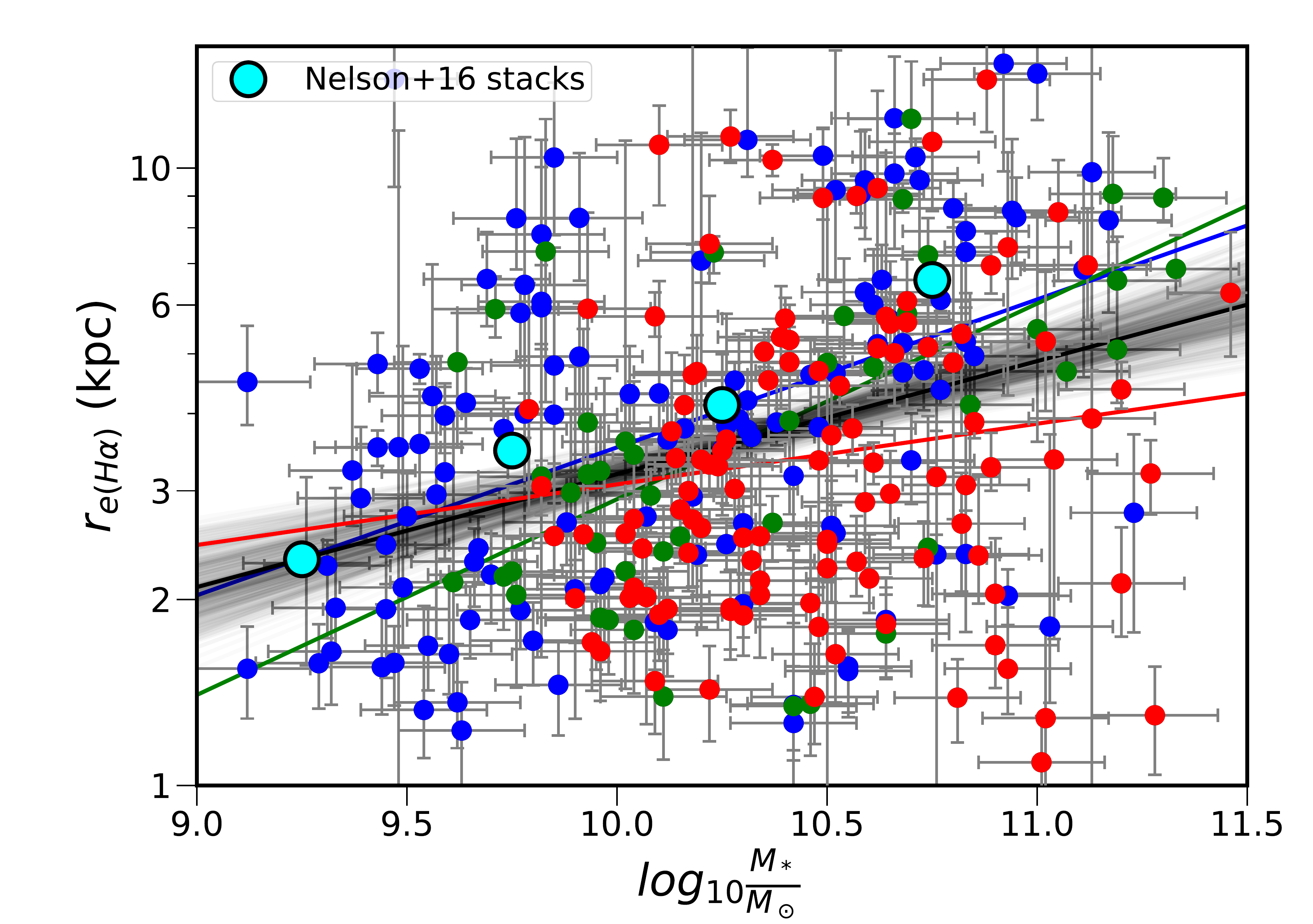}
}
\centerline{
  \includegraphics[width=0.5\textwidth]{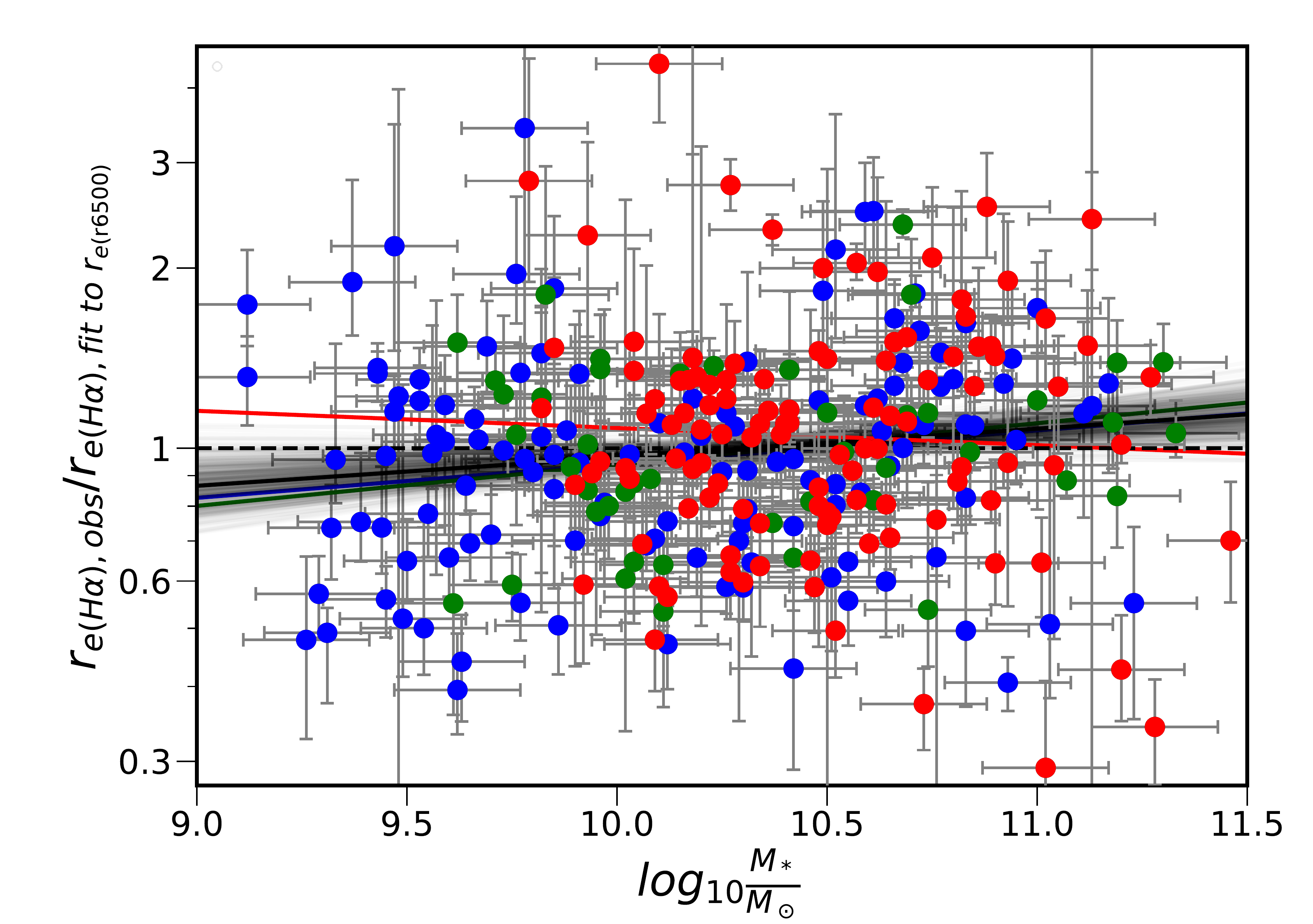}
  \includegraphics[width=0.5\textwidth]{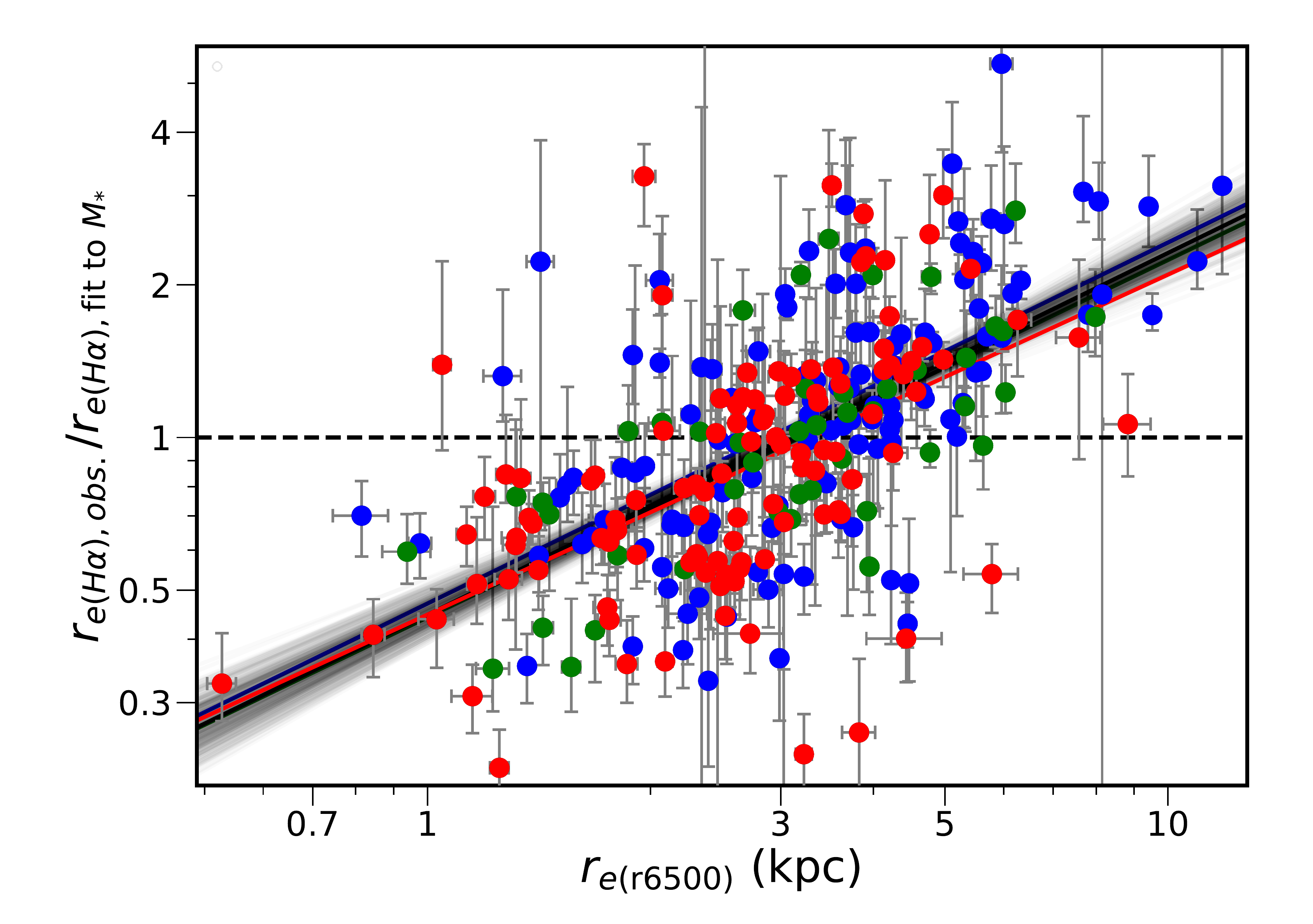}
}
\caption{Top panels: $\ha$ galaxy size plotted against: (left panel)
  galaxy size measured at 6500\AA\ rest-frame band (from F160W data corrected with the \citet{Kelvin12} function) and
  (right panel) galaxy stellar mass. The best fit and sample fits 
  from MCMC are shown with the black solid line and fainter grey lines. Galaxies
  are divided into three redshift bins (corresponding to the KMOS band
  for $\ha$ observation) with data points and fits to each
  redshift bin shown with different colours. Points from the stacked,
  circularized $\ha$ size measurements of \citet{Nelson16a}, 
  assuming an average ellipticity of $\epsilon=0.4$, are shown in
  bins of stellar mass. Bottom panels: Residual $\ha$ size after
  subtraction of the best fit relation above, plotted against the other
  parameter. The dashed black horizontal line shows the level at zero residuals. The $\ha$ size of a galaxy is 
  more tightly correlated with its continuum size than with its
  stellar mass. 
  }
  \label{fig:reha_cont_mass}
\end{figure*}

At high redshift, the high gas masses, densities and accretion rates of $\ktd$
galaxies means that their star-forming gas is mostly molecular.
Well-defined global galaxy relationships
such as that between total star-formation and stellar mass (the star-forming main sequence) exist primarily because total star
formation rates, to first order, smoothly track accretion rates, which
themselves depend mainly upon the global halo potential and growth \citep[e.g.][]{Bouche10,Lilly13}.
It remains unclear to what extent a strong evolution in mass and mass growth should be reflected in changes in galaxy size and size growth. 
Therefore we begin by asking whether the $\ha$ size of galaxies is better correlated with stellar size or total stellar mass.

\begin{figure*}
\centerline{
  \includegraphics[width=0.5\textwidth]{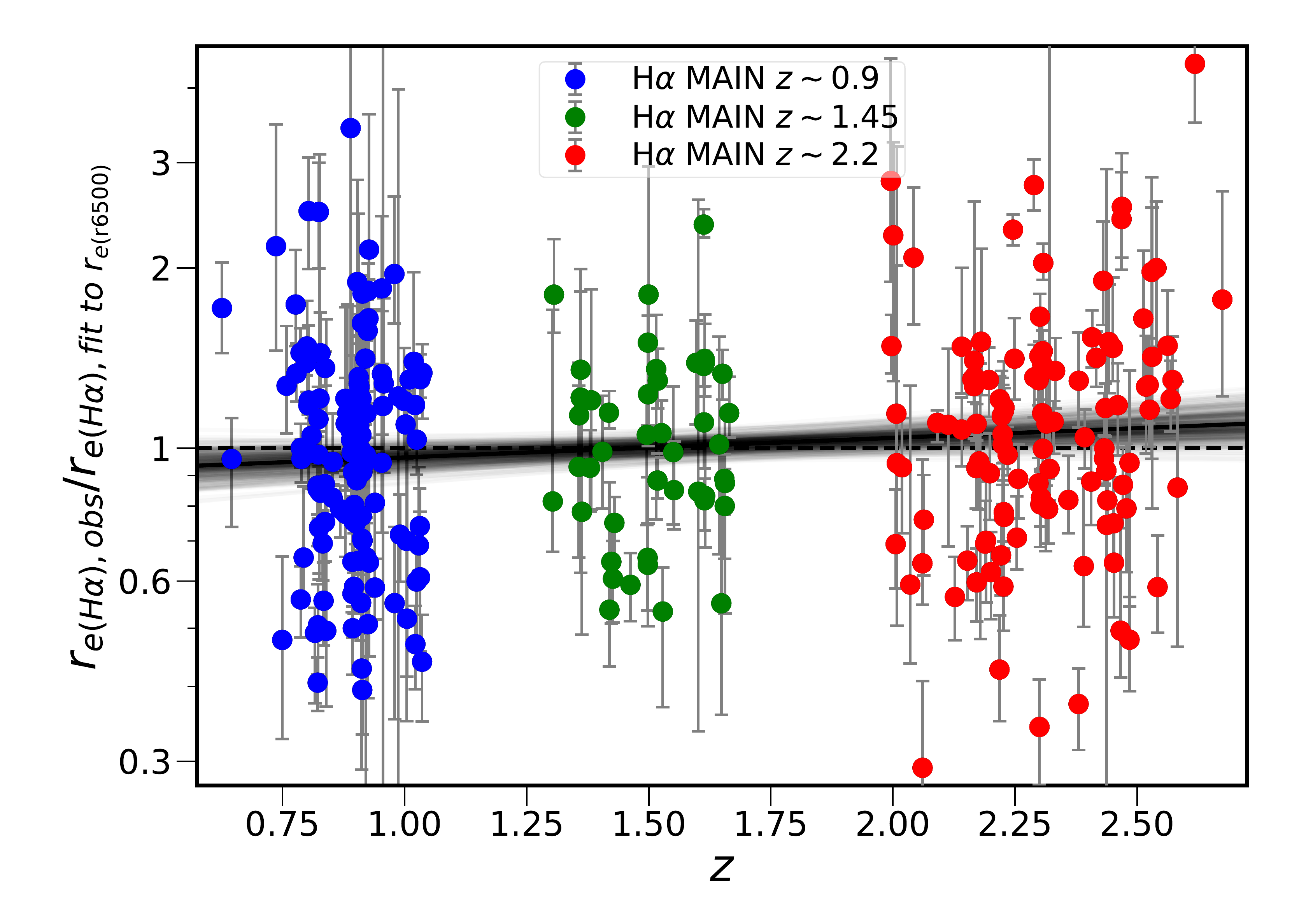}
  \includegraphics[width=0.5\textwidth]{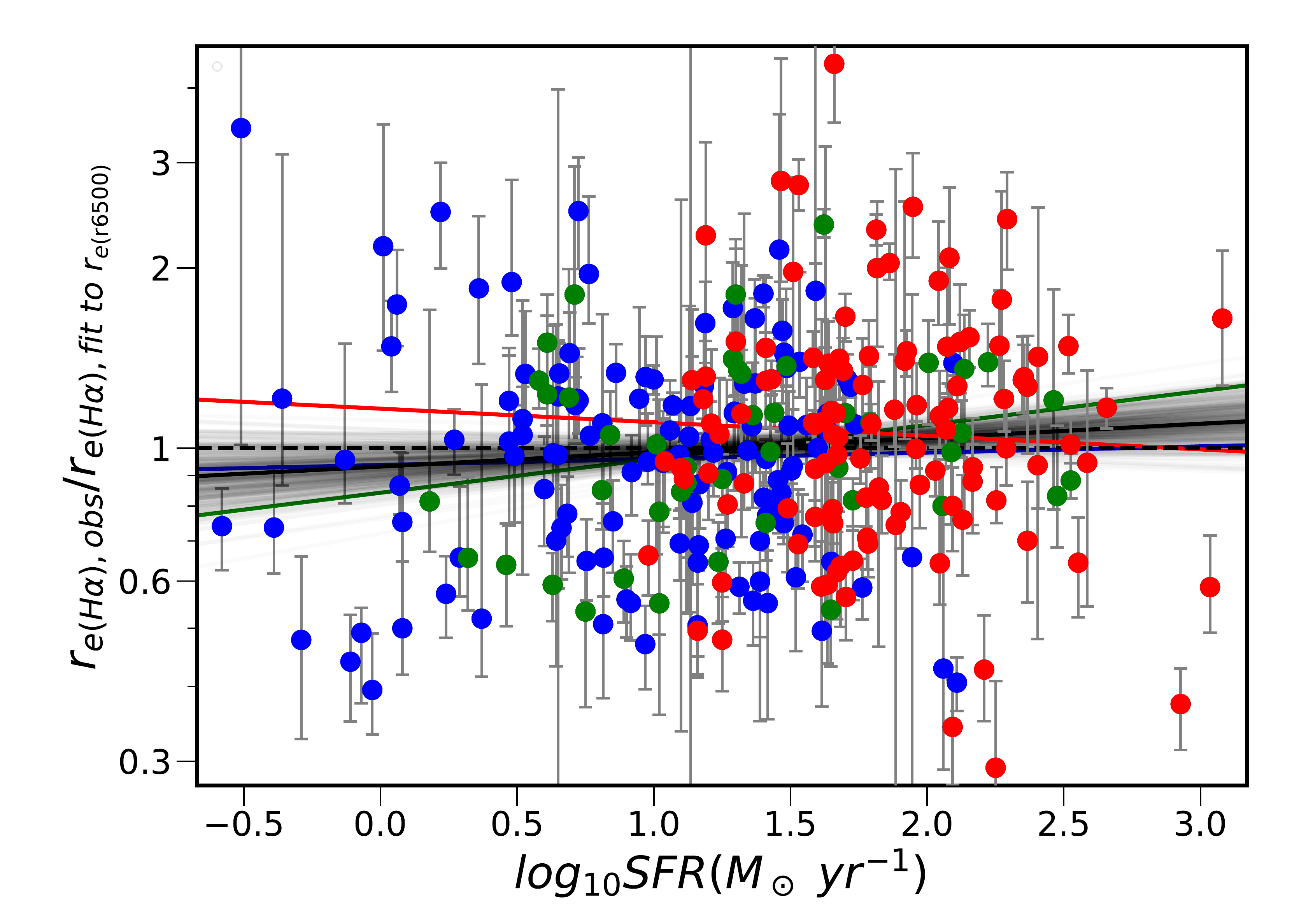}}
\centerline{
  \includegraphics[width=0.5\textwidth]{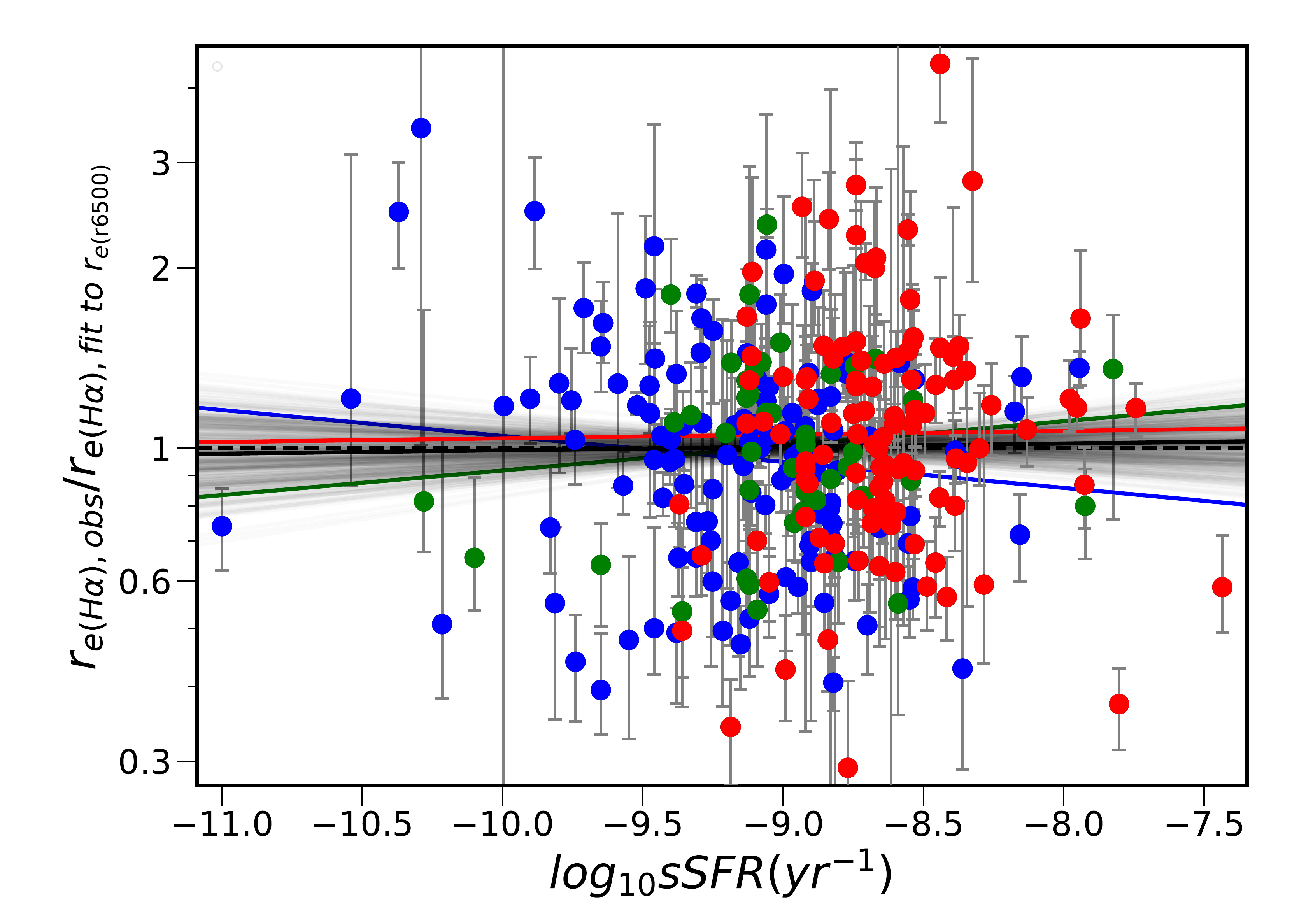}
  \includegraphics[width=0.5\textwidth]{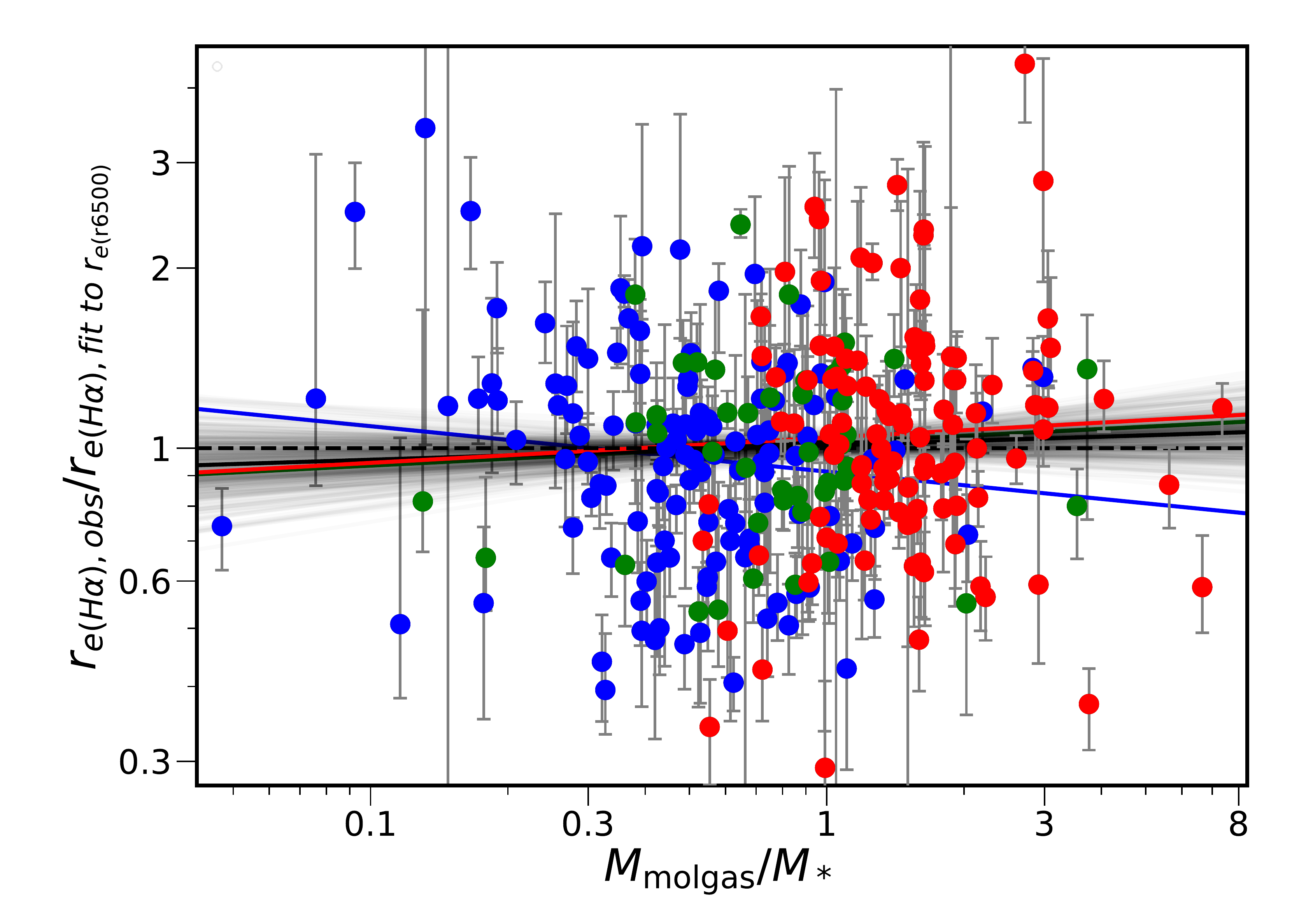}}  
\centerline{
    \includegraphics[width=0.5\textwidth]{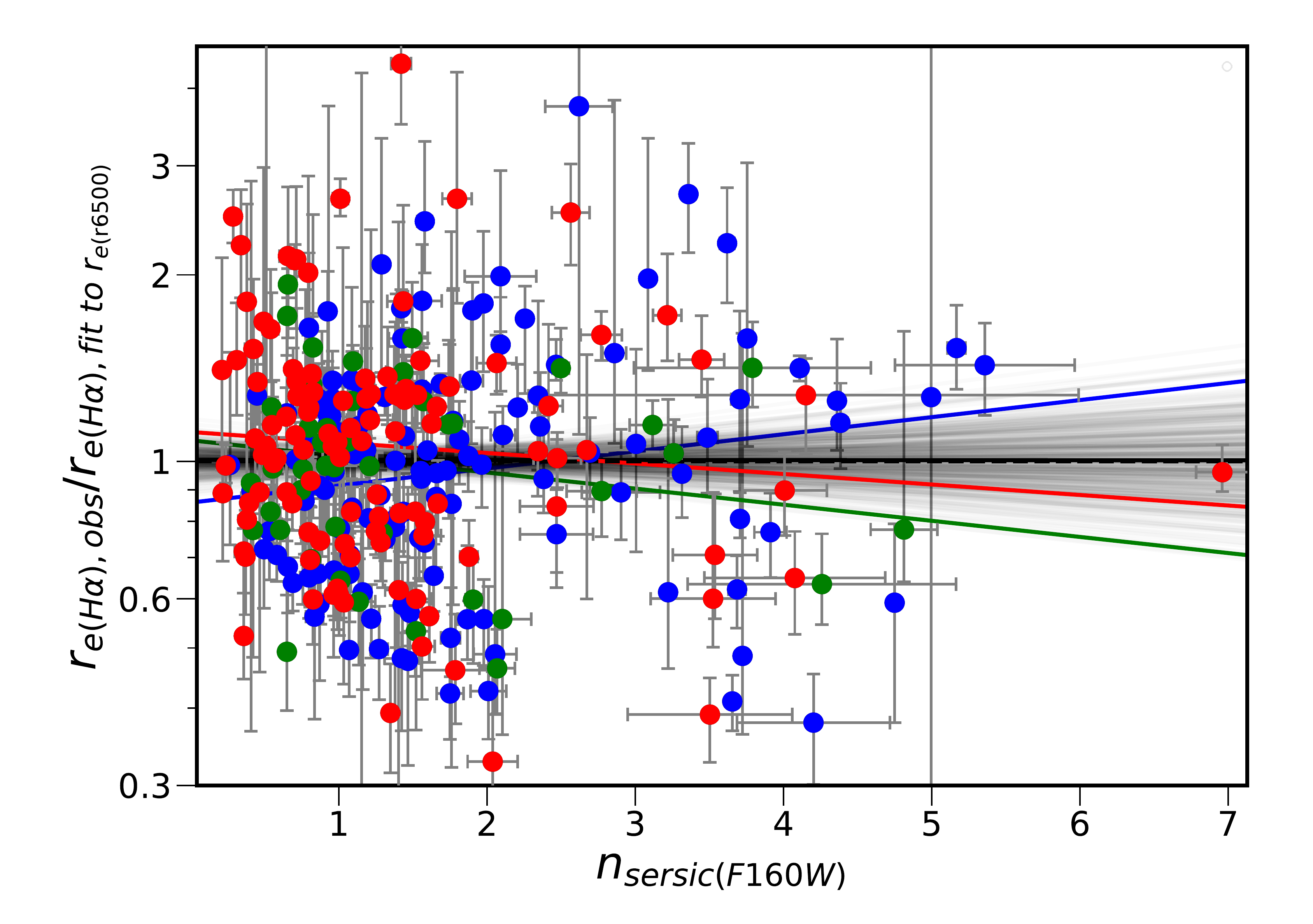}
  \includegraphics[width=0.5\textwidth]{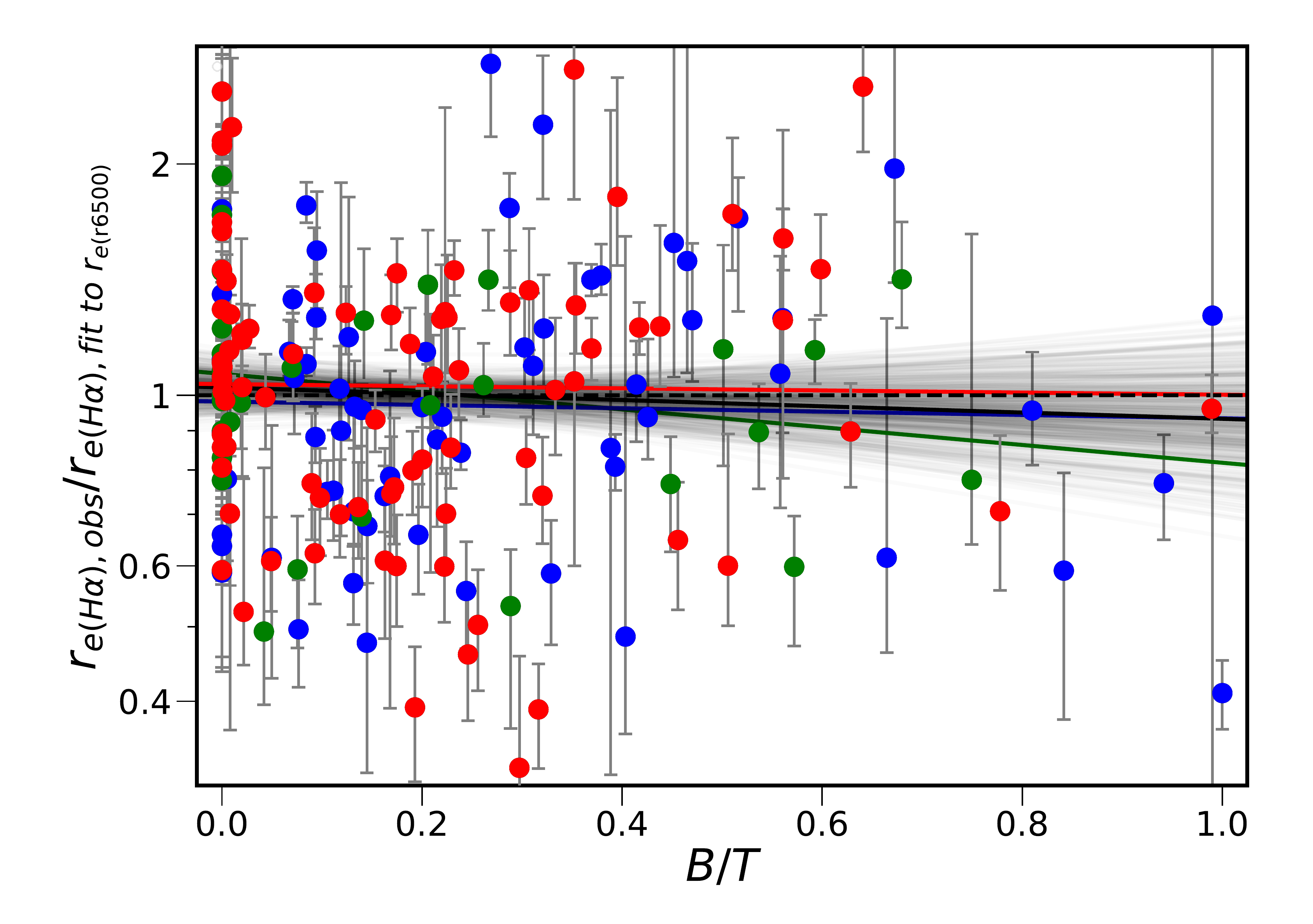}}

\caption{Residual $\ha$ size with respect to the best fit
  versus continuum size (see
  Figure~\ref{fig:reha_cont_mass} top-left panel) versus redshift
  (top left), star-formation rate (top right), specific star-formation rate (middle left), molecular gas to stellar mass ratio (as inferred from \citet{Tacconi18}, middle right), Sersic parameter measured in the F160W band (bottom left), and bulge to total ratio (as measured by \citet{Lang14} for a subset of our $\ha$ MAIN sample, bottom right). The best fit
  and sample fits from MCMC are shown with the
  black solid line and fainter grey lines (and coloured lines for the
  three independent redshift bins). The black dashed line shows the level at zero residuals. }
\label{fig:reha_res_ssSFR_z}
\end{figure*}

Figure~\ref{fig:reha_cont_mass} conclusively answers this question. In
the upper panels, we plot the $\ha$ size of $\ha$ MAIN sample
against continuum size (left panel) and stellar mass (right
panel). The positive correlation between $\ha$ and continuum size is
strong and tight, a positive correlation exists also 
between $\ha$ size and stellar mass, but with much larger scatter.

We fit these relations using the {\it linmix} package for
python\footnote{\url{https://github.com/jmeyers314/linmix}} which
follows the Bayesian framework described by \citet{Kelly07} and
encorporates measurement errors on x- and y- axes as well as an
additional component of intrinsic scatter into the fit. This code makes use of
Monte-Carlo Markov Chains (MCMC) to fit a linear relation (in this case
in the log-log plane and so corresponding to a power law relation
between linear quantities). 
A random uncertainty of 0.15 dex in $\log_{10}(M_*/\rm{M_\odot})$ is included in the analysis, consistently with \blue{vdW14}.

In the size-size plane, we find a mildly sub-linear best fit relation
with a slope of $0.85\pm0.05$ and intrinsic scatter of $43\pm3\%$ (0.15 dex) at
fixed continuum size, significantly smaller than that of continuum size versus stellar mass (56\%, or 0.19 dex \blue{vdW14}). This is shown by the black solid line in Figure~\ref{fig:reha_cont_mass}, while the grey shaded area shows a wider range of draws from the posterior.  The correlation with mass has a
slope of $0.18\pm0.03$ and intrinsic scatter of $68\pm4\%$, and is
roughly consistent with the stacked $\ha$ size measurements of
\citet{Nelson16a} who provide fits to stacked and circularised $\ha$
profiles from the 3D-HST survey in stellar mass bins (outlined cyan
circles, scaled to major axis sizes assuming an average ellipticity
$\epsilon=0.4$).

We now examine the importance of the second parameter in the lower
panels. In each case we plot the residual of $\ha$ size with respect to
the best fit relation from the upper panel against the second parameter
(i.e. left, versus stellar mass; right, versus continuum size) and then
fit relations for these residuals. Here it is clear to see that the
stellar mass adds nothing to the prediction of $\ha$ size once the
continuum size has been taken into account: the residual relation fits
a slope consistent with zero and the intrinsic scatter drops only by 1\% 
to $42\pm3\%$. On the other hand, the continuum size
correlates well to the residual of $\ha$ size at fixed stellar mass,
with a slope of $0.71\pm0.05$ and an intrinsic scatter of $46\pm3\%$,
down from $68\pm4\%$ when fitting versus stellar mass only.

The stellar mass only has any relevance because it is correlated with the
continuum size: once the correlation with continuum size is removed
then there is no residual relation of $\ha$ size with stellar mass. In
other words, star-formation, on average, spatially tracks existing
stars, but at fixed continuum size the global amount of stars has no relevance. 
The scatter of $\ha$ size with stellar mass has a larger contribution from the $55\%$ scatter between continuum size and stellar mass\footnote{This value derived from our sample is fully consistent with the 56\% found by \blue{vdW14}.} than from the $43\%$ scatter between $\ha$ size and continuum size.

We divide the $\ha$ MAIN galaxies into three sub-samples of redshift,
according to the KMOS band in which we observe the $\ha$ emission line.
These are $0.58 \leq z < 1.04$ (blue points, YJ band), $1.27 \leq z <
1.62$ (green points, H band) and $1.98 \leq z < 2.68$ (red points, K
band). We also derive best fit relations for each of these
sub-samples (coloured lines). Best fit slopes to the continuum size -
$\ha$ size relation of $0.93\pm0.08$ (YJ), $1.00\pm0.11$ (H) and
$0.75\pm0.09$ (K) are consistent with our combined best fit
relation within 2$\sigma$, and with close to a
linear relation (with the possible exception of the highest redshift bin which is $2.8\sigma$ away).

The fit intercept corresponds to the typical $\ha$ size of a galaxy at
a particular continuum size: for a near linear relation this is a near
constant ratio. We derive this ratio at the median continuum size of
$3.23\kpc$ to be $1.18\pm0.03$, i.e. a median $\ha$ size which is
$18\%$ larger than the continuum size. Folding in the measured intrinsic scatter, this corresponds to a mean $\ha$ size which is $26\%$ larger than the continuum size\footnote{The mean of a
  log-normal distribution $=$ median$+0.5\sigma^2$.}. This compares to the typical
size ratio of $\sim1.3$ found by \citet{Nelson12} in highly star
forming 3D-HST galaxies at $z\sim1$, and the median ratio from stacked
profiles of normal star-forming 3D-HST galaxies
\citep[$\sim1.1$][]{Nelson16a}. Within our wide redshift range, there is no
evidence for evolution in this value, with consistent best
fit values of $1.13\pm0.05$, $1.17\pm0.06$ and $1.20\pm0.05$ in the three redshift bins defined above. So while $\ha$ sizes do track existing stellar sizes they are, in a median (mean)
sense, larger by $\sim 18\%$ ($\sim 26\%$) over our full redshift range. That star-formation
sizes are larger than stellar sizes is a pre-condition
for in-situ size-growth, and we shall return to this topic in
Section~\ref{sec:sizegrowth}. 

\subsection{Which other parameters influence $\ha$ size?}\label{sec:otherpars}

Figure~\ref{fig:reha_cont_mass} demonstrates that, at fixed galaxy
continuum size, there is no significant residual dependence on 
redshift (when split into three bins).
This is confirmed by directly fitting the residual to the $\ha$
size -- continuum size best fit against redshift
(Figure~\ref{fig:reha_res_ssSFR_z}, top left panel). 
We find only a marginal dependence on redshift at $<2\sigma$ level, 
and the intrinsic scatter of the size-size relation does not drop when
redshift is included as a third parameter. 

In Table \ref{table:restrendstats} we show the best fit slope $\beta$\footnote{With the symbol $\beta$ we simply refer to the slope of the power-law fit and this parameter has no relation with the UV $\beta$ slope.}, the 
fraction of Monte-Carlo realizations with $\beta$ greater than zero
($P(\beta > 0)$), the Spearman rank correlation
coefficient ($\rho$) and the probability that it is 
consistent with the null (no-correlation)
hypothesis ($P(\rho |null)$) for the residuals of 
the size-size relation versus several other parameters.

\begin{table*}
\begin{center}
\caption{Correlation of offset from $\reff(\ha) - \reff(\rm{r}6500)$ relation (Figure~\ref{fig:reha_cont_mass} top-left panel) with other parameters.}
\vspace{0.1cm}
\begin{tabular}{cccccc}
  \hline\hline
  \noalign{\smallskip}
  {\sc Parameter, X} & {\sc $\beta$} & {\sc $P(\beta > 0)$} & {\sc $\rho$} & {\sc $P(\rho |null)$}\\
  \noalign{\smallskip}
  \hline
  \noalign{\smallskip}
  z & $0.03\pm0.02$ & $0.978$ & $0.115$ & $0.05$\\
  $\log_{10}({\Mstellar}/{\Msol})$ & $0.04\pm0.02$ & $0.974$ & $0.114$ & $0.05$\\
  $\log_{10}(SFR(M_\odot~yr^{-1}))$ & $0.02\pm0.02$ & $0.921$ & $0.088$ & $0.14$\\
  $\log_{10}(sSFR(yr^{-1}))$ & $0.01\pm0.02$ & $0.591$ & $0.037$ & $0.53$\\
  $\log_{10}(\deltaMS)$\footnote{$\deltaMS = {SFR(\Msol.yr^{-1})}/ {SFR_{MS,z,\Mstellar}(\Msol.yr^{-1})}$ with the parametrizations of the MS from \citet{Whitaker14}.} & $-0.01\pm0.03$ & $0.424$ & $-0.015$ & $0.80$\\
  $\log_{10}(\fgas)$\footnote{Inferred via $\log_{10}(\fgas) = -1.25 + 2.6log_{10}(1+z) + 0.53log_{10}(\deltaMS) -0.36log_{10}(\Mstellar)$ \citep[][best fit relation]{Tacconi18}.} & $0.02\pm0.03$ & $0.767$ & $0.075$ & $0.21$\\
  $n_{Sersic,F160W}$ & $0.00\pm0.01$ & $0.496$ & $-0.005$ & $0.93$\\
  $B/T,F160W$\footnote{Restricted to galaxies with valid B/T measurements in the F160W band from \citet{Lang14}.} & $-0.04\pm0.06$ & $0.249$ & $-0.020$ & $0.79$\\
  $A_V$ & $-0.03\pm0.02$ & $0.059$ & $-0.08$ & $0.18$\\
  $\log_{10}(L_{H\alpha,erg.s^{-1}})$ from fit\footnote{\label{footnote:fit}$L_{H\alpha,erg.s^{-1}} = 2\pi.h^2.I0.(1.-\epsilon)$} & $0.11\pm0.02$ & $1.0$ & $0.31$ & $<10^{-5}$\\
  $\log_{10}(\frac{SFR(\Msol.yr^{-1})}{L_{H\alpha,erg.s^{-1}}})$ from fit\footref{footnote:fit} & $-0.10\pm0.02$ & $0.0$ & $-0.22$ & $0.0001$\\
  $\log_{10}(\frac{SFR(\Msol.yr^{-1})}{L_{H\alpha,erg.s^{-1}}})$ from fit\footref{footnote:fit} and
  dust-corrected\footnote{\label{footnote:W13}Using the differential
    dust recipe of \citet{Wuyts13}: $\rm\,log_{10}(L_{H\alpha,erg.s^{-1},dust.cor}) = log_{10}(L_{H\alpha,erg.s^{-1}}) + 0.4 (1.9 A_{cont} - 0.15 A_{cont}^2)$} & $-0.11\pm0.03$ & $0.0$ & $-0.23$ & $0.0001$\\
  $(U-V)_{\rm rest}$  & $-0.09\pm0.06$ & $0.080$ & $-0.072$ & $0.25$ \\
  $(V-J)_{\rm rest}$  & $-0.09\pm0.05$ & $0.031$ & $-0.091$ & $0.13$ \\
  $\epsilon$\footnote{Galaxy ellipticity, $\epsilon$ from fit to F160W band.} & $-0.13\pm0.14$ & $0.171$ & $-0.122$ & $0.04$\\
  $\log_{10}({\rm [NII]}/{\rm H\alpha})$\footnote{Restricted to galaxies in the $\ha$ MAIN sample for which the skyline residual contamination at the wavelength of the red [NII] emission line is not strong.} & $0.02\pm0.07$ & $0.594$ & $0.074$ & $0.34$ \\
  $\log_{10}(\delta_{0.75})$\footnote{Environmental overdensity in 0.75\Mpc\ apertures from \citet{Fossati17}.} & $-0.01\pm0.03$ & $0.36$ & $-0.07$ & $0.37$ \\
  $\log_{10}({M_{h|cen,50\%}}/{\Msol}$)\footnote{50\%ile of probability distribution function for halo mass assuming the galaxy is the central of its halo from \citet{Fossati17}.} & $0.05\pm0.03$ & $0.96$ & $0.09$ & $0.12$\\
  $\rm\,P_{sat}$\footnote{Probability galaxy is a satellite, as calibrated by \citet{Fossati17}.} & $-0.08\pm0.05$ & $0.07$ & $-0.07$ & $0.23$\\
  \noalign{\smallskip}
  \hline\hline
\end{tabular}
\vspace{0.1cm}
\begin{scriptsize}
  \begin{minipage}{18cm} {\sc Notes.} Best fit between the
    parameter given in the first column (X) and the residual $\ha$ size relative to the best fit versus continuum size for the full $\ha$
    MAIN sample ($Y=\frac{r_{e(H\alpha)\rm{,obs}}}{r_{e(H\alpha)\rm{,fit}}}$). Fits are of the form $Y = 10^{\alpha + \beta.X}$, with
    the slope $\beta$ and $1-\sigma$ errors given in column 2. In
    column 3, $P(\beta>0)$ is estimated from the fraction of MCMC 
    realizations with a positive slope, so values
    around $0.5$ are random, while values of $0$ or $1$ indicate a
    significant negative or positive slope respectively.  Column 4
    contains the Spearmann rank correlation coefficient, and column 5
    indicates the probability (p-value) of such a value assuming
    no correlation between the two parameters. 
\label{table:restrendstats}
\end{minipage}
\end{scriptsize}
\end{center}
\end{table*}

Having explored the role of continuum size, stellar mass and redshift, we now turn at examining the dependence on star-formation rate, and other quantities which are known to correlate with it. This is done in Figure~\ref{fig:reha_res_ssSFR_z}, where we plot
the size-size relation residuals vs the star-formation rate in the top right panel and specific star-formation rate in the middle left panel. In the middle right panel, we look at the residuals vs the ratio of inferred molecular gas mass to stellar mass, $M_{\rm{molgas}}/\Mstellar$, estimated using the relation of \citet[][their BEST sample as in their Table~3b]{Tacconi18} which depends on z, $\rm\,\deltaMS$ (logarithmic offset from the \citet{Whitaker14} MS relation) and $\Mstellar$. 
There is no significant trend in residual size with SFR, sSFR, $\delta(MS)$ or inferred $M_{\rm{molgas}}/\Mstellar$, nor any notable decrease in
scatter, suggesting that -- at least for normally star-forming main
sequence galaxies -- the star-forming gas traces the stars in exactly
the same way independent of the relative amount of star-forming gas or of star formation efficiency within the limits to which we can measure it.

In the bottom panels of Figure~\ref{fig:reha_res_ssSFR_z} we examine the dependence on galaxy
morphology parametrized by the Sersic index of our fit in the F160W band, and the bulge to total ratio $B/T$ from the fits of \citet{Lang14}). The sample of \citet{Lang14} partially overlaps with the $\ha$ MAIN sample, with only 60\% of the galaxies having a valid value of $B/T$. Nonetheless this subsample is large enough to derive statistically robust conclusions. 
As demonstrated in the figure and in the table, there is no significant dependence.  his contrasts with the situation in the local Universe, where normally star-forming (gas rich)
galaxies with little or no bulge tend to have very similar $\ha$ and
continuum sizes, whereas those with more significant bulges have
relatively larger $\ha$ sizes \citep{Fossati13}. In galaxies with more
bulge, the half-light size of the combined bulge$+$disk is less than
that of the disk, and the $\ha$ emitting gas usually shows little sign
of a bulge component. Therefore it is perhaps surprising that our
$\ktd$ galaxies with higher Sersic index or higher $B/T$ measurements
show no indication of relatively larger $\ha$ sizes. Morphology does
not seem to play a role in driving the relative size of the $\ha$ disk 
in the high redshift Universe, at least not measurably within our 
star-forming sample.

In Table \ref{table:restrendstats} we test even more parameters, including those related to the dust extinction, the galaxy color, the ratio of SFR to $\ha$ luminosity, the ratio [NII]/$\ha$, and the galaxy environment (parametrized by the local galaxy overdensity, halo mass, and probability of being a satellite galaxy from \citet{Fossati17}). 

Our analysis shows that most parameters show little or no correlation with the residual $\ha$ size, and the only 
significant trends are those related
to dust extinction which we discuss further in
Section~\ref{sec:dust}. 
None of these results are significantly
changed if we restrict ourselves to the $\ha$ BEST sample, eliminate galaxies hosting AGN or broad line emission, 
or limit to the stellar mass range of our highest redshift bin ($\log_{10}({\Mstellar}/{\Msol})\gtrsim10$).

\subsection{Caveats}\label{sec:caveats}

In detail, the picture is more complicated. While much of the $\ha$ emission locally traces the ionizing star-formation, there can be a more diffuse component, additional sources of ionization, and much of the emission can be obscured by dust, especially in dense and infrared-bright starbursting regions. For example, two very high mass star-forming galaxies in our sample (U4\_20704 and U4\_36247 at $\rm
\Mstellar > 10^{11}\Msol$) which have been observed by ALMA at $\rm
870\mu m$ \citep{Tadaki17} to have significantly smaller
submillimeter sizes than their $\ha$ or F160W sizes.  

\begin{figure}
\centerline{\includegraphics[width=0.5\textwidth]{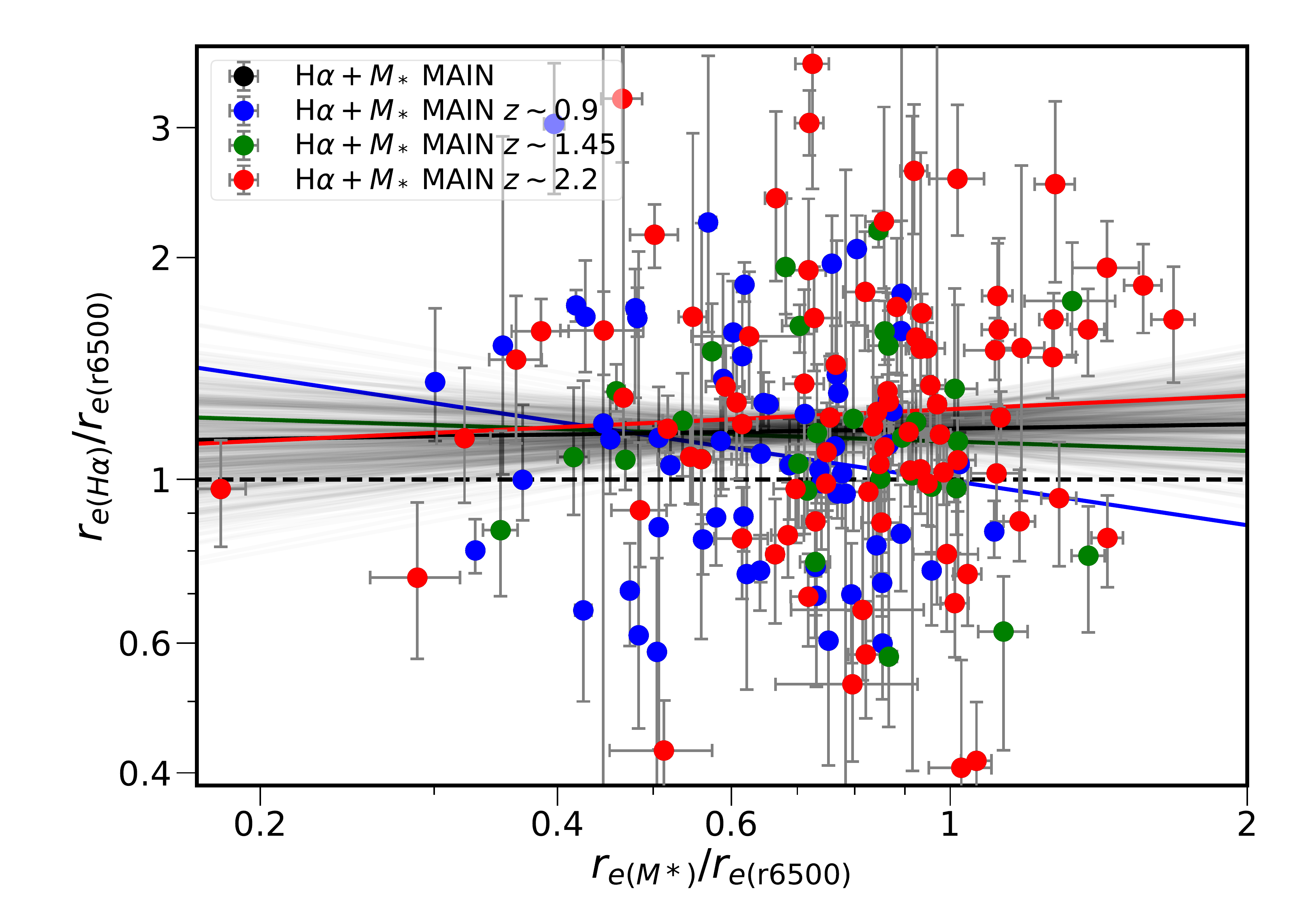}}
\caption{Ratio of the half-mass (from \citealt{Lang14}) to the half-light size versus the ratio of the $\ha$ size to the half-light size. Points and lines are color coded as in Figure \ref{fig:reha_res_ssSFR_z}. The lack of a significant correlation demonstrates that whatever drives variations in $\reffmstar/\reffcorr$ is not driving variations in $\reffha/\reffcorr$. }
\label{fig:massize_correlations}
\end{figure}

Also the continuum emission, which we have corrected to rest frame 6500\AA\ to remove the effects of a variable rest frame wavelength for galaxies at different redshifts, does not perfectly trace the stellar mass in the galaxy. We examine the Sersic profile fits to stellar mass maps produced by \citet{Lang14} for $0.5<z<2.5$ and $\Mstellar>10^{10}\Msol$ galaxies in our sample.
We find for $n_{\rm Sersic}<2$ galaxies that half-mass sizes are on average $\sim75-80\%$ of the rest frame 6500\AA\ half-light sizes with a $<5\%$ dependence on redshift, consistent with \citet[][their Figure~11]{Wuyts12}. 
The ratio $\reffmstar/\reffcorr$ decreases for high $n_{\rm Sersic}$ sources in our lower redshift bin, possibly demonstrating the influence of a higher mass-to-light ratio bulge in such galaxies. However, as shown in Figure \ref{fig:massize_correlations}, we find no correlation between $\reffmstar/\reffcorr$ and $\reffha/\reffcorr$, demonstrating that the driver of variations in $\reffmstar/\reffcorr$, such as large bulges in some galaxies at lower z, is not driving variations in $\reffha/\reffcorr$. 
This suggests that, in our observed continuum band and for the normally star-forming galaxies in our sample, we are predominantly tracing the disk and we are measuring disk sizes, even in galaxies with significant bulges.

The colour gradients in late-type galaxies appear likely to result
predominantly from gradients in dust extinction \citep{Pastrav13}. In this case, any dust gradient will effect the extinction gradient for $\ha$ emission as
well as the continuum. Under the foreground screen approximation, these
are equally affected, and the ratio of half-light sizes should be
unaffected. However, as we shall see in Section~\ref{sec:dust}, there is excess extinction of $\ha$ emission associated to dust embedded in
the star-forming HII regions, well described on average by differential
extinction laws such as \citet{Wuyts13}. The dustier part of the galaxy
(typically the centre, \citealt{Wuyts12,Nelson16}) will have extra
extinction in $\ha$ compared to continuum implying that, while both
observed sizes will be larger than the true size, the $\ha$ size should
be increased by a larger amount. In other words, the ratio of $\ha$ to continuum size should be an upper limit for the ratio of the size of the star-forming disk to that of the stellar mass disk.

\begin{figure*}
\centerline{\includegraphics[width=0.5\textwidth]{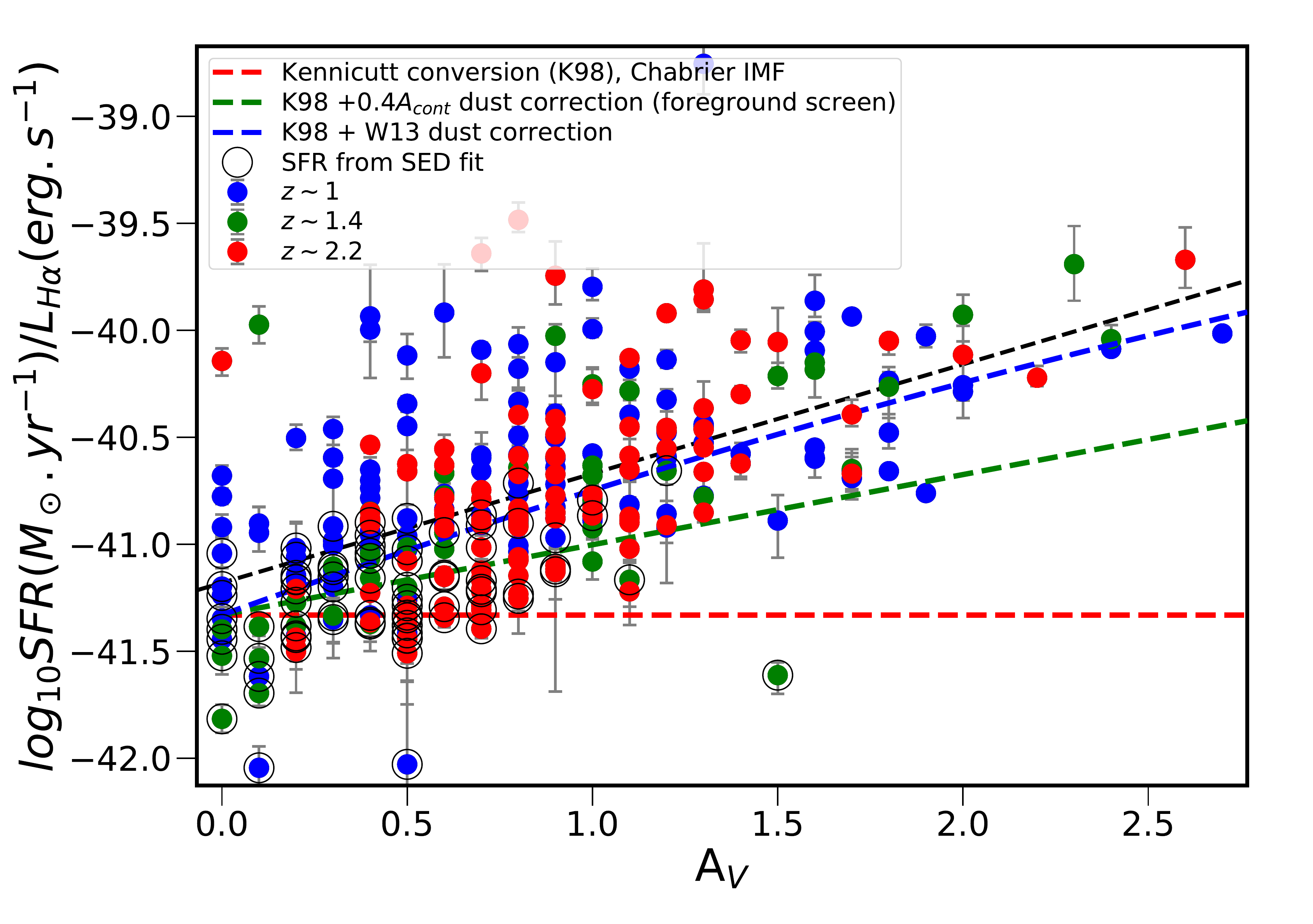}
  \includegraphics[width=0.5\textwidth]{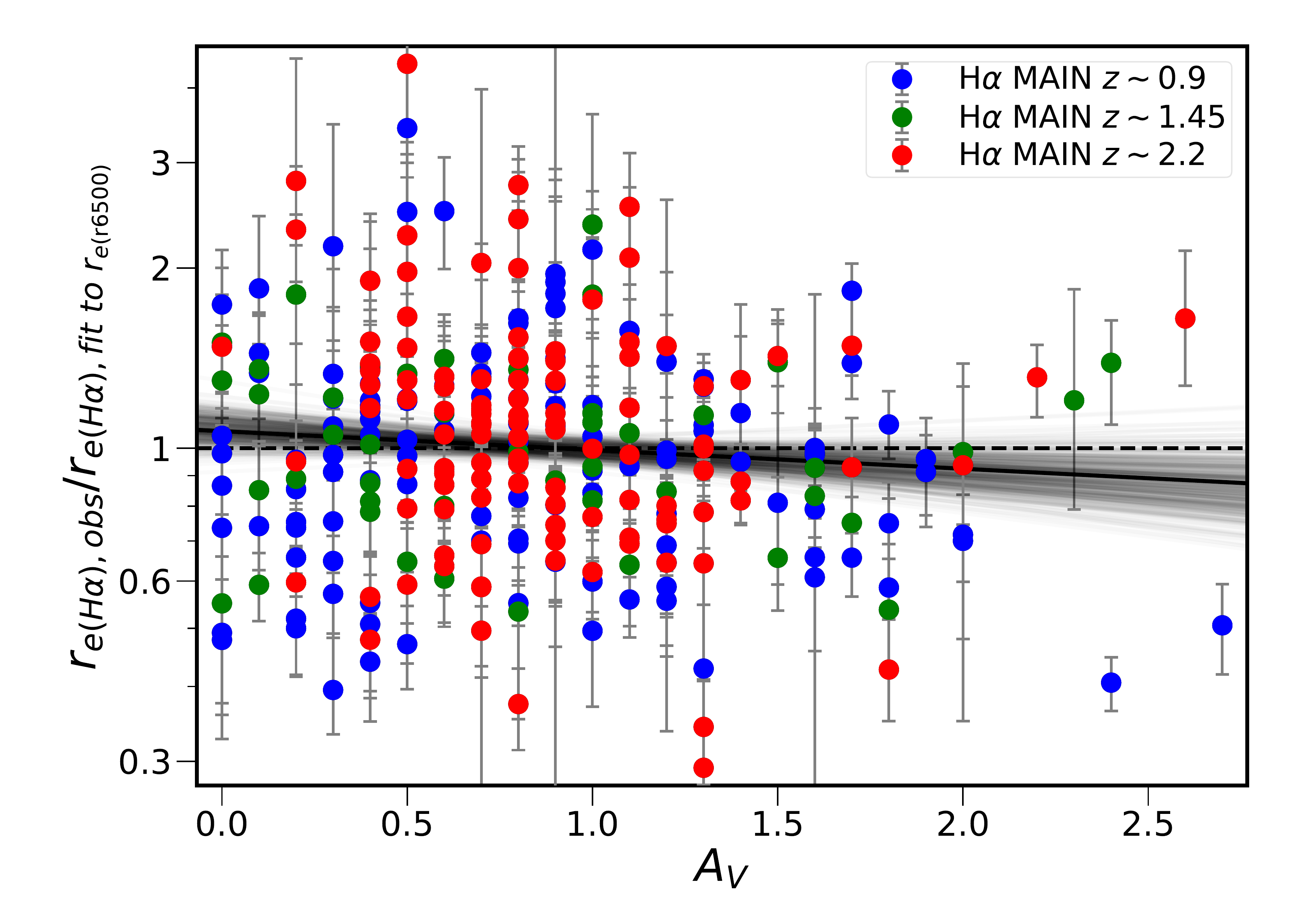}}
\caption{Left: Compares the integrated dust obscuration of the stars with that of $\ha$ emission tracing SFR. The x-axis value $A_V$ is estimated assuming a
  foreground dust screen from SED fitting \citep{Wuyts11}, while the
  y-axis is the ratio of the total estimated SFR (from IR$+$UV where
  available or SED otherwise) to the total $\ha$ luminosity from the exponential fits. The
  horizontal dashed red line denotes the canonical calibration of SFR
  to $L_{\ha}$ from \citet{Kennicutt98}, assuming a Chabrier initial
  mass function (IMF) of stars. Assuming star-formation is obscured to
  the same degree as older stars, we get the dashed green line. An
  additional component of extinction as empitically calibrated by
  \citet{Wuyts13} is shown as the blue dashed line. This comes very
  close to our best fit relation (black dashed line). Galaxies
  with no IR detection and therefore with SFR from SED fits are
  outlined with black circles: clearly these are the less dusty
  objects. Right: The dependence of the residual $\ha$ size at fixed continuum size on
  $A_V$ is weak and not highly significant.}
\label{fig:dustcor}
\end{figure*}

\subsection{Dependence on dust}\label{sec:dust}

The left hand panel of Figure~\ref{fig:dustcor} examines the dust
correction. On the x-axis we
plot the best fit $A_V$ extinction estimated as part of the
SED fits to multiwavelength photometry \citep{Wuyts11}. On the y-axis we plot
the ratio of the total SFR to the total $\ha$ luminosity from KMOS. The
SFR is estimated by \citet{Wuyts11} and includes obscured star-formation
as seen via infrared emission from {\it Herschel}/PACS or {\it Spitzer}/MIPS where
detected, and is based on SED fits where there is no infrared detection. The total $\ha$ luminosity is estimated by integrating the
\imfit\ exponential disk fit out to infinity. Similar results are
obtained by integrating the $\ha$ image (these are merely noisier and
truncated at the edge of the KMOS field of view).\footnote{Note: these
  estimates are both in excellent agreement with the aperture values
  presented by \blue{W19}, for which the clearest difference is
  the expected dependence on the ratio of effective radius to aperture
  size.}  The y-axis value can therefore be interpreted as a conversion
factor from $\ha$ luminosity to total SFR including the dust
correction.  A standard conversion {\it not including dust} from
\citet{Kennicutt98}, shifted to a Chabrier initial mass
function (IMF), describes the lower envelope of the data well
(horizontal dashed red line). As $A_V$ increases, so does the dust correction to this conversion factor
(effectively the mass of stars formed per number of detected $\ha$ photons). If the $\ha$ extinction was equivalent to the continuum
extinction at the wavelength of $\ha$, $A_{cont} = 0.82 A_V$
  from \citet{Calzetti00} as in \citet{Wuyts13}, (as appropriate in the
case of a foreground dust screen) we would get the dashed green line.
Our best fit to the data (black dashed line) implies a steeper
dependence on $A_V$, and is in good agreement with the best
fit polynomial from \citet{Wuyts13} (blue dashed line), implying an
excess extinction $A_{extra}$ for the $\ha$ emission in HII regions such that $A_{\ha} = A_{cont} + A_{extra}$ with $A_{extra} = 0.9A_{cont} - 0.15A_{cont}^2$.

In the right hand panel of Figure~\ref{fig:dustcor} we show that there
is only a weak, barely significant negative correlation between the residual $\ha$ size at fixed continuum size
and $A_V$, implying little or no correlation between the integrated
continuum obscuration and the extent of $\ha$ emission at fixed continuum size in the galaxy, as confirmed in Table~\ref{table:restrendstats}. Similarly weak
correlations are found with rest frame galaxy colours.
The lack of a stronger correlation contrasts with simple expectations. Galaxies exhibit a centrally peaked dust extinction profile \citep[e.g.][]{Wuyts12,Nelson16} and an excess integrated extinction at $\ha$ increasing with increasing $A_V$ as observed in the left panel of  Figure~\ref{fig:dustcor}. This implies that we expect extinction effects to drive a flattening of the light profiles (larger half-light radius observed than the true one, see e.g. \citealt{Pastrav13}), and that this applies more in galaxies with higher extinction ($A_V$) and more in $\ha$ than in continuum ($A_{\ha} > A_V$). 
As a result, the ratio of observed $\ha$ to continuum size should be larger than the true one and, naively, one might expect that this effect should increase with increasing $A_V$.
Instead we see only a very weak -- and negative -- correlation of observed $\ha$ to continuum size ratio (at fixed continuum size) with $A_V$.

\begin{figure}
 \centerline{\includegraphics[width=0.5\textwidth]{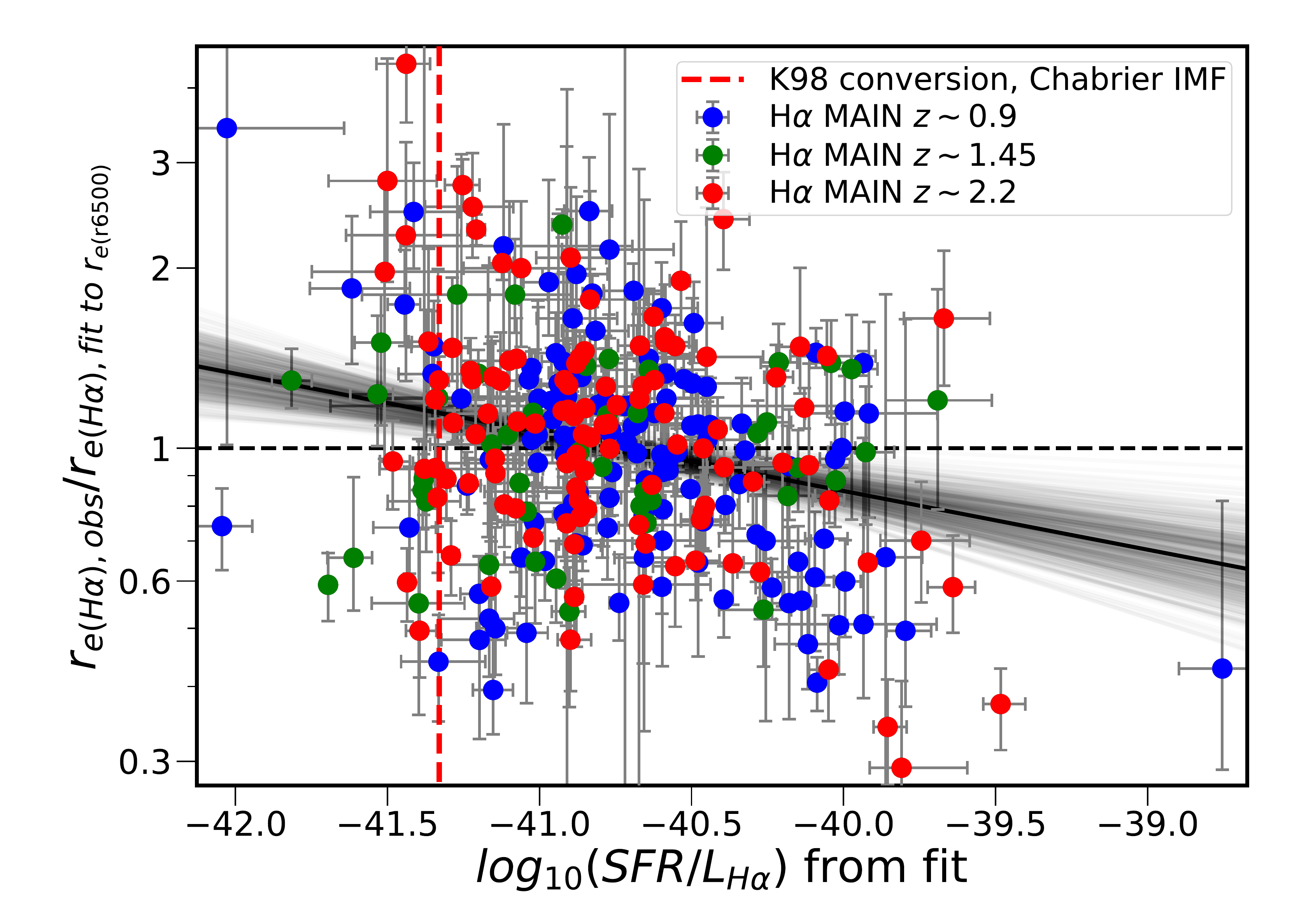}}
 \centerline{\includegraphics[width=0.5\textwidth]{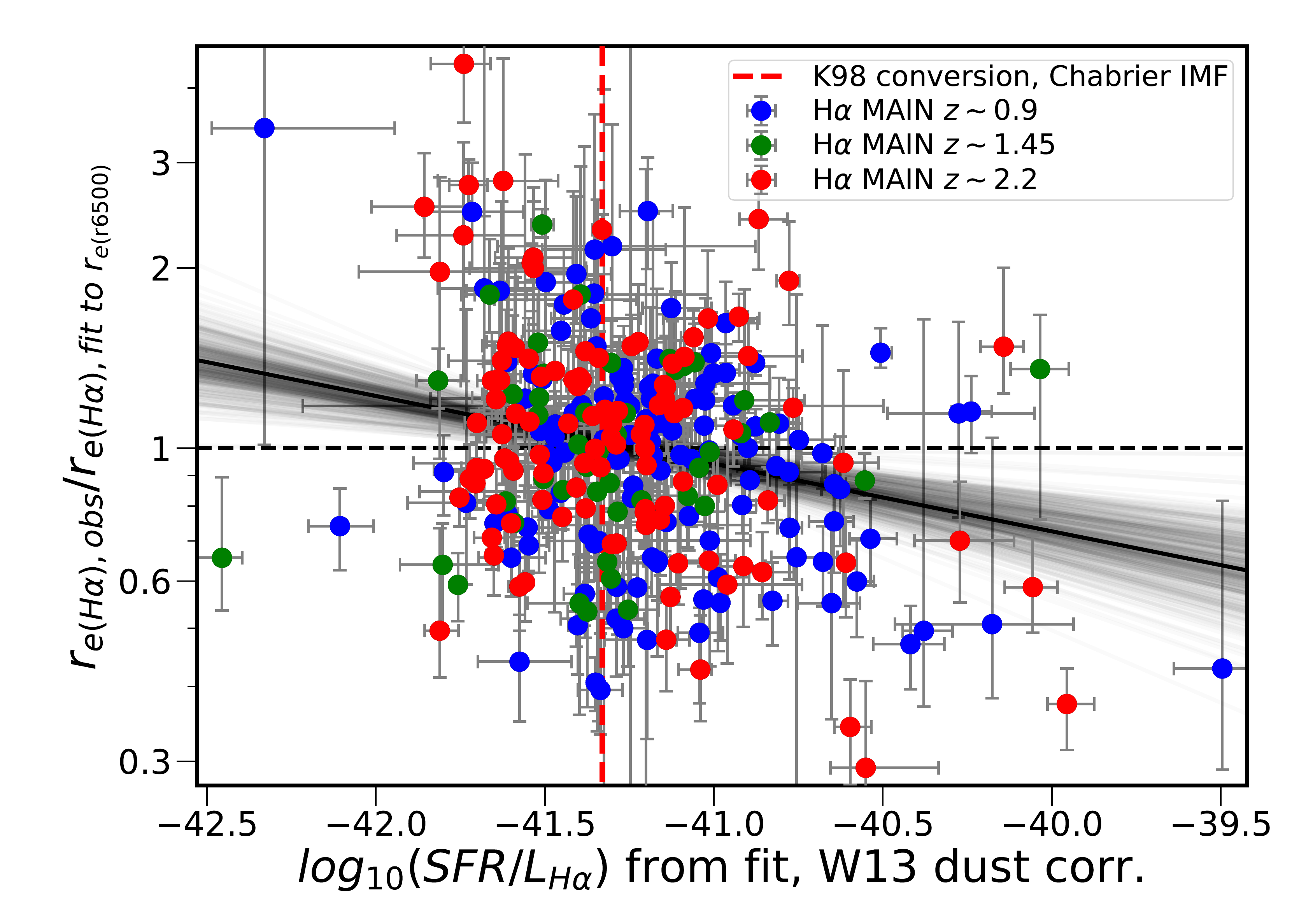}}
 \caption{Residual $\ha$ size at fixed continuum size versus 
   the amount of $\ha$ dust obscuration. The upper panel  
   shows the relation to the total SFR normalized by total $\ha$
   luminosity (y-axis value in Figure~\ref{fig:dustcor}, left panel). The bottom 
   panel corrects this $\ha$ luminosity for
   extinction as a function of $A_V$ using the differential extinction
   law of \citet{Wuyts13} (the blue line from
   Figure~\ref{fig:dustcor}) and examines the correlation of residual $\ha$ size at fixed continuum size with the residual $\ha$ extinction. Even this residual dust extinction
   correlates significantly with the residual $\ha$ size, indicating that galaxies
   with less $\ha$ extinction at fixed $A_V$ can be visibly more extended in
   $\ha$ relative to their stars. 
   Our best fit relations between
   parameters are shown (black solid line), as well as individual Monte-Carlo realizations (fainter grey lines). }
 \label{fig:dusttrends}
\end{figure}

In contrast, Figure~\ref{fig:dusttrends} shows how the residual
$\ha$ size at fixed continuum size does depend upon the parameters specifically
related to the dust obscuration of the $\ha$ emission. The upper panel shows the variation as a function of the total obscuration of $\ha$ emission, as parameterized via the ratio of the total
SFR to $\ha$ luminosity (the y-axis parameter from Figure~\ref{fig:dustcor}). In the lower panel we correct the $\ha$ luminosity for the average $\ha$ dust
extinction using the well-fit prescription from \citet{Wuyts13}. Now the
x-axis values describe the {\it residual dust-correction} to $\ha$
emission after applying this average correction.
In both panels of Figure~\ref{fig:dusttrends}, the vertical dashed green line refers to the \citet{Kennicutt98} conversion between SFR and (extinction-corrected) $\ha$ luminosity.

The typical residual $\ha$ size (at fixed continuum size) increases significantly for less total extinction of $\ha$, even after correction for the average global obscuration, $A_{\ha}(A_V)$ based on the \citet{Wuyts13} recipe. The lack of dependence of residual $\ha$ size on continuum obscuration (Figure~\ref{fig:dustcor} right panel) implies that, moving from the top to the bottom panel of Figure~\ref{fig:dusttrends}, the trend is barely reduced by a global $A_V$-based correction. In other words, the relative $\ha$ size at fixed continuum size must be largely independent of the total amount of foreground dust obscuration $\propto A_V$. Instead our results must be interpreted to mean that the ratio of $\ha$ to continuum obscuration $A_{\ha}/A_V$ decreases for galaxies which are relatively more extended in $\ha$ relative to stars. 

We have seen no notable dependence of residual $\ha$ size at fixed continuum size on global parameters $A_V$, SFR or inferred gas fraction. That the residual $\ha$ size {\it decreases} with increasing $A_{\ha}/A_V$, must originate in the internal geometries of dust differently affecting young and old stars in galaxies. Fitting a dust model to explain correlations with the resolved extinction maps of low redshift MaNGA galaxies, \citet{Li19} derive a best fitting model in which the fraction of dust in a foreground screen increases with galactic-centric radius, with the rest of dust assumed to live in $\ha$-emitting HII regions. Such a model implies that $\ha$ photons escape relatively more easily at large galactic-centric distances than in the centre of galaxies, and that galaxies with intrinsically large $\ha$ disk sizes or steep dust obscuration profiles are less subject to extra obscuration of $\ha$ emission.

This would put such galaxies below the average relation in the left panel of Figure~\ref{fig:dustcor} (blue dashed line), closer to a the pure foreground screen model, and also to the left of both panels in Figure~\ref{fig:dusttrends}. 
These results are also tabulated in Table~\ref{table:restrendstats} and we find equivalent trends at similar significance if we use the total $\ha$ luminosity from integrated $\ha$ maps instead of the integrated exponential fit. This demonstrates that the trends are not driven by covariance in the fitting parameters. 

Finally we note that despite the details of these trends, the overall effect of dust obscuration is still to increase the observed $\ha$ sizes more than that of the stars, with no difference only in the limiting case of no embedded dust (pure foreground screen). In other words, the measured mean (median) size ratio of 1.26 (1.18) is likely to be an upper limit on the true average ratio of the star-forming disk size to the stellar disk size. Indeed, in extreme cases from within our sample we know that there are highly star-forming
massive galaxies hosting strong central star-formation \citep{Tadaki17} which is almost completely obscured in $\ha$. 
However, it appears that the $\ha$ sizes of such galaxies are nonetheless 
similar to that of their stars. Indeed, we
see a relatively unobscured and sub-dominant component of star
formation which is associated to the stellar disk while the majority of
star-formation occurs in highly optically thick central region of the galaxy, with little or no escape of ionizing photons and subsequent $\ha$ emission.

\section{Interpretation}\label{sec:interp}

In the previous Section we have found that over the large redshift and stellar mass range probed by our observations, the ionized gas size is on average 1.26 times larger than the continuum size (tracing older stellar populations), with negligible dependence on other  fundamental parameters. We now interpret these results in terms of gas accretion and its angular momentum.

\subsection{Analytic considerations}
A simplified prediction for disk sizes can be derived by combining equations 2 and 12 of \citet{Mo98}:
\begin{equation}
    R_d \propto H(z)^{-2/3} \cdot \lambda^\prime \cdot M_{halo}^{1/3}  
\label{equ:Motheory}
\end{equation}
for disk size $R_d$, galaxy spin parameter $\lambda^\prime = \lambda \cdot \frac{j_d}{m_d}$, halo spin parameter $\lambda$, a fraction of halo angular momentum in the disk $j_d$ and of halo mass in the disk $m_d$ (such that $j_d/m_d = 1$  where there is no difference in the specific angular momentum of disk and halo). $R_d$ can then be related to the disk mass $M_d$ via $\rm\,M_{halo} = (M_d/m_d)$. 

The predicted dependence on $\rm\,M_{halo}^{1/3}$ originates with the predicted proportionality to the circular velocity of the halo, $\rm\,V_c$, and thus $\rm\,M_{halo}^{1/3}$ from the Virial theorem. 
This is steeper than the observed dependence on stellar mass in the Tully-Fisher relation ($\sim 1/3.75 \sim 0.27$ \citealt{Lelli16}), which is more in line with the slope of the stellar mass - size relation ($\sim 0.22$, \blue{vdW14}). A slope of less than $1/3$ may be accounted for by variations in $m_d$ with mass.
The predicted dependence on $H(z)$, on the other hand, suggests a strong evolution in galaxy sizes at fixed mass. Observations are at potential odds with one another about the rate of evolution. \blue{vdW14} find that observed median sizes of star forming disk galaxies at fixed stellar mass evolve as $H(z)^{-2/3}$ but \citet{Suess19} claim to see little evolution at fixed stellar mass of half-mass sizes, once they account for radial gradients in the mass to light ratio. Both cases appear surprising, as disk sizes are likely to be set at the epoch of formation (not observation) but still evolving as they grow through star formation. That \citet{Suess19} find a much flatter dependence on stellar mass also begs questions about the expected dependence of size on mass or circular velocity.

More straightforwardly, the measured intrinsic scatter in galaxy sizes of $\rm 1\sigma \sim 0.16-0.19$ dex is very consistent with the scatter in continuum galaxy size at fixed specific angular momentum \citet{Burkert16}, and with the scatter in halo spin parameter from simulations. The latter similarity suggests that most of the scatter in galaxy sizes originates as scatter in the halo spin.

Equation~\ref{equ:Motheory} can also be applied to examine the star-forming disk size, noting that the parameters $j_d$ and $m_d$ effectively describe the efficiency of angular momentum and mass transfer from halo to disk. The term $\rm\,M_{halo}^{1/3}$ in equation~\ref{equ:Motheory} describes the gravitational potential at the time of formation of the relevant component. Therefore we can predict the ratio of star-forming disk size to stellar disk size:
\begin{equation}
    \frac{R_{d,SF}}{R_{d,*}} \propto \bigg(\frac{H(z_{SF})}{H(z_{*})}\bigg)^{-2/3} \cdot \frac{\lambda^{\prime}(z_{SF})}{\lambda^{\prime}(z_{*})} \cdot
    \bigg( \frac{M_{halo,z_{SF}}}{M_{halo,z_{*}}}\bigg)^{1/3} 
\label{equ:sizeratioprediction}
\end{equation}
where the Hubble parameter, specific angular momentum, and halo mass should be evaluated at times appropriate to the star-forming and stellar components. Naively, equation~\ref{equ:sizeratioprediction} suggests that the star-forming disk should be larger than the stellar disk by an amount depending upon the Hubble parameter at their relative times of formation with $z_{SF} << z_{*}$, with modifications that can relate to the growth in halo mass, and to changes in the halo spin parameter over time.

We found that the size in $\ha$ emission correlates
strongly with the continuum size, with an intrinsic scatter 
smaller than that of continuum size -- and of inferred halo spin 
parameters -- at fixed mass. This demonstrates the stability over time of the spin parameter, with less variation in time than variation between halos. The intrinsic scatter of $43\pm3\%$ combines any short term temporal variation in $\lambda$ with the changes in the efficiency of the angular momentum transfer from halo to disk scales including the process of star-formation, as well as the effects of dust on both size measurements.  
We evaluate the ratio $R=[H(z_*)/H(z_{SF})]^{2/3}$ appearing in equation \ref{equ:sizeratioprediction} at redshifts 
$z_{SF}=z_{obs}=1$ and 2, upon which our observations are concentrated. Then with the assumed cosmology we obtain $R=1.33$ for $z_{*}=1.5$ and $z_{obs}=1$, and $R=1.59$ for $z_{*}=3$ and $z_{obs}=2$, where the $z_{*}$ values are selected to have a 
roughly constant lookback time of 1 Gyr from the observation redshifts. On this timescale a galaxy would double its 
stellar mass, since $M_*/SFR \lesssim 1$ Gyr for $z>1$. We then evaluate the
last term of equation \ref{equ:sizeratioprediction}, $P=[M_{halo,z_{SF}}/M_{halo,z_{*}}]^{1/3}$ using the halo
growth factor $(1+1.11z) \cdot \sqrt{\Omega_M.(1+z)^3 + \Omega_\Lambda}$ given by \citet{Fakhouri10}. 
We get $P=0.8 (0.7)$ with the same choice of $z_{*}$ and $z_{obs}$ made above, respectively.
In summary, equation \ref{equ:sizeratioprediction} gives $R_{d,SF}/R_{d,*}\sim 1.11 (1.06) \cdot {\lambda^{\prime}(z_{SF})}/{\lambda^{\prime}(z_{*})}$, respectively. The simplest
interpretation of the observed lack of evolution in the ratio of star-forming size (as traced by $\ha$) 
to that in stars (as traced in the rest frame 6500 \AA), as seen in our data, is therefore that the 
specific angular momentum of star-forming material is stable across many Gigayears of cosmic
time, resulting in an almost constant ${\lambda^{\prime}(z_{SF})}/{\lambda^{\prime}(z_{*})}$ ratio.
This evidence can be physically related to the complex interplay of cooling, accretion, star-formation 
and feedback which might regulate the angular momentum of star-forming material and its evolution.

\subsection{A toy model for evolution in size and mass}\label{sec:sizegrowth}
Observed star-forming galaxies have star-formation rates and stellar sizes
which depend upon their stellar mass and redshift, with log-normal
scatter around the observed relations. In this section we construct a
simple toy model to predict how star-formation in galaxies should lead
them to evolve in the size-mass plane.

Our model predicts the evolution in a galaxy's stellar mass by assuming
it to grow purely via star-formation according to the main sequence relation
between SFR and stellar mass from \citet{Whitaker14}. We include
  the log-normal scatter $\sim0.3$ dex \citep{Noeske07} to get estimates of
  mean SFR rather than median SFR, which should be appropriate assuming
  individual galaxies scatter above and below the main sequence. This
is offset by mass loss from stars computed according to the Flexible
Stellar Population Synthesis (FSPS) code \citep{Conroy09}.
Table~\ref{table:massgrowth} provides the stellar mass at $z=2$ for galaxies evolved in this way and ending with fixed masses of $\log_{10}({\Mstellar}/{\Msol})=10,10.5,11,11.5$
at $z=1$. We also compute the much lower $z=2$ stellar mass for a star-formation only recipe (no mass loss, and thus much more rapid mass evolution). Galaxies below
$\Mstellar\sim10^{10.8}\Msol$ at $z=1$ had stellar masses below our approximate $\ktd$ limit of $10^{10}\Msol$ at $z=2$.

\begin{table}
\begin{center}
\caption{}
\vspace{0.1cm}
\begin{tabular}{ccc}
  \hline\hline
  \noalign{\smallskip}
    {\sc $log_{10}\frac{\Mstellar}{\Msol}$ at $z=1$} & \multicolumn{2}{c}{\sc  $log_{10}\frac{\Mstellar}{\Msol}$ at $z=2$}\\
    & SF & SF $+$ mass loss \\
    \noalign{\smallskip}
    \hline
   \noalign{\smallskip}
    10 & 8.02 & 8.63\\  
    10.5 & 8.76& 9.38 \\
    11 & 9.95 & 10.47 \\
    11.5 & 10.96 & 11.20 \\
  \end{tabular}
\vspace{0.1cm}
\begin{scriptsize}
  \begin{minipage}{0.45\textwidth} {\sc Notes.} $z=2$ progenitor galaxy mass of a given $z=1$ galaxy in our toy model. Stellar mass growth via star-formation occurs along the main sequence \citep{Whitaker14}. In the third column we include stellar mass loss in the calculation, which offsets star-formation such that the mass growth is reduced. Galaxies below $\Mstellar\sim10^{10.8}\Msol$ at $z=1$ had stellar masses below $10^{10}\Msol$ at $z=2$, and below our approximate $\ktd$ limit.
\label{table:massgrowth}
\end{minipage}
\end{scriptsize}
\end{center}
\end{table}

Motivated by the consistent ratio of $\ha$ to continuum
sizes $\reffha/\reffcorr$ in our data, and in particular the lack of significant 
mass or redshift dependence, we assume a constant value in the ratio of star
forming to stellar size $\reffsf/\reffmstar$ which we call the size
growth factor. A galaxy starts with an exponential profile at high
redshift. This profile is evolved self-consistently over many small
steps in time. At each step, the newly formed stars are generated with
an a radial distribution described by an exponential profile 
and a half-mass size equal to the current
stellar half-mass size multiplied by the size growth factor. Stellar
mass loss is self-consistently tracked as a function of the stellar
age, such that mass is removed from the radii at which it was added
when those stars formed i.e. stars are assumed to remain on their
initial, circular orbits.  The profile evolves in this
way, driving growth in the stellar
disk with time. The evolving profile retains a roughly exponential form but with slowly changing scale-length with radius ($n_{\rm Sersic}-1 > 0$ but is small ) such that the inner ($r<<\reff$) profile, dominated by old stars, is consistent with an exponential with half-light radius $=\reffmstar$, and the outer ($r>>\reff$) profile, dominated by young stars, is consistent with an exponential with half light radius equal to $\reffmstar \times (\reffsf/\reffmstar)$. 

\begin{figure*}
  \centerline{\includegraphics[width=0.98\textwidth]{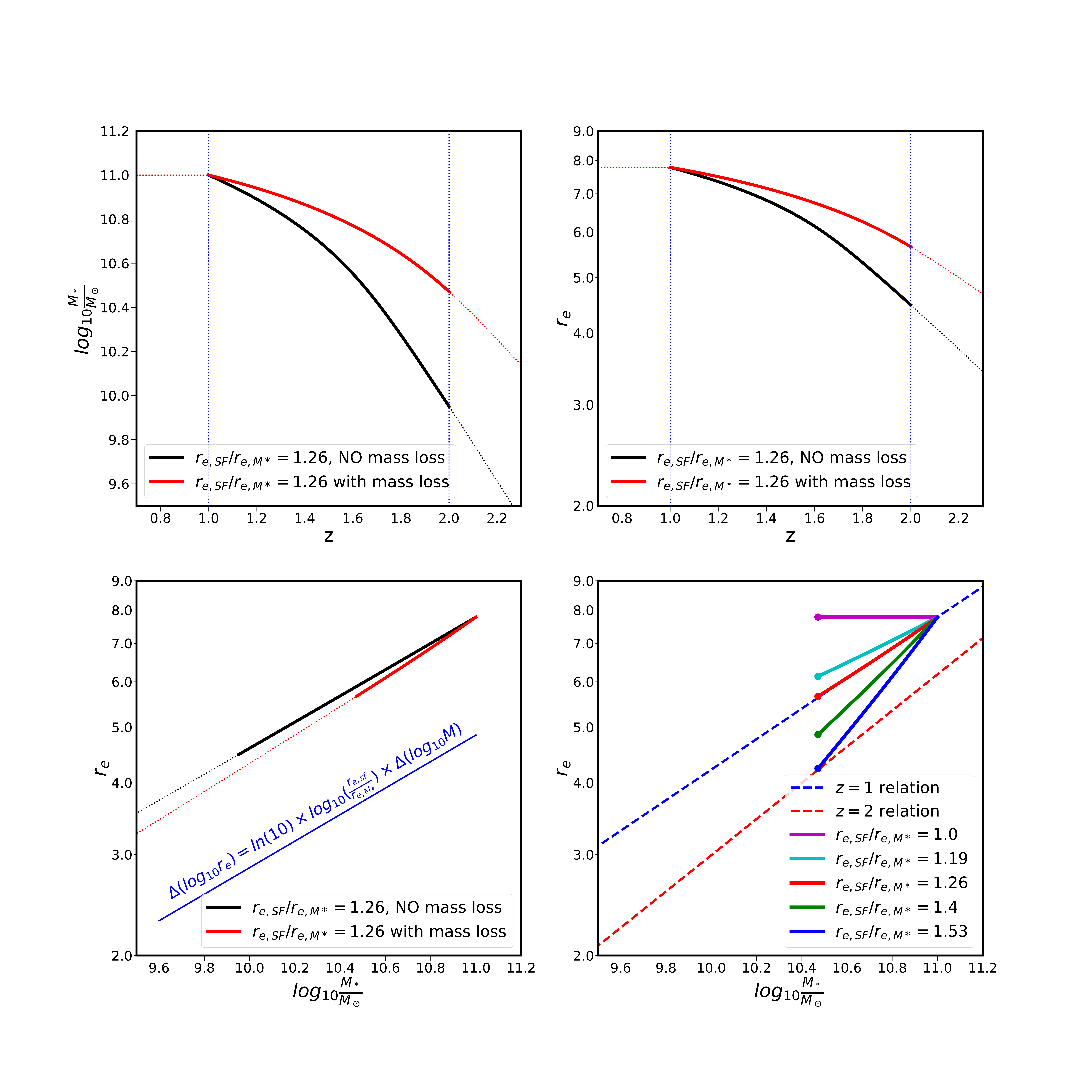}}
  \caption{Toy model evolution of a galaxy with size growth factor
    ${\reffsf}/{\reffmstar}=1.26$ in stellar mass (top-left panel),
    stellar size (top-right panel) and the evolutionary track in
    stellar mass vs size (bottom-left panel).  Tracks are shown for
    models with (red) and without (black) mass loss. Thicker, solid
    lines refer to the evolution between $z=2$ and $z=1$. Including
    mass loss significantly reduces the rates of growth in both mass
    and size, without much changing the evolutionary track in mass vs
    size.  A good analytic approximation for the size evolution is shown in the bottom-left panel (blue line). 
    Bottom-right panel: Toy model growth from $z=2$ to $z=1$ as in the previous panel, but for different values of the size growth factor
  $\reffsf/\reffmstar$. Galaxies are evolved using our toy model 
  and end by design on the $z=1$ 
  size-mass relation of star-forming (LT) galaxies from
  \blue{vdW14} corrected from F160W to rest-frame 6500 \AA\ using the \citet{Kelvin12} recipe (blue dashed  line). The evolution of galaxy size
  depends upon the size growth factor $\reffsf/\reffmstar$ as depicted
  by the solid evolutionary track lines. A very large size growth factor
  $\reffsf/\reffmstar\sim1.53$ is required to reproduce the observed
  evolution, from the $z=2$ LT size-mass relation of \blue{vdW14}
  (red dashed line), while if we assume size growth factors equivalent
  to our measured mean (median) observed values
  $\reffha/\reffH=1.26$ $(1.19)$, these are barely enough to evolve a
  galaxy {\it along} a non-evolving size-mass relation. }
  \label{fig:masssizeevol}
\end{figure*}

Figure~\ref{fig:masssizeevol} shows the toy model evolution of a galaxy
with size growth factor $\frac{\reffsf}{\reffmstar}=1.26$ and final
stellar mass $\log_{10}({\Mstellar}/{\Msol})=11$ at $z=1$. 
Recall that dust considerations in Section~\ref{sec:dust} indicate that $\frac{\reffsf}{\reffmstar}=1.26$ should be an upper limit.
We examine the evolution in stellar mass and size (focused on redshifts $1<z<2$)
and in the stellar mass vs size plane\footnote{For simplicity in this section the simple notation $\reff$ is frequently used to refer to the half mass size $\reffmstar$.}.  
A simple empirical estimate suggests the fractional mass evolution goes as the specific star-formation rate, while the fractional radial evolution goes as the specific star-formation rate times the size growth factor (ignoring mass-loss and the mildly non-exponential nature of evolving profiles) . With this recipe, the rate of change of log size with respect to the change in log mass is the natural logarithm of the size growth factor: i.e. $\rm\, \Delta(log\reff) \sim 
ln(\frac{\reffsf}{\reffmstar})\times\Delta(log\Mstellar)$
(blue line in Figure~\ref{fig:masssizeevol}). This approximation is
close to our numerically derived slope, especially in the no mass loss
case (age-dependent mass loss leads to mildly non-linear effects).

In the bottom-right panel of Figure~\ref{fig:masssizeevol} we examine how the toy model evolution
with different size growth factors compares to the measured evolution
of the size-mass relation for late type (star-forming) galaxies, from \blue{vdW14}, interpolated
  between the tabulated midpoints of redshift bins at $z=1$ and $z=2$, and corrected to rest-frame 6500\AA\ for consistency with our observed sample. 
This choice of observed relation is discussed further in Section~\ref{sec:justifyvdW}.

Our evolutionary tracks are fixed such that
they result in a galaxy of $\log_{10}({\Mstellar}/{\Msol})=11$ at
$z=1$ and with a size defined by the $z=1$ size-mass relation of
\blue{vdW14}. Evolving up to this fixed point, we find that a size
growth factor close to our upper limit value of 1.26 is
required merely to evolve galaxies {\it along} the size-mass relation.
With smaller values, the sizes of star-forming galaxies at fixed
stellar mass would actually {\it decrease} with increasing cosmic time.
These results suggest that growth via star-formation is unlikely to grow 
galaxies enough in size to do more than retain a non-evolving size-mass relation 
and almost certainly not explain such a strong evolution as observed by \blue{VdW14}.

\begin{figure}
  \centerline{\includegraphics[width=0.5\textwidth]{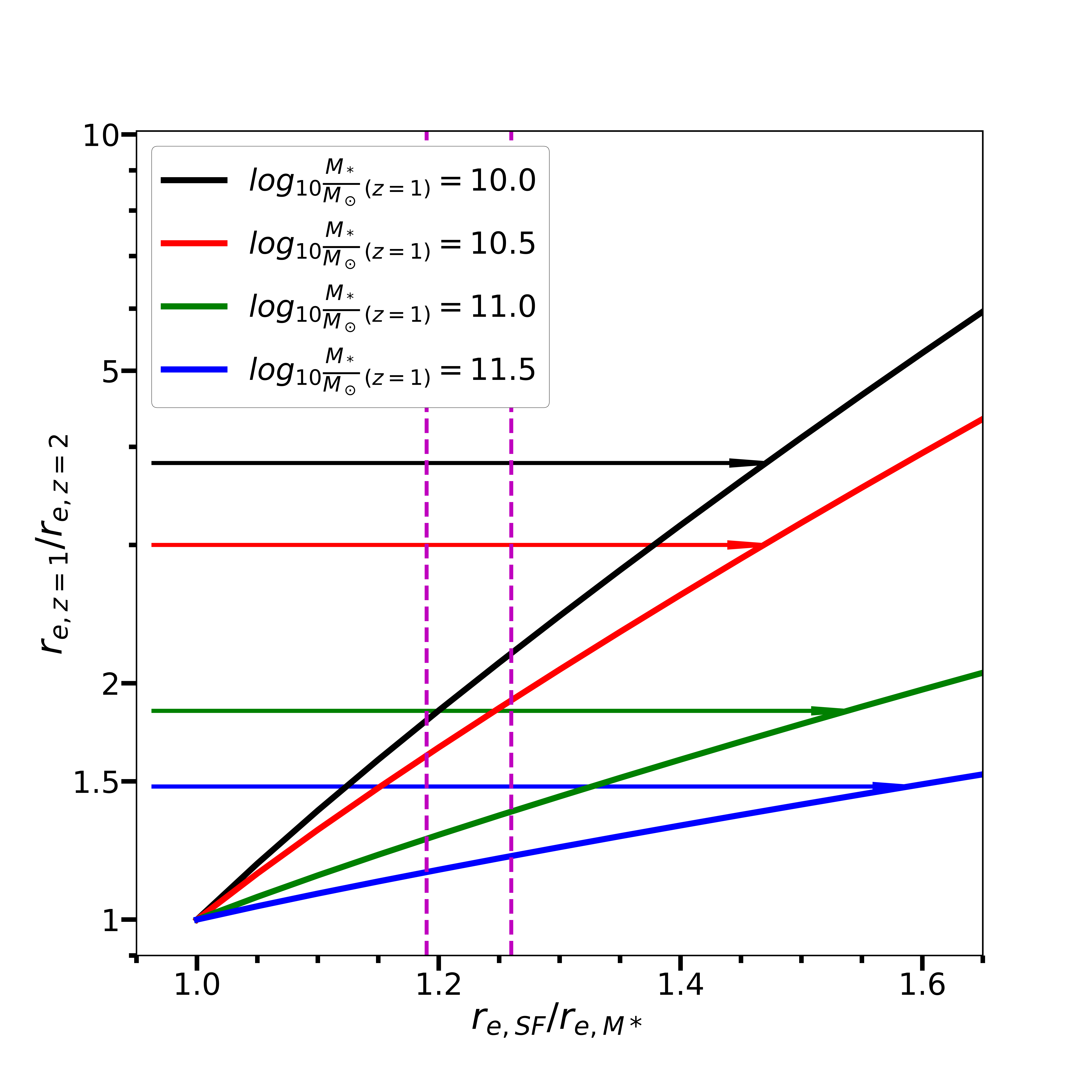}}
  \caption{Total size growth in our toy model from $z=2$ to $z=1$ for
    galaxies with four values of final stellar mass at $z=1$ (coloured lines, see legend).  
    Sizes are assumed to grow exclusively via the formation
    of new stars with a constant size growth factor
    $\reffsf/\reffmstar$ (x-axis), and with a mass evolution
    following the star-formation main sequence.
    The arrows denote the equivalent size growth
    of the same galaxy over the same redshift interval 
    from the size-mass relation of \blue{vdW14}, and point to the place where this 
    growth is reached by our size growth model. These values are much larger than those measured in
    our sample (vertical, dashed magenta lines show the median and mean values at 1.19 and 1.26). 
    This suggests that star-formation driven
    size growth is not enough to explain the observed size evolution of
    star-forming galaxies.}
  \label{fig:sizemassgrowth}
\end{figure}

In Figure~\ref{fig:sizemassgrowth} we study the ratio of 
$z=1$ to $z=2$ size as a function of the size growth factor for 
galaxies resulting in a galaxy of a particular mass
at $z=1$ and with a size defined by the $z=1$ size-mass relation of
\blue{vdW14}. The arrows point to the location where the model size growth matches the observed 
growth for the average star-forming galaxy. We examine four
final ($z=1$) galaxy masses $10 \leq
\log_{10}({\Mstellar}/{\Msol}) \leq 11.5$.
Although the progenitor galaxies at $z=2$ would not make our $\ktd$ sample for the lower two
mass bins (Table~\ref{table:massgrowth}), it is still useful to examine
predictions from our model under the assumption that the mass and
redshift independence of the size growth factor extends to lower
mass. The two vertical lines correspond to the median (1.19) and mean (1.26)
values of $\reffha/\reffcorr$ from our analysis, respectively.

In all mass bins considered a very large
size growth factor is required for galaxies to grow sufficiently
to evolve from the $z=2$ size-mass relation onto the $z=1$ size-mass
relation, assuming the \blue{vdW14} late type relations. 
This also increases with mass, from $\rm\, \reffsf/\reffmstar \sim
1.50$ at $\log_{10}({\Mstellar}/{\Msol})=10$ to $\rm\, \sim 1.60$ at
$\log_{10}({\Mstellar}/{\Msol})=11.5$. 

\subsection{Considerations and constraints on evolution}\label{sec:justifyvdW}

To understand our constraints, we should consider a couple of apparent contradictions, and how to resolve them.

\begin{enumerate}
    \item \blue{vdW14} determine an evolution in late type galaxy sizes at 5000\AA\ at fixed stellar mass of $\sim H(z)^{-2/3}$, and an (almost un-evolving) dependence on stellar mass of $\Mstellar^{0.22}$. \citet{Suess19} find little dependence on stellar mass or redshift in {\it half-mass} sizes, accounting for gradients in the stellar mass to light ratio. Little redshift or mass-dependent evolution in sizes indicates little or no growth in size. 
    \item Without star formation driven size growth, there would be no age-gradients in galaxies, contradicting the idea of larger sizes at shorter wavelengths due to the increasing importance of younger stellar populations. Instead, our results {\it do} support the idea that newly forming stars populate a slightly larger disk than older stars.
    \item A constant size growth factor $\frac{\reffsf}{\reffmstar}$ as a function of size, mass and redshift would imply a scale-free growth such that, while younger stars can be found on average at larger galactic-centric distances than older stars, the ratio of the two is independent of redshift and galaxy mass: we do not detect significant deviations from this constant growth. If true, there should be no epoch at which age gradients should disappear without invoking complex age-dependent radial migration. This appears to contradict the lack of $M/L$ gradients at $z=2$ seen by \citet{Suess19}. 
    
    We do also find a greater difference between mass and light sizes in galaxies with higher sersic index, but only at $z\sim1$ (not $z\sim2$). This suggests a role for bulges, more prevalent at lower redshift, and driving greater $M/L$ gradients in some galaxies. We note that in general star-forming disks may grow while overall galaxy sizes stay the same due to an increasing bulge contribution: due to the differing contributions of bulge and disk to light and mass, the bulge is more likely to dominate in mass, while more size growth may be seen in light.
    \item On the other hand in Figure~\ref{fig:massize_correlations} we found no correlation of $\reffha/\reffcorr$ with $\reffmstar/\reffcorr$. This strongly suggests that whatever drives the gradients in $M/L$ (such as age and bulge contributions) is not the dominant factor driving variations in $\reffha/\reffcorr$. We argue that these are driven instead mainly by variations in the gradient of dust and in particular embedded dust in star forming disks.
    Continuum light and $\ha$ are therefore more closely tied than mass and $\ha$, and the ratio should remove the effect of a foreground dust screen but not of embedded dust. 
\end{enumerate}

Taking account of these considerations, we selected to compare to the simpler, light-based \blue{vdW14} relations. Considering our results, we cannot match the observed evolution of size-mass relations from \blue{vdW14}. This would suggest that other physical processes for growth of star forming galaxies might be at play, and we will discuss candidates in the next section. However the results of \citet{Suess19} suggest that evolution might not be so steep if we assume there is also a role for evolution in the mass to light gradients (even if this is driven in part by bulge formation). Such milder evolution can be consistent with our upper limit of $\sim 26\%$ for the size growth factor. 

\subsection{Physical origins of galaxy size growth}\label{sec:physicsofgrowth}

Equation~\ref{equ:Motheory} conveniently separates the dependencies on redshift and mass in the derivation of disk size for star-forming galaxies. In this context the mass-dependence is simply an imprint of the dependence of galaxy size on $\big(\frac{M_d}{m_d}\big)^{1/3} = M_{halo}^{1/3} \propto V_c$ and thus an imprint of the Tully-Fisher relation between galaxy mass and circular velocity. 
As star-formation appears to drive evolution approximately along this relation, this implies that the integrated growth in size and mass can be described by Equation~\ref{equ:Motheory}, with $\rm\,R_d \propto (M_d/m_d)^{1/3}$ and thus with $\rm\, M_{halo}^{1/3}$, but with galaxy sizes scaled to the Hubble parameter at the epoch of observation, $H(z_{obs})$ such that there is no redshift dependence of the size growth factor (Equation~\ref{equ:sizeratioprediction}). The measured size evolution of the size-mass relation by \blue{vdW14} on the other hand, scales as $H(z_{obs})^{-2/3}$ (or alternatively as $(1+z)^{0.75}$. 
Although this evolution might be overestimated in light-weighted sizes (see Section~\ref{sec:justifyvdW}), if real it would suggest another form of growth not associated to star-formation -- i.e. that the stellar component of star-forming galaxies does not retain its initial size, but rather evolves in size as predicted by Equation~\ref{equ:Motheory} due to angular momentum transfer with the surrounding material (gas and dark matter, see e.g. \citealt{Struck17}). In addition, as we discuss below, {\it any} measured evolution of the size - mass relation is likely to be an overestimate, given that many of the more compact massive galaxies have their star-formation quenched between epochs. 

The constant size growth factor, with no obvious dependence on redshift, size, and stellar or gas mass, implies that the model star
formation driven size growth as seen in Figures~\ref{fig:masssizeevol} and~\ref{fig:sizemassgrowth}
applies under widely varying conditions. While this may simply reflect the halo mass growth, the small intrinsic scatter ($43\%$) also implies that the halo spin parameter $\lambda$ remains very stable over time, and that the specific angular momentum transfer ($\frac{\lambda^\prime}{\lambda} = \frac{j_d}{m_d}$ in Equation~\ref{equ:Motheory}) is also quite insensitive to a wide variety of halo growth rates and physical conditions. Such stability can be achieved if disk growth is regulated, e.g. via feedback. 
For example, \citet{Pezzulli17} describe gas accretion from a rotating hot corona gas in the halo.  The accretion of such gas -- with its associated angular momentum -- is expected in a
galactic fountain model in which stellar winds interact with the corona
gas before falling back onto the disk.  Such models explain the
rotation lag of extraplanar gas in the Milky Way galaxy \citep{Marinacci11}, and can help explain a slow growth in the size of the star-forming disk, strictly linked to the stellar disk size with small scatter.

Our results and modeling suggest that individual galaxies evolve almost parallel to the size-mass relation
with a maximum evolution at $\drdM\sim 0.26$ for $\reffsf/\reffmstar = 1.26$ (Figure~\ref{fig:masssizeevol}), in agreement with the models from \citet{Nelson19}. 
The robustness of our measurement is also supported by comparisons to a completely independent estimate coming from expectations when comparing the sizes of Milky-Way progenitor galaxies selected at different redshifts assuming a constant cumulative co-moving number density \citep[$\drdM\sim0.27$ from][]{van-Dokkum13}, and is slightly shallower than the slope $\sim0.3$ discussed by \citet{van-Dokkum15} and $\sim0.4$ for simulated galaxies with realistic wind models \citep{Hirschmann13}. In contrast to models with no winds, the efficient removal of low angular momentum material at high redshift leads to much larger sizes for high redshift galaxies and shallower evolution, emphasising the role of feedback in the regulation of angular momentum in galaxies.

Some of the massive, star-forming $z=2$ galaxies will have had their star-formation quenched by the time they reach $z=1$.
There are different theories of how such quenching proceeds and why it returns passive galaxies more compact than the coeval star-forming population: galaxies with low ($\lambda\lesssim0.05$) spin parameters can become unstable and contract before rapidly being quenched \citep{Dekel14}; galaxies reach a threshold stellar surface density, velocity dispersion, central surface density, stellar mass or bulge to total ratio before quenching \citep[e.g.][]{van-Dokkum15}, or galaxies evolve along the main sequence and the effective quenching of massive galaxies is a gradual process, with older, earlier forming galaxies with higher density and smaller sizes departing first from the main sequence \citep{Abramson18,Lilly16}. \citet{Abramson18} argue that the distribution of galaxies in the size-mass plane is not inconsistent with such a scenario in which galaxies evolve at constant surface mass density ($\drdM=0.5$). Our constraints show that such a steep evolution is only possible if the stellar sizes of galaxies grow via mechanisms other than star-formation. As shown -- for very different models -- by \citet{van-Dokkum15} and \citet{Abramson18}, such a strong apparent evolution can happen even if the evolution of individual galaxies is relatively weak, so long as the densest galaxies fall out of the star-forming population first and become passive. Our weaker measured star-formation driven evolution in size vs mass suggests that a more aggressive quenching is required, resulting in passive galaxies which are particularly dense and compact. 

\section{Summary and Conclusions}\label{sec:conclusions}

This paper utilises data from the $\ktd$ survey to measure the star-formation driven size growth in individual galaxies, and understand the physical processes driving this evolution at $\zrange$, spanning the time when most of their stars were formed.

$\ktd$ targeted the $\ha+[NII]$ emission line complex in 740 galaxies
at $\zrange$ with 75 nights of observation using the multiplexing NIR
IFU instrument KMOS on the VLT. Datacubes and associated bootstrap
cubes are released with an associated data release paper \blue{W19}.
In this paper we derive galaxy sizes in $\ha$
emission, tracing ongoing star-formation. 
Our initial goal was to demonstrate the
accurate measurement of galaxy half-light sizes with
ground-based data at these redshifts, and in particular $\ha$ sizes tracing star
formation.  With
the investment of a significant calibration effort we have achieved our goal, with galaxy sizes measured
with a typical accuracy of $\sim 20\%$, almost independent of the
absolute size.  Our analysis resulted in a sample of 281 galaxies for which we have accurate $\ha$ size measurements with associated errors. It is representative of the overall star-forming population in terms of SFR,
stellar mass, colours and continuum size.  We publish sizes and
associated errors for this sample.

We then examined how the size of the star-forming gas, traced
by $\ha$, relates to other galaxy properties. Our results can be
summarized as follows:

\begin{itemize}
\item{$\ha$ sizes depend primarily on the continuum size of a galaxy,
    with a near-linear relation, a median (mean) $\ha$ size $=\reffha$
    $18\pm3\%$ ($\sim 26\%$) larger than $\reffcorr$, the continuum size 
    at rest-frame 6500\AA, with $43\pm3\%$ intrinsic log-normal scatter. This is
    much tighter than other correlations (e.g. $68\pm4\%$ intrinsic
    scatter in $\ha$ size vs stellar mass) and explains most of the
    variation in $\ha$ size. It is also smaller than the intrinsic
    scatter in continuum size vs stellar mass ($\sim 56\%$).}
\item{The dependence of $\ha$ size on continuum size shows no residual
    dependence on stellar mass, redshift, star-formation activity, galaxy
    morphology, or (indirectly) vs gas mass. }
\item{There is a significant residual dependence of $\ha$ size on dust
    extinction properties affecting the $\ha$ emission. This dependence
    does not arise from the continuum extinction $A_V$. Instead, the size ratio depends
    primarily on the amount of extinction in the stellar birth clouds $A_{\ha}$ compared to
    $A_V$. For galaxies with larger $\ha$ disks the ratio of these two extinction measurements 
    tends to be lower, closer to a pure foreground screen approximation. This is in line 
    with models suggesting that most obscuration of $\ha$ emission far from galactic centres 
    takes place in a diffuse, foreground component (equally affecting the continuum), while 
    in galactic centres more obscuration of $\ha$ is caused by dust embedded in HII regions \citep[e.g.][]{Li19}.}
\end{itemize}

Based upon these results, we surmise:

\begin{itemize}

\item{The tight correlation between the sizes of the star-forming and stellar components in star-forming, high redshift galaxies infers that the spin parameters of galaxies and their gas are tightly linked to the halo in which they live \citep[see also][]{Burkert16}, and  are highly stable over long periods of cosmic time. Such stability not only requires stable halo growth but also stability in the transfer of specific angular momentum from halo to disk scales, including the processes of cooling, accretion, star-formation and feedback. Simulations and models suggest that feedback helps to modulate the disk growth \citep[e.g.][]{Hirschmann13} and might help regulate the growth and its relation to the existing stars via galactic fountains \citep[e.g.][]{Pezzulli17}. Such regulation applies consistently across a wide range of physical conditions as characterised by e.g. redshift, stellar or gas mass, star-formation rates, morphology, environment, and global extinction by dust, $A_V$.}

\item{Star-formation drives size growth in galaxies, but is unlikely
    to evolve galaxies more steeply in size vs mass than the observed
    relation between those parameters at fixed redshift. We model this
    process including stellar mass loss. Excess dust extinction of $\ha$ in galaxy centres means that our mean value of $\reffha/\reffH=1.26$ is likely an upper limit on the size growth factor $\reffsf/\reffmstar$. A growth of $\reffsf/\reffmstar=1.26$ moves galaxies along a locus with a slope $\drdM\sim 0.26$, consistent with the slope of the mass-size relation and with the predicted evolution of Milky Way progenitors based on the observed sizes of galaxies selected at a constant cumulative co-moving number density in the Universe \citep[$\drdM\sim0.27$][]{van-Dokkum13}. A steeper evolution of the sizes of star-forming galaxies with mass, required to explain the roughly $H(z)^{-2/3}$ evolution in stellar sizes \blue{(vdW14)}, may be partially accommodated by accounting for the quenching of star-formation in the more compact galaxies \citep[e.g.][]{van-Dokkum15,Abramson18} but likely requires other physical processes such as minor merging and angular momentum exchange with the halo.  }

\end{itemize}

\section*{Acknowledgments}

D.J.W. and M.F. acknowledge the support of the Deutsche
Forschungsgemeinschaft via Projects WI 3871/1-1, and WI
3871/1-2. M.F. has received funding from the European Research Council (ERC) under the European Union's Horizon 2020 research and innovation programme (grant agreement No 757535). E.W. and J.T.M. acknowledges support by the Australian Research Council Centre of Excellence for All Sky Astrophysics in 3 Dimensions (ASTRO 3D), through project number CE170100013. PL acknowledges funding from the European Research Council (ERC) under the European Union’s Horizon 2020 research and innovation programme (grant agreement No. 694343).
G.B.B. acknowledges support from the Cosmic Dawn Center,
which is funded by the Danish National Research Foundation.
We thank the ESO and Paranal staff for the excellent support over the course of our KMOS observations. This work is also based on observations taken by the 3D-HST Treasury Program (GO 12177 and 12328) and by the CANDELS Multi-Cycle Treasury Program with the NASA/ESA HST, which is operated by the Association of Universities for Research in Astronomy, Inc., under NASA contract NAS5-26555.

\appendix

\section{Size Measurements: Accuracy and Consistency}\label{sec:sizeconsistency}

The main goal of this paper is to present measurements of the 
size in $\ha$ for $\ktd$ galaxies. With complimentary continuum
measurements this provides the means to examine the spatial growth of
high redshift galaxies via star-formation. This section provides a
characterization and tests of the accuracy of our
measurements with natural seeing KMOS data for both continuum and
$\ha$.

In Figure~\ref{fig:fitsf160w} we have seen that our \imfit\ -based
sizes measured with the F160W WFC3 band are consistent with those
measured by \blue{vdW14}. Moreover, the right panel
shows a remarkable consistency between our higher resolution WFC3 sizes
and those derived from the KMOS continuum. 

\subsection{Definition of Galaxy Size Errors}\label{sec:sizeerrorsdef}

The accuracy of measured half-light size of galaxies is sensitive to
sources of noise (systematic and random), and to the spatial
resolution. 

Our bootstrap cubes randomly sample most sources of systematic and
random noise, so we use these to assess asymmetric $1-\sigma$ errors in
size. {\bf Bootstrap size errors} encompass the range between the
16$^{th}$ and 84$^{th}$ percentiles of the sizes measured from the
bootstrap cubes. 

In some cases the median size measured from the bootstrap cubes differs
from the size measured using our total combine cube. To be
conservative, we also define {\bf statistical size errors} such that
the negative and positive errors are each the maximum of the difference
between the median bootstrap or total combine, and the 16$^{th}$ and
84$^{th}$ bootstrap percentiles respectively.

Sizes based on the bootstrap cubes do not account for any uncertainty
on the PSF. Therefore in Section~\ref{sec:sizeerrorspsf} we assess the
impact of PSF uncertainty on galaxy size measurements, assigning a
minimum size error based on the uncertainty due to the PSF.  Our {\bf
  final size errors} are the maximum of the statistical errors, and
this minimum error based on the PSF uncertainty.

In the following discussion we shall examine our bootstrap, statistical
and final errors. Our final results use final errors in all cases.

\subsection{Errors on Galaxy Sizes: Continuum}\label{sec:sizeerrorscont}

\begin{figure*}
  \centering
  \includegraphics[width=1.00\textwidth]{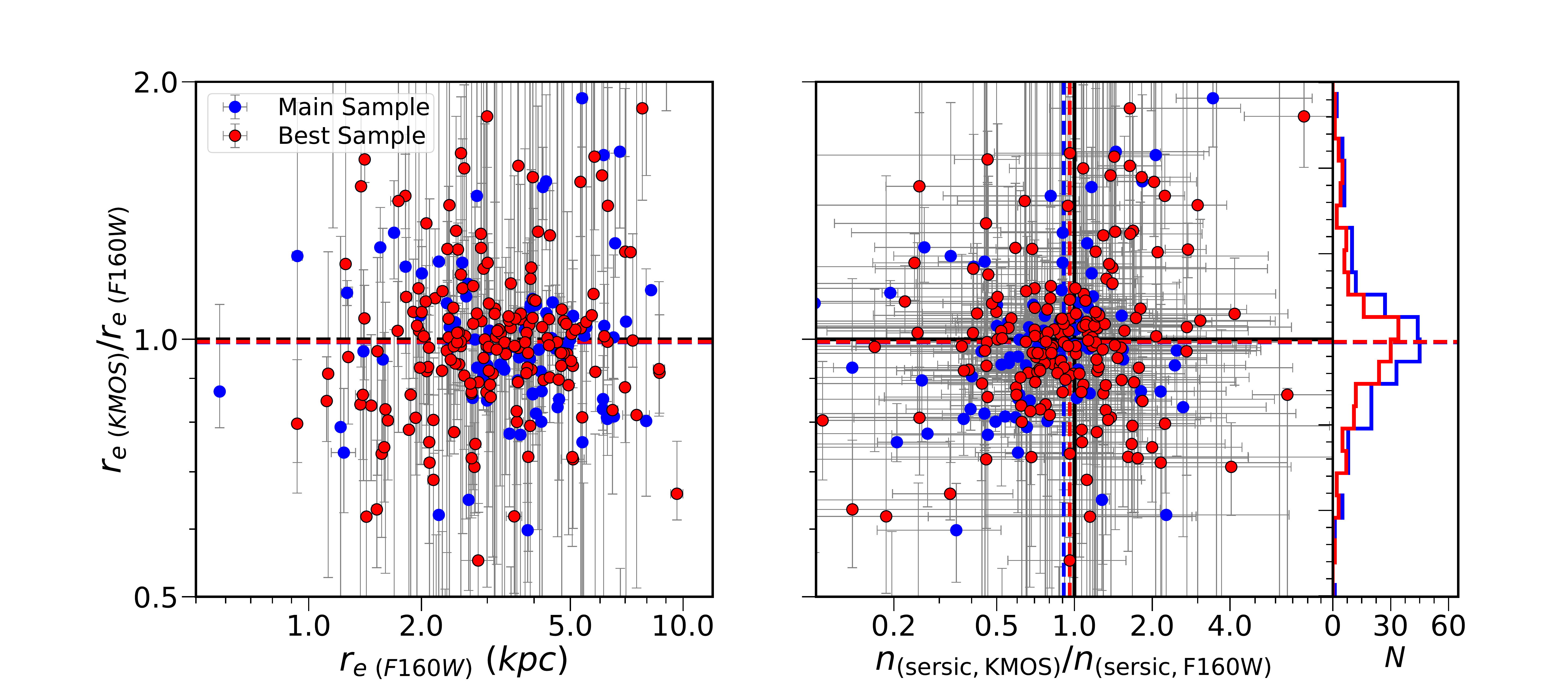}
  \caption{Left: Ratio of KMOS continuum based effective radius
    ($\reff$) to that from CANDELS (F160W), as a function of the
    CANDELS F160W $\reff$ for galaxies in the continuum MAIN (blue) and
    BEST (red) samples, and its distribution (narrow panel to right).
    Errors on the size ratio are derived from bootstrap errors on KMOS
    continuum sizes. The median ratio is 0.99 for both samples (dashed
    horizontal lines) and does not notably depend upon the galaxy size,
    demonstrating the well recovered continuum sizes of galaxies from
    KMOS data. \newline Right: Compares the
    ratio of parameters -- size ($\reff$) and Sersic index ($\nsers$)
    -- from fitting to the KMOS continuum data to those from fitting
    the CANDELS F160W-band images, with bootstrap error bars, and their
    medians (horizontal and vertical dashed lines).  Both parameters
    are well reproduced with fits to the KMOS continuuum data.
    Galaxies are divided into the continuum MAIN and BEST samples.  }
  \label{fig:fitskmoscont}
\end{figure*}

We now examine the accuracy of KMOS continuum sizes, assessed via
comparison with the higher resolution sizes measured on CANDELS WFC3
data.  In the left panel of Figure~\ref{fig:fitskmoscont}, we find that
the galaxy sizes fit to the KMOS continuum image are equivalent to
those from fits to the higher resolution and signal to noise CANDELS
F160W-band images, with a median offset of just $\rm 1\%$ for both
continuum MAIN and BEST samples, and no apparent dependence on galaxy
size. The right panel extends this to the joint parameter space of size
and Sersic index: Sersic indices are also compatible, with median
offsets of $\rm <10\%$. 

\begin{figure*}
  \centerline{\includegraphics[width=0.95\textwidth]{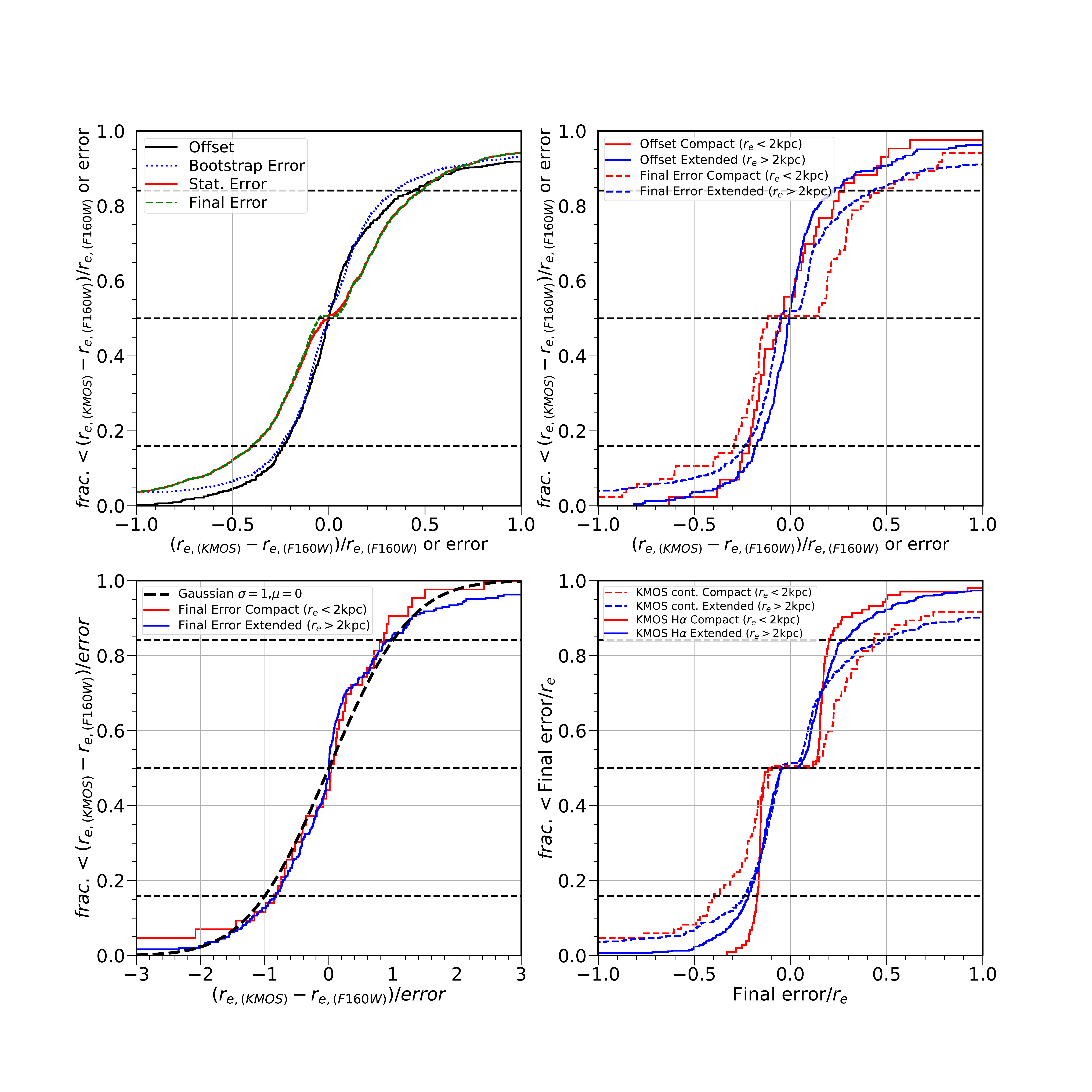}}
  \caption{ {\bf Top left:} Cumulative distribution of the relative
    difference (offset) between KMOS continuum and CANDELS F160W based size
    measurements ($\rm \frac{\reff(KMOS)-\reff(F160W)}{\reff(F160W)}$,
    solid black line) from the continuum MAIN
    sample.  This is compared to the cumulative
    distribution of the expected size error distribution
    for these galaxies: we show bootstrap (dotted), statistical (dashed)
    and final (solid) error distributions (see
    Section~\ref{sec:sizeerrorsdef} for error definitions). 
    {\bf Top right:} Offsets and final error cumulative distributions, divided into compact (red) and extended (blue) sources
    at $\rm \reff=2\kpc$. 
    Remarkably, fractional
    offsets are similar for compact and extended galaxies, and while
    the final error (which accounts for PSF uncertainty) inevitably
    gives larger fractional errors for compact galaxies, this converges
    to similar values for larger fractional errors. \newline 
    {\bf Bottom Left:} Cumulative distribution of the error-normalized size
    offset: $\rm
    \frac{\reff(KMOS)-\reff(F160W)}{\sigma(\reff(KMOS)-\reff(F160W))}$
    again divided into compact and extended sources at $\rm \reff=2\kpc$. 
    Galaxies offset negatively from their F160W
    sizes are normalized by the negative error and vice versa.
    With accurate errors, this should descibe a normal
    distribution with a mean of 0 and scatter of 1, the cumulative
    version of which is the equivalent error function (black
    dashed-line). This provides a remarkably good match to both compact
    and extended sources, indicating that our final errors are
    accurate, not a function of size, and applicable to galaxies
    sampled with KMOS resolution with our best guess KMOS PSF.
    {\bf Bottom Right:} Cumulative distribution of fractional (final) size errors 
    for KMOS continuum and $\ha$, selected respectively from the continuum MAIN and $\ha$
    MAIN samples, and divided into compact and extended galaxies at
    $\rm \reff=2\kpc$ (independently for KMOS continuum and $\ha$
    measurements). While the relative
    errors on KMOS continuum sizes are slightly larger for compact
    galaxies, the opposite is true for the $\ha$
    case: compact galaxies have smaller relative errors on average.
    }
            \label{fig:fitserrors}
\end{figure*}

Figure~\ref{fig:fitserrors} examines the distribution of these
size offsets compared to our derived errors. In
the top-left panel we show the cumulative distribution of size offsets
normalized by the size itself ($\rm
\frac{\reff(KMOS)-\reff(F160W)}{\reff(F160W)}$) and of our estimated
size errors, also normalized by size. We measure more small fractional
offsets in size than would be predicted by our measurement errors, 
suggesting some errors are slightly overestimated. This is mostly due to the difference between bootstrap and statistical errors -- i.e. to account for differences between the median bootstrap realisation and the best estimate.
The top-right panel examines this fractional (final) error and offset distribution separately for the most
compact galaxies ($\reff < 2\kpc$) and for more extended galaxies, demonstrating little difference in the accuracy of sizes or of size errors, perhaps because compact sources tend to be
brighter, compensating for the lack of resolution with higher signal to
noise.
In the bottom-left panel
the size offsets are normalized by the size error. Especially using our
final errors, this describes something very close to an error function
with a dispersion of 1 and median of 0, as would be expected in the
case that the errors are accurate. Therefore we consider our size
errors (at least for KMOS continuum sizes) to be well calibrated.

\subsection{Errors on Galaxy Sizes: $\ha$}\label{sec:sizeerrorsha}

The bottom-right panel of Figure~\ref{fig:fitserrors} now examines the errors for both KMOS
continuum sizes and KMOS-based $\ha$ sizes.
This shows the cumulative distribution of fractional errors on KMOS
continuum and $\ha$ sizes, from the respective MAIN samples and divided
into compact and extended sub-samples (at $2\kpc$). 

While the error distribution for KMOS continuum and $\ha$ is very
similar for extended sources, the compact $\ha$ galaxies have
significantly smaller errors than the compact KMOS continuum galaxies.

We have shown that the errors on
the KMOS continuum sizes are well described by (systematic and random)
signal to noise variations as traced primarily by the bootstrap errors.
Figure~\ref{fig:fitserrors} demonstrates that the compact (typically
high signal-to-noise) $\ha$ sources can have very small size errors.
This motivates a more thorough examination of the effects of
uncertainty on the PSF in Section~\ref{sec:sizeerrorspsf}, resulting in
the final size errors. We note here, based on a comparison of bootstrap
/ statistical, and final errors in Figure~\ref{fig:fitserrors}, that
this inflates the smallest size errors for compact galaxies, but that
there are nonetheless fewer compact galaxies with large fractional size
errors in $\ha$ than for extended galaxies or for KMOS continuum,
implying that this is robust and likely a consequence of the high
signal to noise data. We also tested the impact of our choice of 
a flat extension of the velocity field beyond the regions where we trust 
kinematic fits in individual spaxels. We modified the velocity in these
extrapolated regions to values 25\% above and below the nominal values.
Thsi simulates either declining rotation curves at large radii \citep{Lang17},
or rotation curves that keep raising to larger velocities. The impact
on the final $\ha$ sizes is negligible, with a scatter with respect 
to the best values of 1-2\% in both our tests. This scatter is much smaller
than the individual measurements errors, which are therefore not affected
by the algorithm used to extend the velocity fields.

\subsection{Effects of PSF Uncertainty on Size Errors}\label{sec:sizeerrorspsf}

As described in Section~\ref{sec:fitimage}, the generation of PSF
images for each combined cube relies on our ability to accurately
measure the shifts between exposures, acquisitions and setups, as
characterised via the simultaneously observed stars, and shifts between
partial combines.  An uncertainty on the PSF (e.g. from arm positioning
errors not accounted for via shifts between partial combines)
translates to a limit to our effective size or size error estimates
which is not accounted for by the bootstrap errors, but which must be
smaller than the typical error on the KMOS continuum size (given those
bootstrap errors are large enough to describe the offset from CANDELS
based sizes, Section~\ref{sec:sizeerrorscont}).

\begin{figure}
  \centerline{\includegraphics[width=0.5\textwidth]{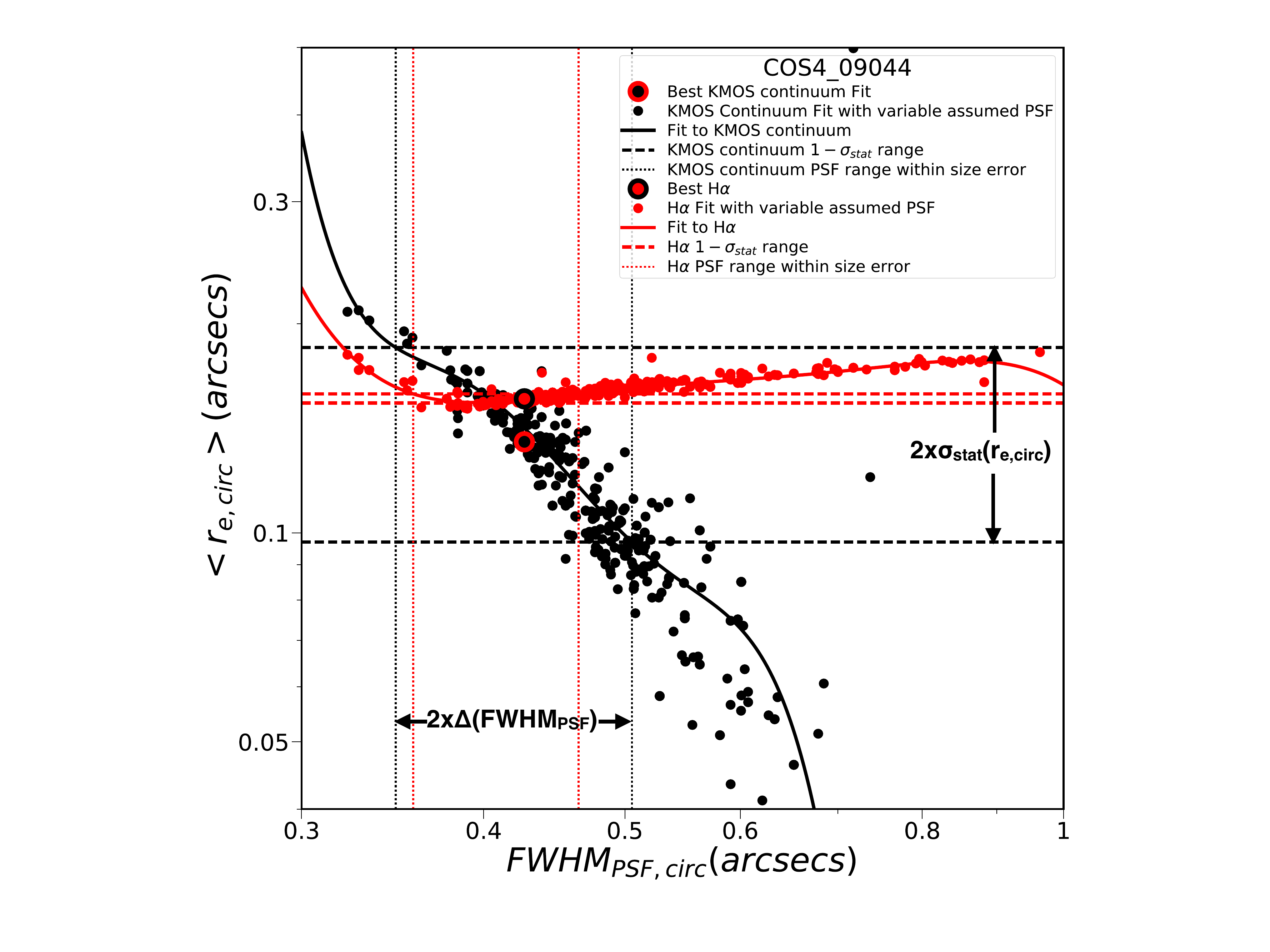}}
  \caption{Best fit circularized effective radius, $<\reffcirc>$
    expressed in angular units, for compact galaxy COS4\_09044 in KMOS
    continuum (black) and $\ha$ (red) as a function of the FWHM of the
    circularized PSF ($\fwhmpsfcirc$) which is assumed for the fit (to
    the same data). The fit with the best guess PSF for this galaxy is
    indicated with the larger, outlined datapoints. With a larger PSF,
    the best fit size to KMOS continuum data decreases to compensate.
    This is representative of the typical behaviour for most galaxies.
    Only if the assumed PSF is well away from our best guess will the
    fit to the KMOS continuum data provide a best fit galaxy size
    outside the range allowed by the statistical errors from bootstrap
    cubes (dashed horizontal black lines) which samples random and
    systematic variations in the input data. We fit the trend with a
    fifth order polynomial including sigma-clipping (black solid line)
    and measure the range of assumed $\fwhmpsfcirc$ accommodated within
    the statistical errors (black dotted vertical lines, the separation
    of which is defined to be $2 \times \deltapsfcirc$). The $\ha$
    fits, in contrast, provide a very stable circularized galaxy size
    {\it almost independent of bootstrap iteration} (red dashed
    horizontal lines) {\it or of the assumed PSF} (lack of variation
    with $\fwhmpsfcirc$, fit indicated as the red solid line). This is
    typical of high signal to noise $\ha$ data for compact galaxies
    ($\reff (\ha) \lesssim 0.25 \arcsec$ or 1.25 pixels) as seen in
    Figure~\ref{fig:fitpsfeffect}. }
  \label{fig:fitpsf_COS4_09044}
\end{figure}

To examine how the size estimates are sensitive to the assumed PSF, we
refit every galaxy in the sample with the PSF (image) as computed for all of the other galaxies in the sample. This
covers a much broader range of PSF than any realistic, residual error
on the true PSF. Figure~\ref{fig:fitpsf_COS4_09044} shows the resultant
variation of best fit galaxy half-light radius with assumed PSF
(circularized) FWHM for one of our compact galaxies, COS4\_09044. The
dependence on the size of the PSF is most closely related to the
circularised size, $\reffcirc$, expressed in angular units
(arcseconds)\footnote{Circularized sizes are defined as $\rm \reffcirc
  = \reff.\sqrt{1-\epsilon}$, $\rm \fwhmpsfcirc =
  \fwhmpsf.\sqrt{1-\epsilon_{PSF}}$.}.

Focusing on the results for the fit to the KMOS continuum (black
points), we see the expected trend as typically seen for most galaxies
(including more extended ones): as a larger PSF is assumed, the fitting
procedure compensates such that the best fit intrinsic galaxy size is
more compact. To illustrate the range of assumed PSF which can be
accommodated within the $\rm 1-\sigma$ size range derived using the
statistical errors from the bootstrap cubes dashed black horizontal
lines -- i.e. the range of PSF for which the statistical errors dominate over any size error induced by the assumption of the incorrect PSF -- we fit the dependence of predicted galaxy size on assumed PSF
FWHM with a fifth order polynomial (green solid line, fit is iterative
with sigma-clipping) and determine the range of PSF FWHM for which this
fit lies within the statistical errors on size (dotted vertical black
lines), with a full range $\rm 2 \times \deltapsfcirc$.

\begin{figure}
  \centerline{\includegraphics[width=0.5\textwidth]{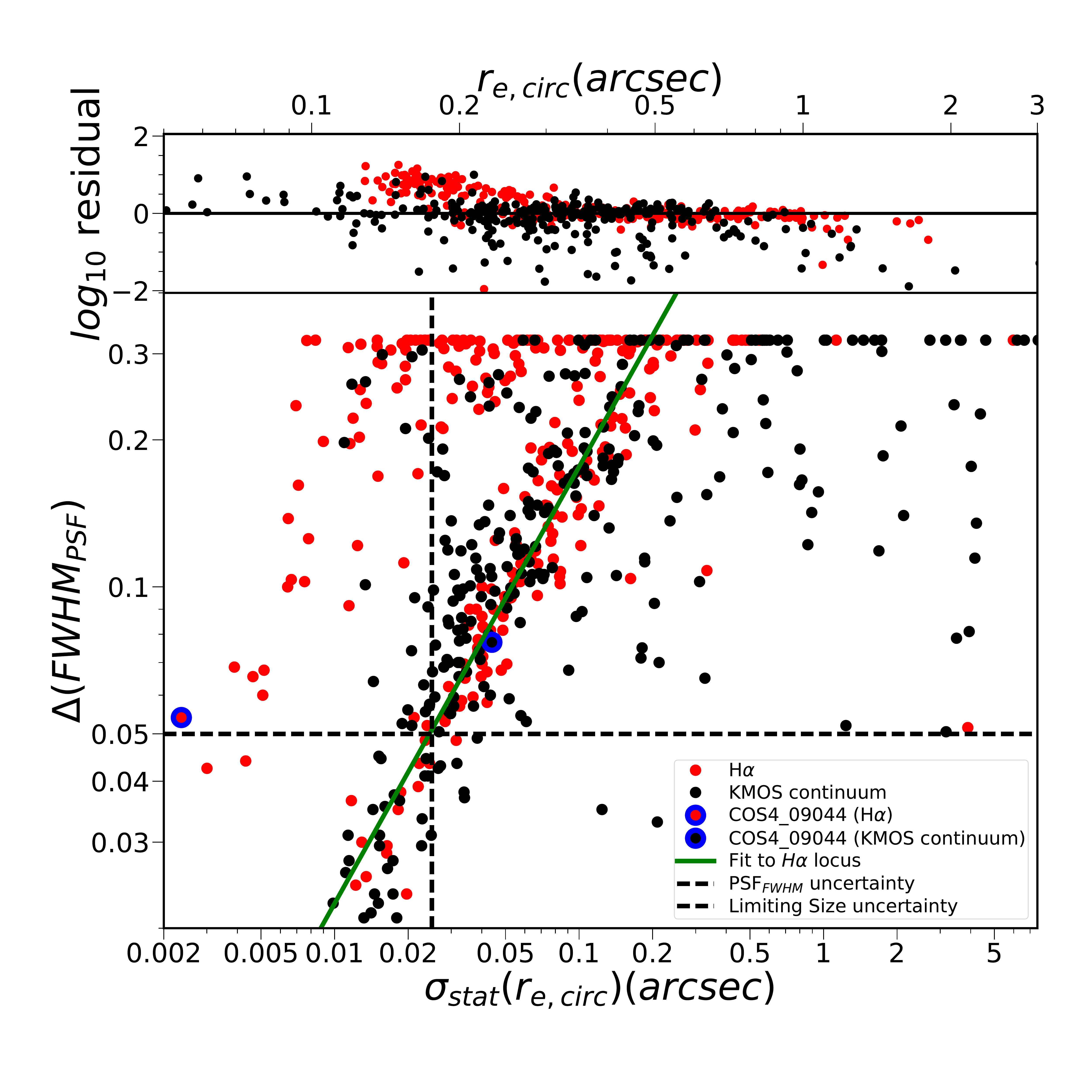}}
  \caption{Each galaxy is fit with the full range of assumed PSF, as
    described in Section~\ref{sec:sizeconsistency} and illustrated by
    Figure~\ref{fig:fitpsf_COS4_09044}. The {\it statistical} error on
    galaxy size, $\sigmastatreff$ is estimated using the fits to
    bootstrap cubes. Here the statistical error on circularized galaxy
    size, $\sigmastatreffcirc$, is shown plotted aginst the half-range
    of PSF circularized FWHM, $\deltapsfcirc$, for which the best fit
    galaxy size lies within the statistical error (i.e. the range of
    PSF error for which the statistical error dominates the error on
    the assumed PSF). All galaxies in the continuum MAIN (black, KMOS
    continuum) and $\ha$ MAIN (red) are shown. The vast majority of
    galaxies in KMOS continuum, and many in $\ha$, lie along a locus of
    decreasing $\deltapsfcirc$ with decreasing $\sigmastatreffcirc$
    (solid green line) -- such that for very small statistical size
    errors the error on assumed PSF can dominate. As described in the
    main text we assume a conservative error on the assumed PSF of
    $\deltapsfcirc=0.05\arcsec$ (horizontal dashed black line), for
    which the error is matched by a statistical error of
    $\sigmastatreffcirc=0.025\arcsec$, or $\frac{1}{8}$ of a KMOS pixel
    (vertical dashed black line). At very compact sizes, $\reffcirc
    \lesssim 2.5\arcsec$, the statistical error on $\ha$ sizes becomes
    very small, and the sensitivity to $\fwhmpsfcirc$ becomes quite
    flat, as seen in Figure~\ref{fig:fitpsf_COS4_09044} for
    COS4\_09044. These galaxies are to the left of the main locus of
    points in this Figure, and are clearly shown with positive
    residuals in the upper panel in which we show the residual of
    $\deltapsfcirc$ with respect to the $\ha$ locus versus galaxy size.
    We choose to apply a conservative lower limit to the circularized
    galaxy size error, $\sigmareffcirc \geq 0.025\arcsec$,
    corresponding to the vertical dashed line.}
  \label{fig:fitpsfeffect}
\end{figure}

In Figure~\ref{fig:fitpsfeffect} we show how $\deltapsfcirc$ depends
upon the statistical error range of the galaxy size
($\sigmastatreffcirc$ in arcseconds, the average of positive and
negative errors ). $\deltapsfcirc$ saturates at a maximum value,
corresponding to the full range of $\fwhmpsfcirc$. For smaller size
errors, there exist some range of assumed PSF which would drive the
galaxy size outside the statistical error: the range of PSF consistent with the measured size and statistical errors
becomes smaller with decreasing statistical error along a locus of
slope $\rm \sim 0.9$ (a power law on linear scales with exponent 0.9).
Repeating the exercise for $\ha$ fits (red points in
Figures~\ref{fig:fitpsf_COS4_09044} and~\ref{fig:fitpsfeffect}) we see
that for most cases (in particular for galaxy sizes $\rm \reff\gtrsim
0.25\arcsec$) the data follow the same trend, for which we perform a
linear fit (green solid line in Figure~\ref{fig:fitpsfeffect}).

We estimate that there can exist a {\it maximum} residual error of $\rm
\sim 1$pixel $\rm = 0.2\arcsec$ on $\fwhmpsfcirc$, as a result of
uncertain manual shifts between partial combines, and residual errors
after the average exposure to exposure shift of PSF stars has been
removed. This sets a conservative upper limit on the error on
$\fwhmpsfcirc$ for a compact PSF ($\rm \fwhmpsfcirc \sim 0.4\arcsec$)
of $\rm \sim \sqrt{ (0.4\arcsec)^2 + (0.2\arcsec)^2} - 0.4\arcsec \sim
0.05\arcsec $. We assume that where $\rm \deltapsfcirc < 0.05\arcsec$
(horizontal dashed black line, Figure~\ref{fig:fitpsfeffect}) the PSF
error overrides the statistical error: this sets a lower limit on the
final error on the size at the point where our locus of points (green
line) intersects this limiting value of $\deltapsfcirc$, such that our
final error on circularized galaxy size, $\sigmareffcirc \geq
0.025\arcsec$ (vertical dashed black line, $\rm =\frac{1}{8}$ of a
pixel).

For the fits to compact galaxies in $\ha$ ($\rm \reff < 0.25\arcsec$),
the statistical errors on size from the bootstraps are often very
small, $\rm \sigmastatreffcirc < 0.01 \arcsec$ or $< 0.05$ pixels and
with a much shallower dependence of best fit size on assumed PSF (and
thus larger value of $\deltapsfcirc$ at fixed $\sigmastatreffcirc$). A
good example is COS4\_09044, highlit in blue in
Figure~\ref{fig:fitpsfeffect}, for which the $\ha$ fit provides a
extremely consistent size almost independent of the bootstrap iteration
or the assumed $\fwhmpsfcirc$ (red points in
Figure~\ref{fig:fitpsf_COS4_09044}). Such galaxies are high signal to
noise and high surface brightness (surface brightness increases with
decreasing size), but the best correlation is with size: the residuals
with respect to the main locus in Figure~\ref{fig:fitpsfeffect} are
plotted against galaxy size in the upper panel. We conservatively set
the errors of all such fits to our adopted minimum value of $\rm
\sigmareffcirc \geq 0.025\arcsec$ (vertical dashed black line). Most
galaxies, especially in the case of KMOS continuum have larger
statistical errors and so these error estimates remain effectively
consistent with the differences between KMOS continuum and CANDELS
sizes. We now define the {\bf final size errors} on (major axis) sizes,
$\sigmareff$, to be the maximum of the statistical error
and the uncertainty on size due to the PSF uncertainty of $\rm 0.025\arcsec \times \sqrt{1-\epsilon} \times D_A$ where $D_A$ is the angular diameter distance at the redshift of the galaxy in $\rm kpc~arcsec^{-1}$.

Finally we note that there are no fits to $\ha$ data which are flagged
as OK for which best fit sizes are below $\reff \sim 0.69$ pixels.
This seems to be the real limit for galaxy $\ha$ sizes in our sample as
there is no reason that smaller sizes should be flagged as bad (even if
perfectly described by the PSF). This corresponds to a minimum physical
size of $\reff \sim 1.1~\kpc$, very similar to the minimum size from
CANDELS F160W continuum imaging (Figure~\ref{fig:fitsf160w}). In
contrast, a few KMOS continuum sizes reach to both much lower and
larger sizes (as seen in Figure~\ref{fig:fitsf160w}): these can be
explained as outliers, and are mostly consistent with the tails in the
difference between KMOS continuum and CANDELS F160W sizes, normalized
by the size errors: i.e.  they are mostly expected
given the errors, with a few possible exceptions.

\section{Examples of H$\alpha$ profiles and exponential fits}\label{sec:profilesgallery}
\begin{figure*}
\centerline{
  \includegraphics[width=0.33\textwidth]{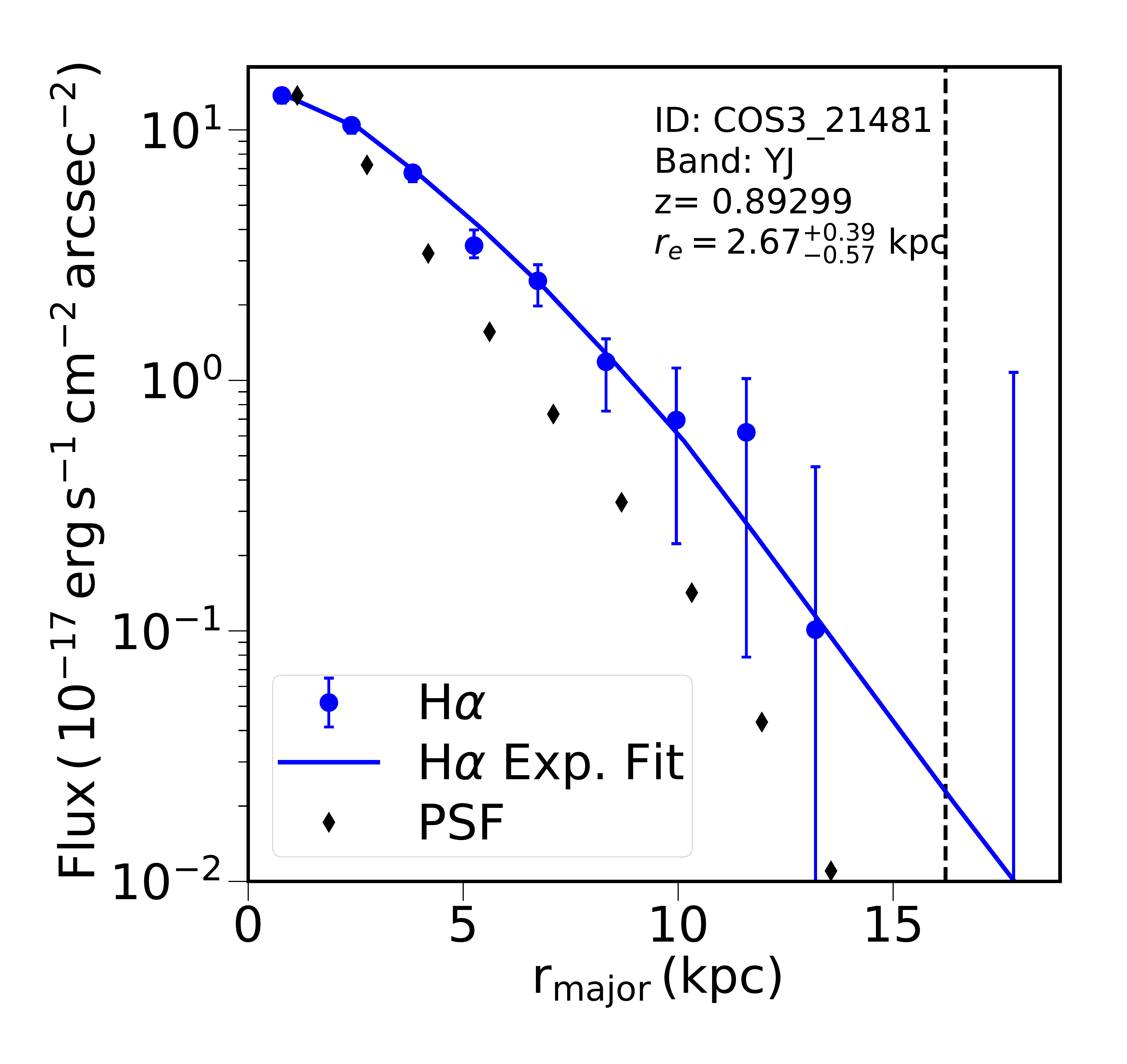}
  \includegraphics[width=0.33\textwidth]{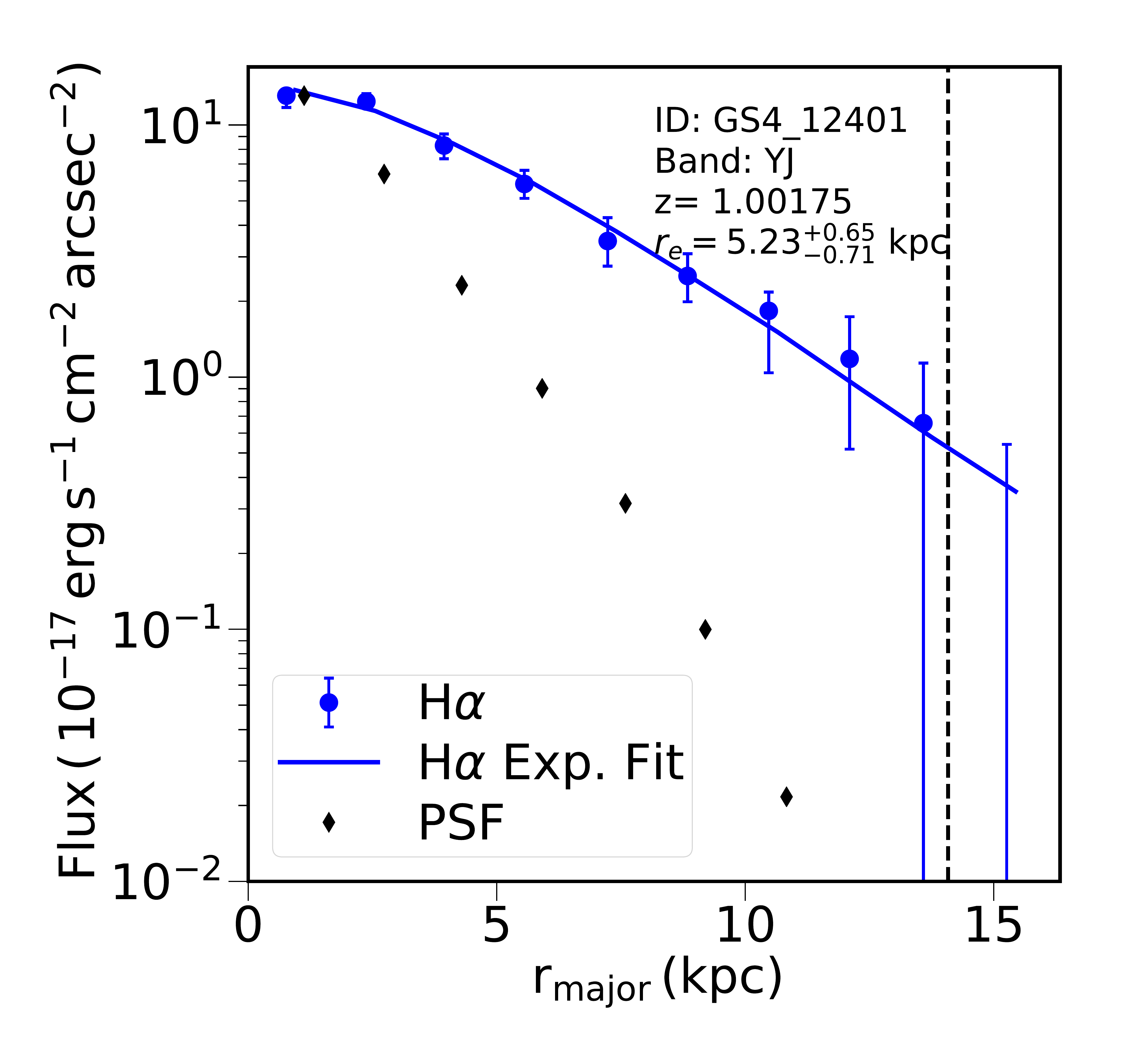}
 \includegraphics[width=0.33\textwidth]{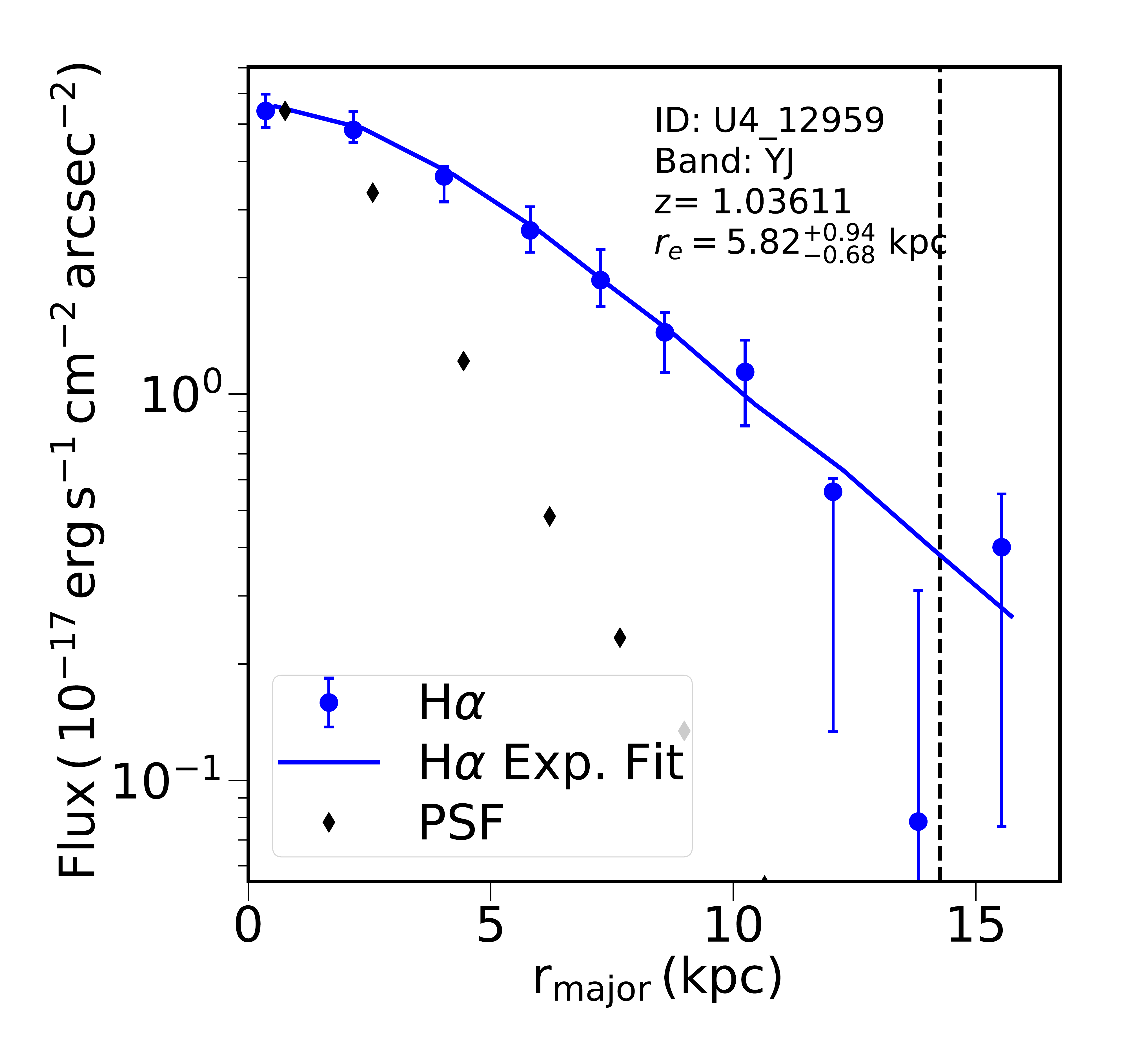}}
\centerline{
  \includegraphics[width=0.33\textwidth]{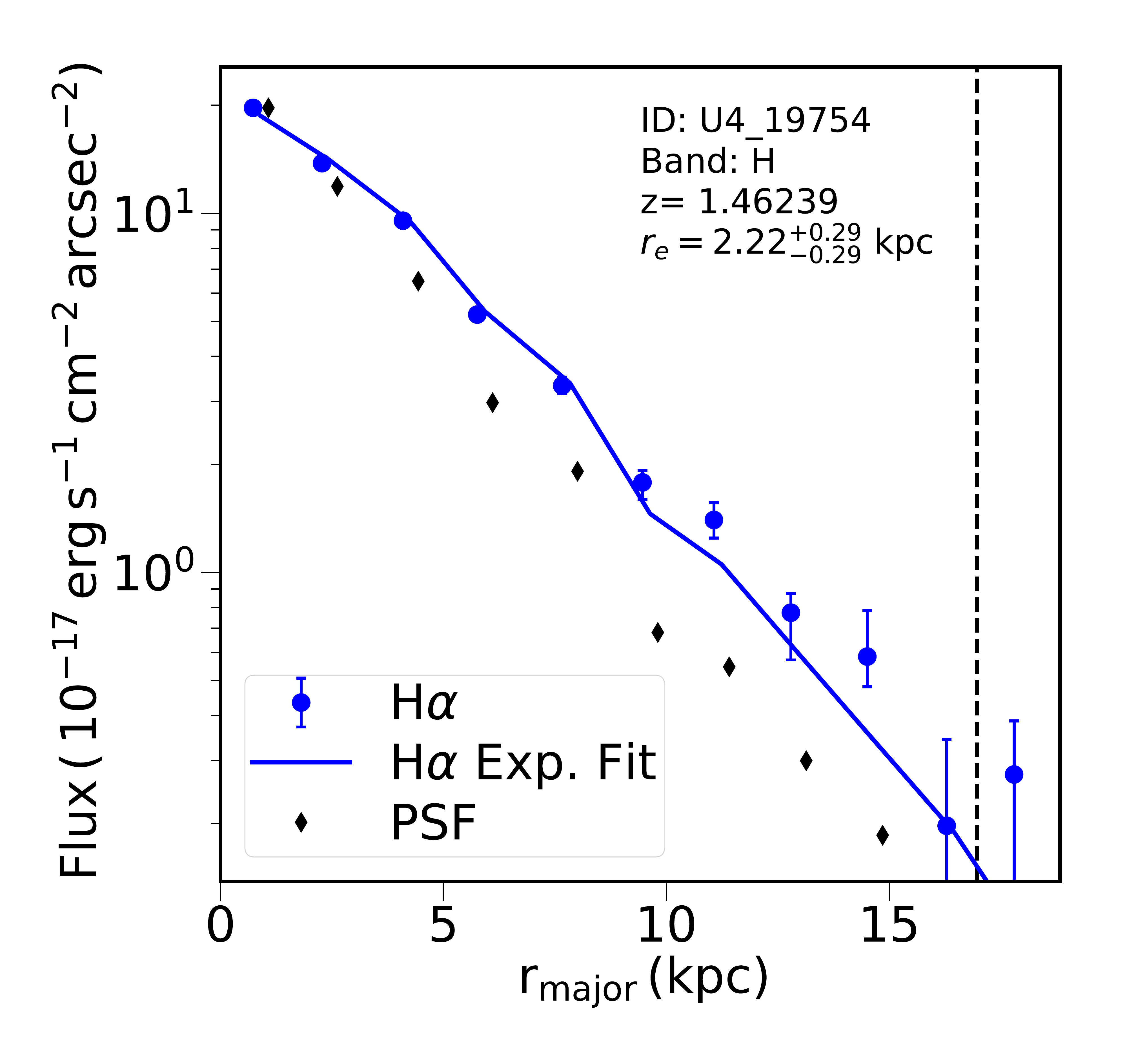}
  \includegraphics[width=0.33\textwidth]{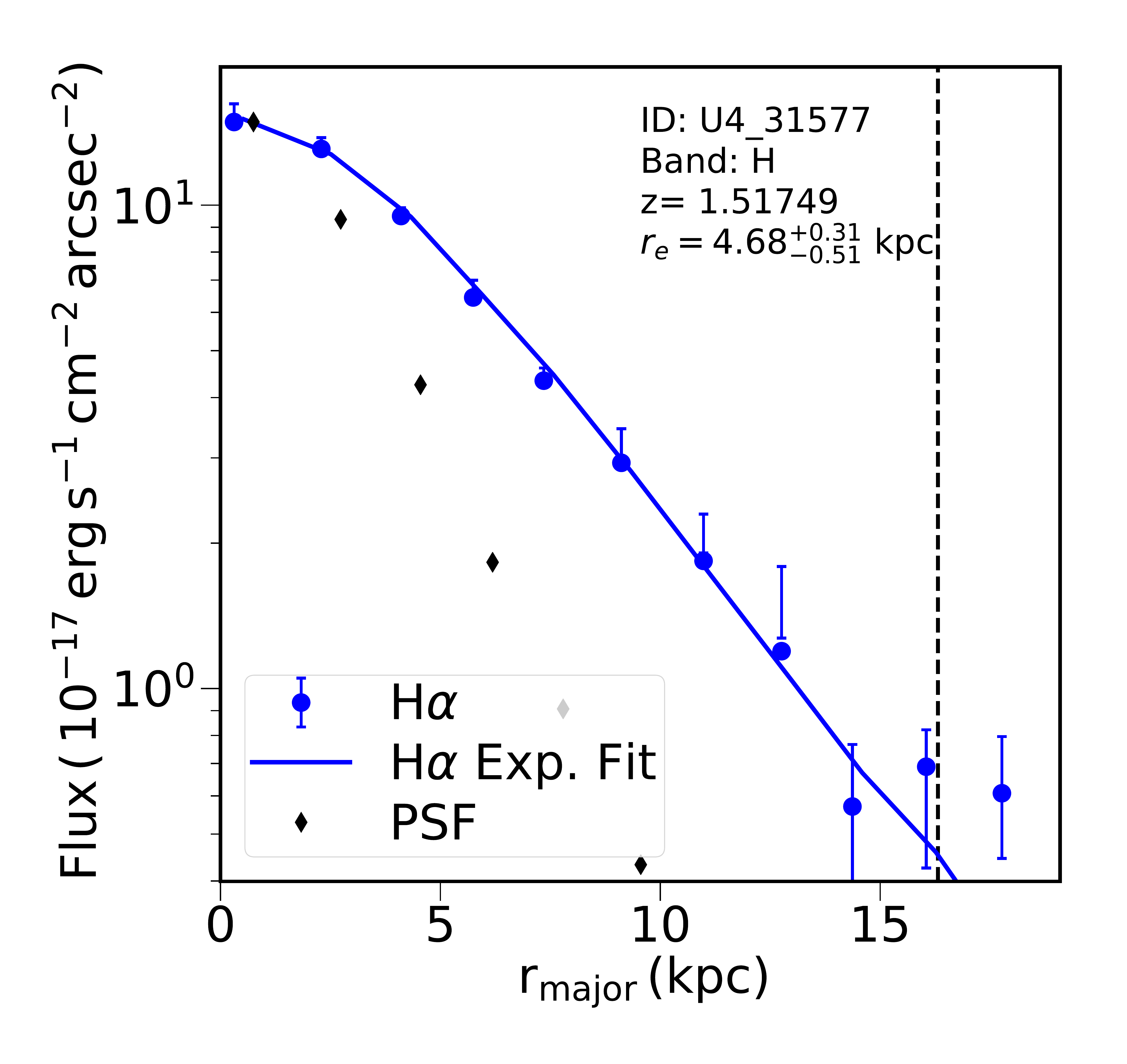}
 \includegraphics[width=0.33\textwidth]{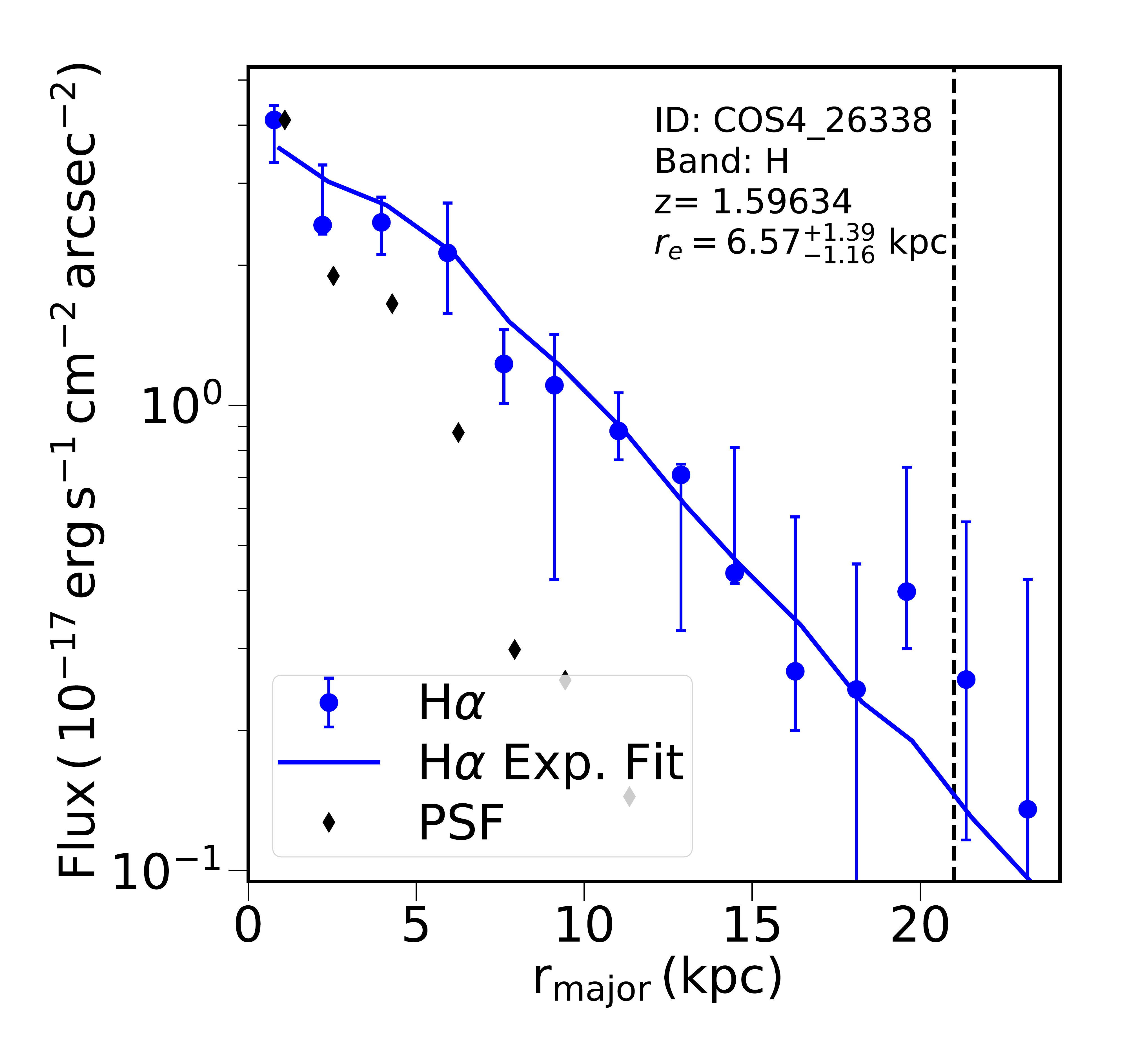}}
 \centerline{
  \includegraphics[width=0.33\textwidth]{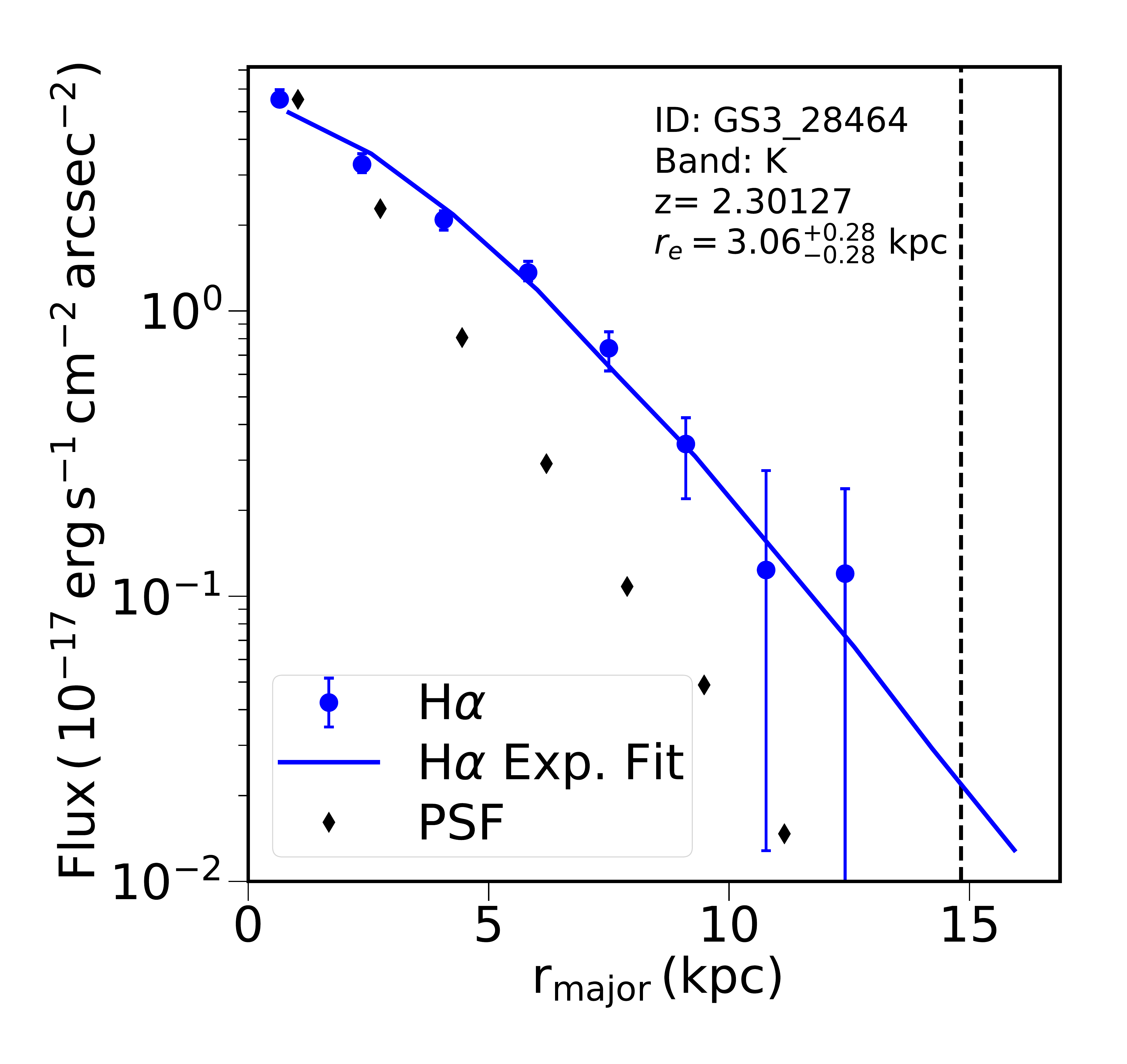}
  \includegraphics[width=0.33\textwidth]{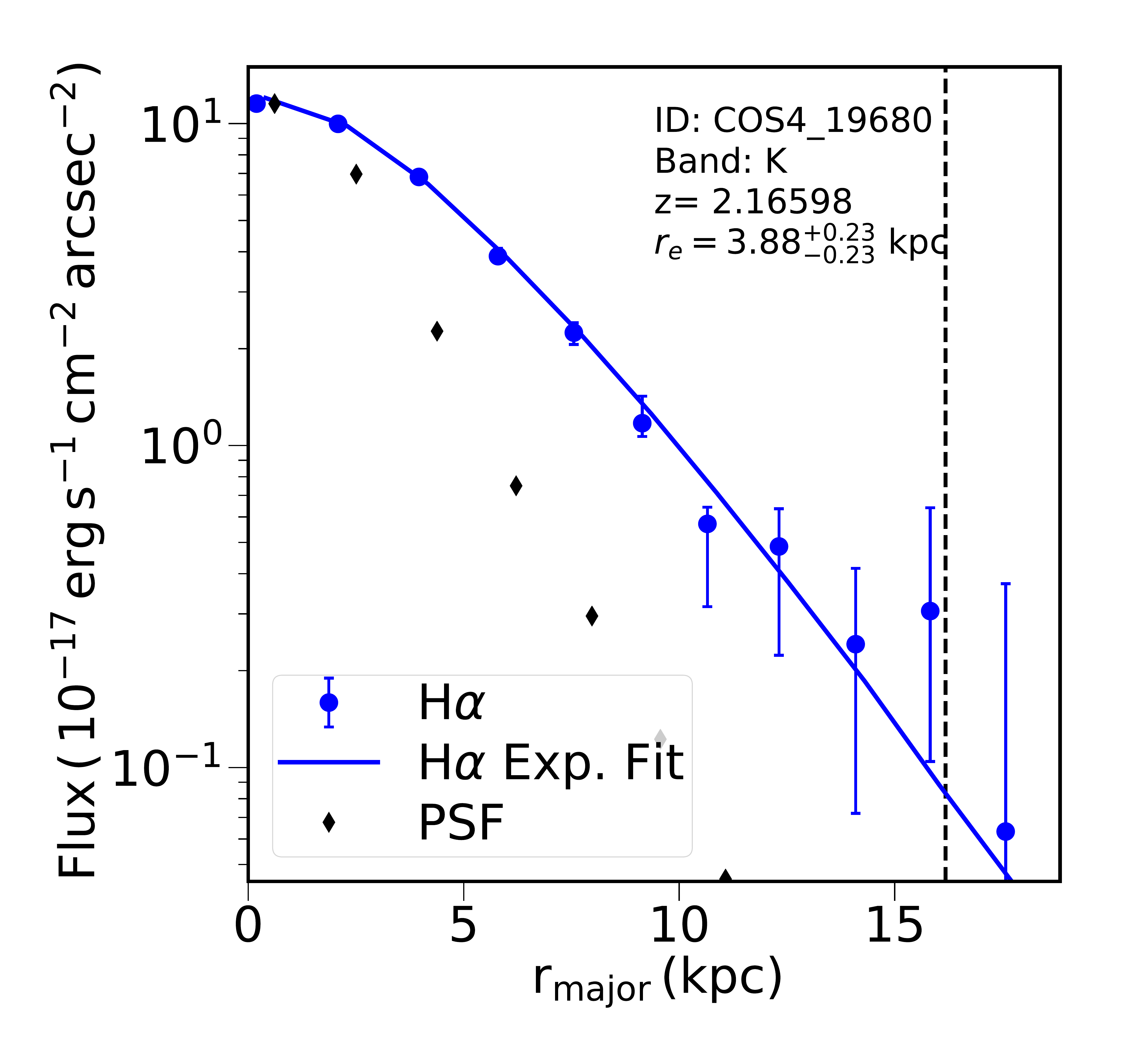}
 \includegraphics[width=0.33\textwidth]{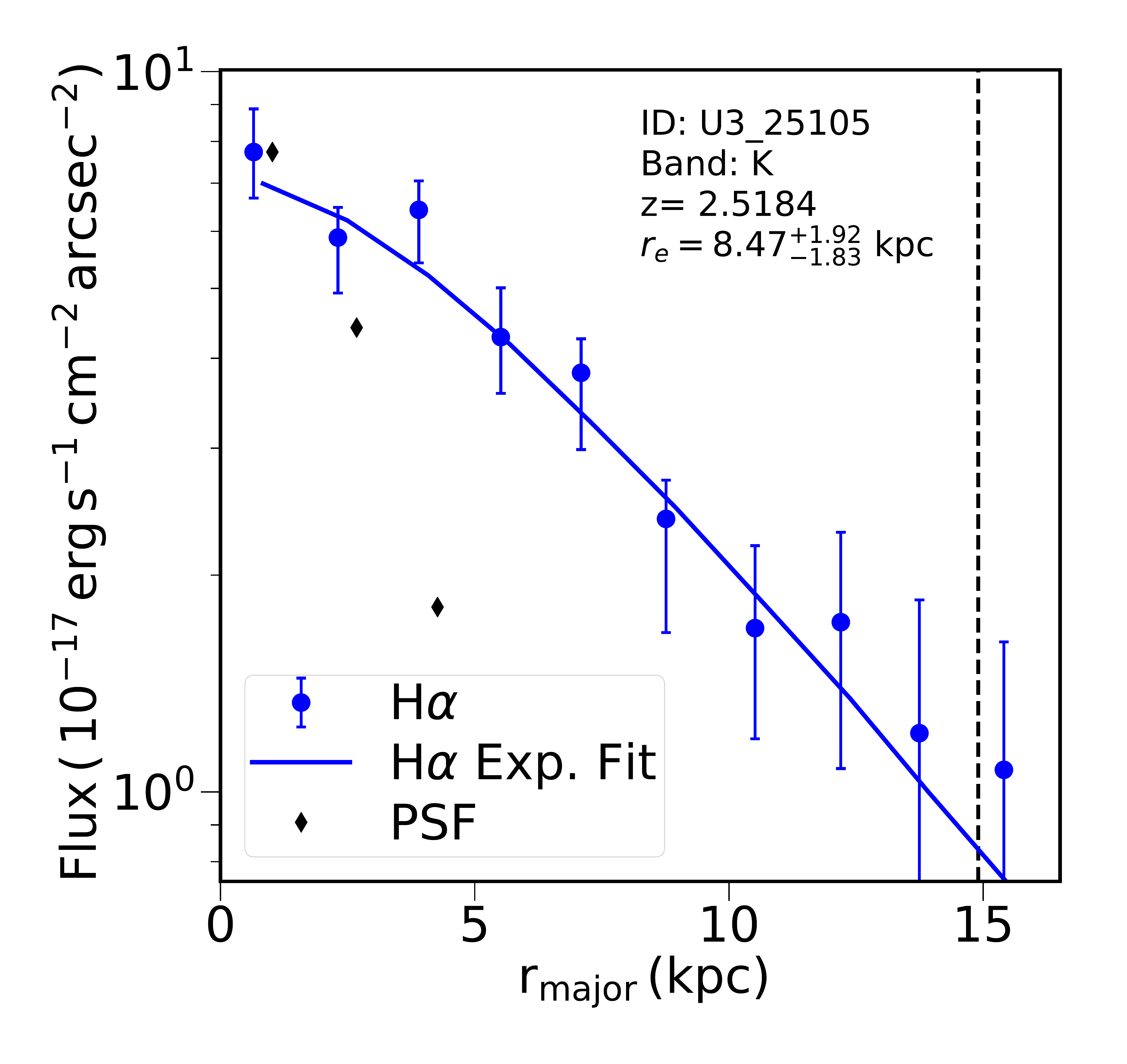}}
\caption{Gallery of radial H$\alpha$ profiles extracted using elliptical apertures from our KMOS data (blue points with $1\sigma$ bootstrap errors) as described in Section \ref{sec:maps}. The blue solid lines show the 1D profiles of the best fit 2D exponential model. For comparison the PSF image is also extracted in the same apertures (black diamonds). The vertical dashed line indicates the radius where the major axis first crosses the edge of the KMOS field of view. The galaxies have increasing effective radii from the left to the right column, while the galaxy redshift increases from top to bottom. }
\label{fig:profilesgallery}
\end{figure*}

In Figure \ref{fig:profilesgallery} we show a gallery of H$\alpha$ profiles from the MAIN sample spanning a range of size, redshift and observed surface brightness, to show the quality of the data and of the fitting procedure. 

\section{Public Release of Size measurements}\label{sec:publicdata}

The size measurements of 281 galaxies in the Main Sample are made available as a Machine Readable Table. Table \ref{releasetableexample} gives an example of the quantities provided with this work and the description of the columns:

\begin{table*}
\begin{center}
\begin{tabular}{c c c c c c c c}
    \hline
       KMOS$^{\rm{3D}}$ ID & KMOS$^{\rm{3D}}$ ID TARGETED & $z_{\rm spec}$ &  $r_{e\ (F160W)}$ & $\sigma(r_{e\ (F160W)})$ & $r_{e\ (H\alpha)}$ & $\sigma_{\rm neg.}(r_{e\ (H\alpha)})$ & $\sigma_{\rm pos.}(r_{e\ (H\alpha)})$  \\
       & & & (kpc) & (kpc) & (kpc) & (kpc) & (kpc)\\
 (1) & (2) & (3) & (4) & (5) & (6) & (7) & (8) \\
    \hline
        COS4\_06327 & COS3\_06511 & 0.80364 & 2.458 & 0.034 & 3.803 & 0.453 & 0.842 \\
        GS4\_34568 & GS4\_34568 & 2.57255 & 3.845 & 0.071 & 5.125 & 0.612 & 0.940 \\
        U4\_09733 & U4\_09733 & 2.28886 & 3.931 & 0.092 & 11.243 & 1.051 & 1.170 \\
        ..... \\
    \hline
  \end{tabular}
  \caption{Example of the size measurements table made available with this work.}
  \label{releasetableexample}
  \vspace{0.1cm}

\begin{scriptsize}
\begin{minipage}{18cm}
{\sc Notes.}
\begin{itemize}
\itemsep0em
\item (1) KMOS$^{\rm 3D}$ ID : Object ID as defined in the data release \blue{(W19)}.
\item (2) KMOS$^{\rm 3D}$ ID TARGETED : the object ID that defined the target at the time of the observations as defined in \blue{W19}. In this paper we refer to these IDs.
\item (3) Spectroscopic redshift based on $\ktd$ emission line detection.
\item (4) $r_{e\ (F160W)}$ effective radius from WFC3/F160W images in kpc.
\item (5) $\sigma(r_{e\ (F160W)})$ symmetric 1$\sigma$ uncertainty on the effective radius from WFC3/F160W images in kpc.
\item (6) $r_{e\ (H\alpha)}$ effective radius from KMOS H$\alpha$ images in kpc. The flux image is derived as described in equation \ref{eq:hafluxmap}.  
\item (7) $\sigma_{\rm neg.}(r_{e\ (H\alpha)})$ asymmetric negative 1$\sigma$ uncertainty on the effective radius from KMOS H$\alpha$ images in kpc. The final error as defined in Appendix~\ref{sec:sizeconsistency}.
\item (8) $\sigma_{\rm pos.}(r_{e\ (H\alpha)})$ asymmetric positive 1$\sigma$ uncertainty on the effective radius from KMOS H$\alpha$ images in kpc. The final error as defined in Appendix~\ref{sec:sizeconsistency}. 
\end{itemize}
\label{table:releasetable}
\end{minipage}
\end{scriptsize}

\end{center}
\end{table*}

\end{document}